\documentclass[11pt]{article}
\pdfoutput=1
\usepackage{jcapmod} 
\usepackage{booktabs} 
\usepackage[english]{babel}
\usepackage{amsmath,amssymb,amsbsy,amstext, amsthm, simplewick}
\usepackage{wrapfig}
\usepackage{hyperref}
\usepackage{graphicx}
\usepackage{amsfonts}
\usepackage{amssymb}
\usepackage{subfig}
\usepackage{enumitem}
\usepackage{upgreek}
 \usepackage{exscale,relsize}
 \usepackage[makeroom]{cancel}
\usepackage{soul}
\usepackage{bbold}
\usepackage{lipsum}
\usepackage{mdframed}
\usepackage{mathtools}
\usepackage{tikz-cd}
\usepackage[export]{adjustbox}
\usepackage[outdir=./]{epstopdf}
\usepackage{placeins}
\usepackage{afterpage}
\usepackage{float}

\usetikzlibrary{shapes,arrows}

\RequirePackage{color}

\usepackage{colortbl}
\definecolor{rp}{cmyk}{0.2, 1, 0.6, 0}
\definecolor{green2}{cmyk}{0, 1, 0.5, 0}
\definecolor{lightgreen}{cmyk}{0.2, 0, 0.2, 0.2}
\definecolor{lightgray}{cmyk}{0.1,0.2,0,0.1}
\definecolor{lightgray2}{cmyk}{0.4,0.4,0,0.8}
\definecolor{black}{cmyk}{1.0,1.0,1.0,1.0}

\allowdisplaybreaks[1]

\usepackage{colortbl}
\definecolor{lightgreen}{cmyk}{0.2, 0, 0.2, 0.2}
\definecolor{lightgray}{cmyk}{0.1,0.2,0,0.1}
\definecolor{lightgray2}{cmyk}{0.1,0.1,0,0.1}

\makeatletter
\newlength{\apb@width}
\newcommand{\autoparbox}[2][c]{\settowidth{\apb@width}{#2}\parbox[#1]{\apb@width}{#2}}

\newcommand{\Cen}[2]{%
  \ifmeasuring@
    #2%
  \else
    \makebox[\ifcase\expandafter #1\maxcolumn@widths\fi]{$\displaystyle#2$}%
  \fi
}
\makeatother

\numberwithin{equation}{section}

\def\beq{\begin{equation}}
\def\eeq{\end{equation}}
\def\bea{\begin{eqnarray}}
\def\eea{\end{eqnarray}}

\def\Beq{\begin{equation}\begin{aligned}}
\def\Eeq{\end{aligned}\end{equation}}

\def\Nf{N_{\rm f}}

\def\S{{\sf S}}

\def\k{{\bf k}}

\def\M{{\sf M}}
\def\R{{\sf R}}

\def\Nf{{N_{\rm{f}}}}

\def\I{\mathbb{1}}

\def\S{\mathcal{N}_s(\sigma/H)^2}

\DeclareRobustCommand{\SkipTocEntry}[4]{}
\DeclareSymbolFont{extraup}{U}{zavm}{m}{n}
\DeclareMathSymbol{\varheart}{\mathalpha}{extraup}{86}
\DeclareMathSymbol{\vardiamond}{\mathalpha}{extraup}{87}

\begin{document}

\hypersetup{pageanchor=false}

\begin{titlepage}

\setcounter{page}{1} \baselineskip=15.5pt \thispagestyle{empty}

\bigskip\

\vspace{1cm}
\begin{center}

{\fontsize{20.74}{24}\selectfont  \sffamily \bfseries  Stochastic Particle Production \medskip in \\ \medskip a de Sitter Background} 

\end{center}

\vspace{0.2cm}

\begin{center}
{\fontsize{12}{30}\selectfont  Marcos A.~G.~Garcia$^{\clubsuit}$\footnote{marcos.garcia@rice.edu}, Mustafa A.~Amin$^{\spadesuit}$\footnote{mustafa.a.amin@gmail.com},\\ Scott G.~Carlsten$^{\varheart}$, Daniel Green$^{\vardiamond}$}
\end{center}
\begin{center}

\vskip 7pt

\textsl{$^{\clubsuit,\spadesuit}$ Department of Physics \& Astronomy, Rice University, Houston, Texas 77005, USA\\
$^{\varheart}$ Department of Astrophysical Sciences, Princeton University, Princeton, NJ 08544, USA\\
$^{\vardiamond}$ Department of Physics, University of California, San Diego, La Jolla, CA 92093, USA}
\vskip 7pt

\end{center}

\vspace{1.2cm}
\hrule \vspace{0.3cm}
\noindent {\sffamily \bfseries Abstract} \\[0.1cm]
We explore non-adiabatic particle production in a de Sitter universe for a scalar spectator field, by allowing the effective mass $m^2(t)$ of this field and the cosmic time interval between non-adiabatic events to vary stochastically. Two main scenarios are considered depending on the (non-stochastic) mass $M$ of the spectator field: the conformal case with $M^2=2H^2$, and the case of a massless field. We make use of the transfer matrix formalism to parametrize the evolution of the system in terms of the ``occupation number",  and two phases associated with the transfer matrix; these are used to construct the evolution of the spectator field. Assuming short-time interactions approximated by Dirac-delta functions, we numerically track the change of these parameters and the field in all regimes: sub- and super-horizon with weak and strong scattering. {\it In all cases a log-normally distributed field amplitude is observed, and the logarithm of the field amplitude approximately satisfies the properties of a Wiener process outside the horizon.} We derive a Fokker-Planck equation for the evolution of the transfer matrix parameters, which allows us to calculate analytically non-trivial distributions and moments in the weak-scattering limit. 
\vskip 10pt
\hrule
\vskip 10pt

\vspace{0.6cm}
 \end{titlepage}

\hypersetup{pageanchor=true}

\tableofcontents

\newpage

\section{Introduction}

The embedding of the inflationary paradigm within ultraviolet completions of particle theories often involves many fields with potentially complicated interactions that may lead to a chaotic evolution as a function of the initial conditions and values of the model parameters. The presence of such a large number of degrees of freedom can also dramatically complicate the dynamics of post-inflationary reheating.

Although the full deterministic description of such models can be highly model-dependent, one might expect that in the limit of many fields/interactions, emergent universal properties may arise. Moreover, the coarse-grained nature of the available cosmological observations is unlikely to shed light on all the microscopic details of the fundamental theory. A theoretical framework that advocates a coarse-grained and approximately model-independent approach, with a focus on universal features,  is at the heart of our efforts here and in \cite{Amin:2015ftc,Amin:2017wvc,Green:2014xqa} (also see \cite{Cheung:2007st,Weinberg:2008hq,Tye:2008ef,LopezNacir:2011kk,Dias:2012nf,McAllister:2012am,Marsh:2013qca,Easther:2013rva,Price:2014xpa,Dias:2016slx,Giblin:2017qjp,Bjorkmo:2017nzd,Guo:2017vxp,Dias:2017gva,Dias:2018koa}).
 
It is plausible that the complex dynamics of fields can lead to repeated non-adiabatic particle production in the inflationary \cite{Green:2009ds,Pearce:2016qtn,Barnaby:2012xt,Pearce:2017bdc,Flauger:2016idt} and post-inflationary \cite{Amin:2014eta,Traschen:1990sw, Kofman:1994rk, Kofman:1997yn, Joras:2001yh, Braden:2010wd,Leung:2012ve, Lozanov:2016pac, DeCross:2015uza,Figueroa:2015rqa,DeCross:2016fdz,DeCross:2016cbs,Lozanov:2017hjm,Iarygina:2018kee,Amin:2018kkg} universe. With sufficient complexity, the strength of the interactions and intervals between them can be treated stochastically; statistical tools can then be invoked without having to rely on detailed model building. In earlier work \cite{Amin:2015ftc, Amin:2017wvc}, we were motivated by the connection between particle production in cosmology and current conduction in disordered wires~ (also see \cite{Bassett:1997gb,Zanchin:1998fj}). Apart from the elegant mathematical correspondence, the primary drive there was that certain universal features, such as {\it Anderson Localization} \cite{anderson1958absence} in one dimension, arise independent of the details of the systems -- motivating a search for similar universality in particle production.

In previous works \cite{Amin:2015ftc, Amin:2017wvc}, the problem of stochastic, non-adiabatic particle production has been formulated exclusively in a non-expanding Minkowski background, for simplicity. In~\cite{Amin:2015ftc} the evolution of the occupation number for a single scalar degree of freedom was studied in detail, in the limit of narrowly localized interactions in time; this allows for a quasi-discrete description of the dynamics by means of the Transfer Matrix formalism. Under the assumption that each scattering can be treated as a perturbation of the transfer matrix, the authors derived a Fokker-Planck equation describing the dynamical evolution of the probability distribution for the occupation number of the scalar field. The results were then generalized for multiple fields by imposing a maximality constraint on the Shannon entropy of the probability distribution; this constraint is known as the {\em Maximum Entropy Ansatz}~\cite{mello2004quantum} (MEA), and results in a dramatic reduction of the effective degrees of freedom that describe the average behavior of the system. 

In \cite{Amin:2017wvc} the Fokker-Planck formalism was extended to the case of multiple statistically {\it inequivalent} fields, with stochastically varying effective masses, cross couplings and intervals between interactions, which, as we will demonstrate later, somewhat mimics the phase scrambling that takes place in an expanding universe. The main results therein were (1) a practical demonstration of the equivalence between the MEA and statistically equivalent interacting fields, and (2) the convergence to the MEA in the limit of large number of (possibly statistically inequivalent) fields.

Particle production in an expanding universe is distinct from the flat space case. An expanding universe introduces a competition between particle production from the interactions and dilution. More importantly, the existence of the Hubble horizon introduces an additional scale into the problem, with qualitatively different behavior of particle production expected in the spectator fields on super-horizon scales compared to the sub-horizon case.  In spite of these complications, we find a surprisingly simple and universal behavior of the non-adiabatically excited fields on sub-horizon (which is expected from earlier work) and on super-horizon scales (which is new to this work). In upcoming work, the results from this manuscript will be used to calculate the curvature fluctuations resulting from the particle production during inflation, and to estimate the efficiency of reheating after inflation. 

For the sake of simplicity we will mostly restrict ourselves to the single spectator field case in de Sitter space. Most of our mathematical framework is valid for a general expansion history and a general mass of the spectator field. However, to contain this already long paper to a manageable size we have limited the detailed discussion to (1) conformal mass ($M^2=2H^2$) and (2) massless spectator fields ($M^2=0$) in de Sitter space ($H={\rm const.}$). These choices are phenomenologically interesting. As an example, in supergravity models with minimal kinetic terms, the large vacuum density during inflation $V\sim H^2M_P^2$, where $M_P$ denotes the Planck mass, typically leads to an induced mass for all scalar fields of order $H$~\cite{Goncharov:1984qm,Evans:2013nka}. As we shall see, setting the constant of proportionality to $\sqrt{2}$ in the conformally massive case greatly simplifies our calculations. Similarly, the massless case can approximate light fields ($M\ll H$), which may be easily perturbed during inflation and source curvature fluctuations~\cite{Kobayashi:2010fm}. While $M=0$ and $M=\sqrt{2}H$ are special in terms of their calculational convenience, they are not special in terms of the physical implications of our results.


Our analysis will be restricted to the linear regime of the spectator field, for which each Fourier mode can be treated independently and the backreaction on the homogeneous expanding background can be ignored; these assumptions can break down when the energy density of the spectator fields becomes sufficiently large. We will show that significant amount of scattering is allowed for a sufficient number of e-folds to make this analysis worthwhile, and relevant for calculating observables. In addition to the above simplifying assumptions, we will also consider for the sake of analytic and numerical tractability that each interaction can be modeled as a Dirac-delta function\footnote{ie. we assume that the physical wavelengths are large compared to the duration of the non-adiabatic interactions.} in time whose amplitude and location are drawn from different distributions. 

The narrow-width interactions allow us to use the Transfer Matrix Formalism quite efficiently since the evolution between scatterings is that of free fields. We will go beyond the assumption of small changes per scattering in our numerical explorations, although our analytical understanding based on a Fokker-Planck equation will be robust in the weak scattering case only.  In distinction with earlier papers, we prefer to follow Fourier modes of the spectator field rather than the occupation number density. This is natural since the occupation number density is ill-defined on super-horizon scales. 
For an application to inflation which we will pursue in an upcoming paper, appropriate combinations of these Fourier modes of the spectator field will serve as a source for curvature perturbations. We can also use these to calculate gravitational wave production from this period.\par\bigskip

\noindent The rest of the paper is organized as follows:\par\medskip

\noindent{\bf Section~\ref{sec:Summary}} provides a bird's eye view of the most important, and simplest to state, results of our analysis. We caution that a lot is left out here; we intend this section to be more of an invitation to explore the analysis in the rest of the text. \par\bigskip

\noindent {\bf Section~\ref{sec:MP}} contains the formalism necessary to study the dynamics of a spectator field excited by a non-adiabatic, stochastic mass term in an expanding background. In Section~\ref{sec:SF} we introduce the effective single-field model in an expanding Universe that we will study. In particular we discuss a conformally massive and a massless field in a de Sitter background. In Section~\ref{sec:TM} we describe the transfer matrix formalism that will allow us to track the evolution of the scalar field and its number density after each consecutive scattering. Section~\ref{sec:FP} contains a brief summary of the Fokker-Planck formalism, with emphasis on single-field models. The general results of this section are applied to the specific cases of the conformally massive (and massless) spectator field in the subsequent sections. \par\medskip

\noindent {\bf Section~\ref{sec:DS}} contains the results for the evolution of a conformally massive scalar field, its occupation number and other transfer matrix parameters in a de Sitter background. Section~\ref{sec:ssds} shows numerical results in the weak- and strong-scattering limits for the field amplitude and the transfer matrix parameters, including their values given individual realizations of an ensemble of location and scattering amplitudes, as well as their probability densities and their lowest moments. In Section~\ref{sec:analconf} we describe the analytical results obtained from the application of the Fokker-Planck formalism, which are valid in the weak-scattering regime, for which the instantaneous change in the transfer matrix can be treated perturbatively. We also demonstrate how these results (for sub-horizon modes) correspond to a natural generalization of the Minkowski result discussed in earlier papers.\par\medskip

\noindent {\bf Section~\ref{sec:MDS}} provides a discussion of the corresponding numerical (Section~\ref{sec:mssds}) and analytical (Section~\ref{sec:analnom}) results for a massless scalar field in a de Sitter background. \par\medskip

\noindent {\bf Section~\ref{sec:SU}} contains a summary of our results and our conclusions. \par\medskip

\noindent In Appendix~\ref{ap:nm} we provide some essential checks for our results and some justification for our focus on certain variables in the main text. In the Appendix~\ref{ap:typ} we discuss the difference between typical and average quantities when the distributions are not normal. In Appendix~\ref{ap:conv}, we verify the approximate independence of our results from the details of the distribution from which the effective mass and time of scattering are drawn. Finally, in Appendix~\ref{app:back} we determine the regime in which the excitation of the field is strong enough to backreact on the expanding background. We show the domain of validity of our results in terms of the strength of the non-adiabatic events and the number of $e$-folds of inflation during which the scalar field is excited.

\section{Summary of the main results}\label{sec:Summary}
Consider a Fourier mode $\chi_k(t)$ of the spectator field $\chi(t,\bf x)$ in de Sitter space satisfying the equation of motion\footnote{While we discuss $\chi_k$ here, we find it more convenient to use the scaled field $X_k=a\chi_k$ in the main text. We also drop the subscript $k$ (ie. the momentum dependence) in denoting $X$ and other related quantities in much of the main text (though we do analyze the behavior with $k$).} (see Section~\ref{sec:MP} for details):
\beq
\ddot{\chi}_{k}+3H\dot{\chi}_k+\left[\frac{k^2}{a^2}+M^2+m^2(t)\right]\chi_k=0\,,
\eeq
where $H$ is the expansion rate, $M$ is the mass of the field and we include a stochastic mass term $m^2(t)=\sum_{j=1}^{N_s}m_j\delta(t-t_j)$  to capture the complicated interaction that this field is undergoing with other fields/background. The masses $m_j$ and locations $t_j$ are drawn from independent distributions. We assume that each $m_j$ is independent and identically distributed (not necessarily Gaussian), with $\langle m_i m_j\rangle= \sigma^2\delta_{ij}$ where $\langle \hdots\rangle$ represents an average over the ensemble. For convenience, we define $\mathcal{N}_s$ as the number of non-adiabatic events per Hubble time and assume that $\mathcal{N}_s\gg 1$.\footnote{In this limit the specific form of the distribution of $m_i$ is irrelevant, see Appendix~\ref{ap:conv}.} The single dimensionless scattering parameter $\S$ is sufficient to determine the statistical behavior of the fields outside the horizon.
\begin{figure}[!t] 
   \centering
   \includegraphics[width=0.7\textwidth]{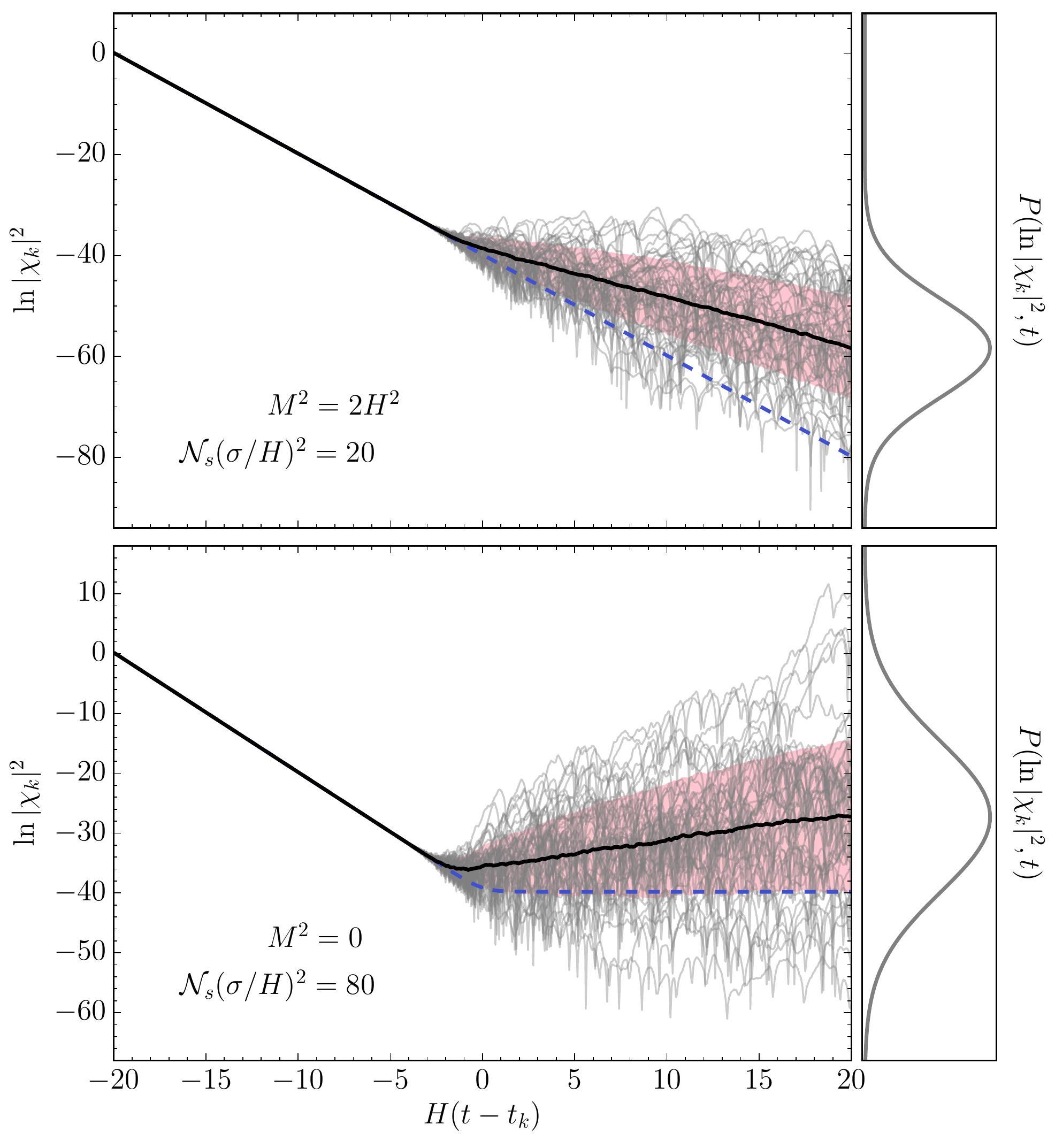} 
   \caption{The behavior of $\ln |\chi_k|^2$ for $M^2=2H^2$ (upper row) and $M^2=0$ (lower row) for different realizations of $m^2(t)$. The grey curves are $\ln|\chi_k|^2$ corresponding to different realizations of $m^2(t)$, the black curves are the ensemble means of $\ln |\chi_k|^2$ and the pink region represents trajectories within one standard deviation of the mean. The instantaneous probability distribution of $\ln |\chi_k|^2$ is shown in the right panels. In the above figure $k$ crosses the Horizon at $t=t_k$. The trajectories of $\ln |\chi_k|^2$ undergo a ``random walk" like behavior, and  have a Gaussian distribution (over the ensemble) at all times, ie. $|\chi_k|$ is log-normally distributed. Note that $\S\gtrsim 1$ is in the strong scattering regime, with the ensemble mean visibly deviating from the corresponding value without scattering (dashed lines). For sufficiently weak scattering $\S\ll 1$, the means of the weakly excited fields track the dashed lines, but the variance still grows linearly with time.
   }
   \label{fig:trajectories_3}
\end{figure}

\begin{figure}[t!]
\centering
    \includegraphics[width=0.9\textwidth]{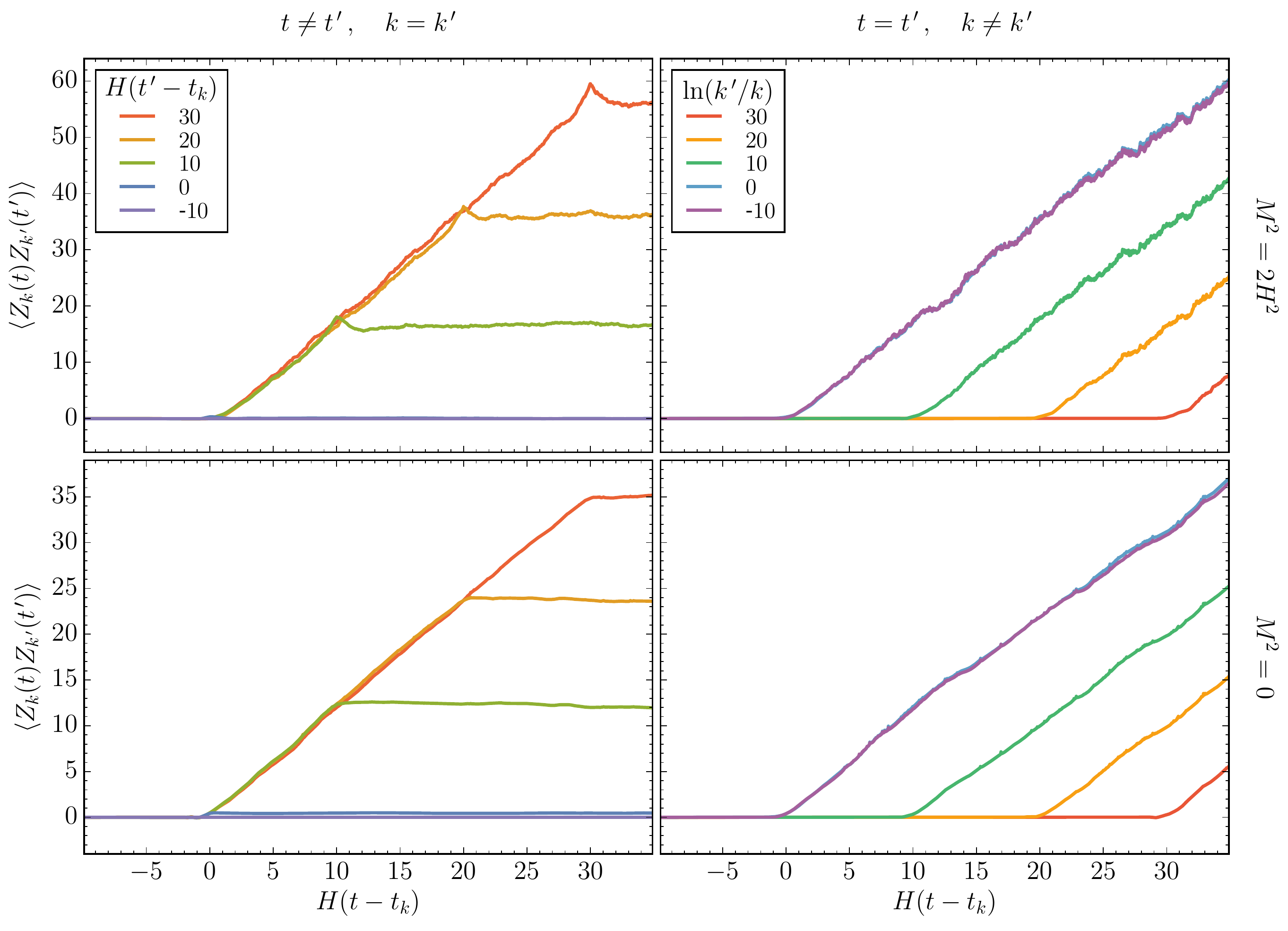}
    \caption{Sub- and super-horizon evolution of the field two-point $\langle Z_{k}(t) Z_{k'}(t') \rangle$ where $Z_k(t)\equiv \ln |\chi_k|^2-\langle \ln |\chi_k|^2\rangle$. We find that $Z_k(t)$ satisfies the requirements of a driftless random walk. We show the two point correlation for equal momenta and unequal time (left), and unequal momenta and equal time (right), in the conformal (above) and massless (below) cases. Here $\mathcal{N}_s(\sigma/H)^2=1$.  The time of horizon crossing for the $k$  mode is denoted by $t_k$. }
    \label{fig:2point_sum}
\end{figure}

The behavior of $\ln |\chi_k(t)|^2$ for different realizations of $m^2(t)$ are shown in Fig.~\ref{fig:trajectories_3}. We provide the most important and simple to understand takeaways from our analysis below:
\begin{enumerate}

\item The $\chi$ field is approximately in its vacuum state {\it sufficiently} inside the horizon (i.e.~$|\chi_k|^2\simeq 1/(2ka^2)$).\footnote{This statement has caveats, in terms of the magnitude of $\mathcal{N}_s$ and $\S$ as well as initial conditions. A large magnitude of these parameters can lead to deviations from the vacuum behavior as one would expect. We explore these caveats and details further in the main text.} Outside the horizon, $\ln|\chi_k|^2$ evolves linearly with cosmic time (in an ensemble averaged sense), with
\Beq
\partial_{Ht}\langle \ln|\chi_k|^2\rangle=\mu_1-2 \qquad {\rm and} \qquad \partial_{Ht}{\rm Var} [\ln|\chi_k|^2]=\mu_2\,.
\Eeq
where the variance and mean are over different realizations of the effective mass $m^2(t)$. The rates $(\mu_1,\mu_2)$ are functions of $\S$. The values of $\mu_1$ and $\mu_2$ as a function of $\S$ are shown in Fig.~\ref{fig:nfrates} (conformal mass) and Fig.~\ref{fig:nfrates_m} (massless).

\item Importantly, $\ln |\chi_k(t)|^2$ is normally distributed on super and sub-horizon scales at all times (as an ensemble over realizations of $m^2(t)$). Equivalently, $|\chi_k(t)|^2$ is log-normally distributed. This means that 
\beq
|\chi_k(t)|_{\rm typ}^2\equiv e^{\langle \ln |\chi_k(t)|^2\rangle}\,,
\eeq 
is a better representative of the ensemble rather than $\langle |\chi_k(t)|^2\rangle$, which will be dominated by the largest values of $\chi_k(t)$ in the ensemble. 

\item On super-horizon scales, $\ln |\chi_k|^2$ satisfies the properties of a drifted random walk. In particular, as mentioned above, the mean and variance of $\ln |\chi_k|^2$ grow linearly with time and for the drift-less variable $Z_k(t)\equiv \ln |\chi_k|^2-\langle \ln|\chi_k|^2\rangle$, we find 
\begin{equation}
\langle Z_{k}(t) Z_{k'}(t') \rangle \;\simeq\;
\mu_2H\,\textrm{min}[t-t_k,t-t_{k'}, t'-t_k, t'-t_{k'}]\,,
\end{equation}
where $t_k$ is the time when the $k$-mode exits the horizon (see Fig.~\ref{fig:2point_sum}). Note that the above condition contains within it the statement that the increments $Z_k(t_4)-Z_k(t_3)$ and $Z_k(t_2)-Z_k(t_1)$ are uncorrelated when the interval $(t_3,t_4)$ does not overlap with $(t_1,t_2)$ on super-horizon scales.

\hspace{5pt} The behavior of  $\ln |\chi_k|^2$ implies that $|\chi_k(t)|$ performs a {\it geometric} random walk. With this understanding, all $n$-point correlation functions for the field magnitude can be computed in terms of the $Z_k$ two-point functions, 
\beq
\langle |\chi_{k_1}(t_1)|^2\cdots |\chi_{k_n}(t_n)|^2 \rangle \;=\; \exp\left[\sum_{i=1}^n \langle \ln|\chi_{k_i}(t_i)|^2\rangle + \frac{1}{2}\sum_{i,j=1}^n \langle Z_{k_i}(t_i)Z_{k_j}(t_j) \rangle \right]\,.
\eeq

\item The phase $\arg \chi_k$ is randomly distributed inside the horizon and converges to an $\S$-dependent value outside the horizon.

\item In the weak scattering limit, using a Fokker-Planck equation, we analytically derive the log-normal probability distribution for $|\chi_k(t)|$ on sub-horizon scales (though we cannot do so yet on super-horizon scales). On super-horizon scales, we can derive the time rate of evolution for $\langle\ln |\chi_k|^2\rangle$ in the weak scattering limit. These results are consistent with our numerical simulations.

\end{enumerate}

While the behavior of the field is easiest to discuss, and perhaps the most useful for future calculations, we found it useful and at times necessary to understand the behavior of the transfer matrix parameters individually (a combination of which yields the field amplitude and phase). These parameters include an ``occupation number density"($n_k$) and two phases $\phi_k$ and $\psi_k$ (see Section~\ref{sec:MP} for definitions):
\begin{enumerate}
\item The occupation number density grows exponentially inside and outside the horizon. The growth rates of the ensemble mean and variance of $\ln(1+n_k)$ are linear in cosmic time outside the horizon, and are determined by $\S$.
\item The phases are uniformly distributed on subhorizon scales, but their distribution will in general depend on $\S$ outside the horizon.
\item The means and variances of $\phi_k$ and $\psi_k$ undergo non-trivial evolution at horizon crossing, but converge to constants (which can be $\S$ dependent) outside the horizon. 
\item In the weak scattering limit, we derive the behavior of the occupation number and phases, including their distribution and evolution rates of the lowest moments, inside the horizon using a Fokker-Planck equation. Outside the horizon, we derive a highly non-trivial distribution of the phase $\phi_k$ as well the evolution rates of the mean of $\ln(1+n_k)$.
\end{enumerate}
We re-iterate that this is a rather cursory summary. Details, caveats and many other relevant results, which are not included here, are discussed for the main text.  

\section{Mathematical preliminaries}\label{sec:MP}

\subsection{Spectator field in an expanding Universe}\label{sec:SF}

Consider a spectator field\footnote{$\chi$ is assumed to contribute a negligible amount to the total energy density of the universe.} $\chi(t,{\bf x})$ of mass $M$ in a homogeneous and isotropic expanding universe. For this field, we use an additional effective mass $m(t)$ to parametrize the coupling of this field to a time-dependent background, especially including random non-adiabatic events arising from complicated interactions with other fields.  The quadratic action for such a (quantum) field is taken to be
\begin{align} \displaybreak[0]
\mathcal{S} \;&=\; \frac{1}{2}\int \sqrt{-g}\, d^4x\, \Big[ \partial_{\mu}\hat{\chi}\partial^{\mu}\hat{\chi}  - \left(M^2 + m^{2}(t)\right)\hat{\chi}^2\Big]\\ \displaybreak[0]
&=\; \frac{1}{2} \int d^3{\bf x}\, d\tau\, a^2\Big[(\partial_{\tau}\hat{\chi} )^2 - (\nabla\hat{\chi} )^2- \left(M^2 + m^{2}(t)\right)a^2\hat{\chi} ^2 \Big]\,,\\ \displaybreak[0]
&=\;\frac{1}{2} \int d^3{\bf x}\, d\tau\, \Big[(\partial_{\tau}\hat{X} )^2 - (\nabla\hat{X} )^2-\left(a^2M^2 + a^2m^{2}(\tau)-\frac{\partial^2_\tau a}{a}\right)\hat{X}^2 \Big]\,,
\end{align}
where $a$ is the scale factor, the conformal time $\tau$ is related to cosmic time $t$ via $dt/d\tau=a$, and where we have defined 
\Beq
\hat{X}\equiv a\hat{\chi}\,,
\Eeq
in going from the second to the third line. In slight abuse of notation, $m^2(\tau)=m^2(t(\tau))$ and $a(\tau)=a(t(\tau))$ above. While it is most convenient to write the necessary equations and formalism in terms of $X$ and $\tau$, some of our results are most naturally written in terms of $\chi$ and $t$. 

The equation of motion for the $\hat{X}$ field is then given by 
\Beq
\label{eq:chieom}
\left[\partial_\tau^2 - \nabla^2 -\frac{a''}{a}+a^2(M^2 + m^2(\tau))\right]\,\hat{X}(\tau,{\bf x}) = 0 \,.
\Eeq
The mode expansion for this field can be written as 
\beq
\hat{X}({\bf x},\tau) \;=\; \int \frac{d^3\k}{(2\pi)^{3/2}} \, e^{-i \k\cdot{\bf x}}\Big[ X_{k}(\tau) \hat{a}_{\k} + X^*_{k}(\tau) \hat{a}^{\dagger}_{-\k} \Big]\,,
\eeq
where $[\hat{a}_{\k},\hat{a}^{\dagger}_{\k'}] = \delta(\k-\k')$, $[\hat{a}_{\k},\hat{a}_{\k'}] = [\hat{a}^{\dagger}_{\k},\hat{a}^{\dagger}_{\k'}]=0$. The mode functions $X_k(\tau)$ satisfy
\Beq
\label{eq:eomTau}
X_k''(\tau) + \left[k^2 -\frac{a''}{a}+a^2(M^2 + m^2(\tau))\right]\,X_k(\tau) = 0 \,,
\Eeq
and are normalized\footnote{For consistency with the canonical commutation relations between $\hat{X}(t,{\bf x})$ and its conjugate momentum.} by means of the Wronskian condition $X_{k}(\tau)X_{k}^{*'}(\tau) - X_{k}'(\tau) X_{k}^*(\tau) \;=\; i\,$. In addition, the physical mode functions are chosen so that in the infinite past the vacuum is of Bunch-Davies type.

In general, the time dependence of the stochastic mass $m(\tau)$ can be complicated. To track the evolution of the field, we will assume that this effective mass consists of localized, non-adiabatic events. In between these non-adiabatic events, the free field solutions to (\ref{eq:eomTau}) will have the form
\beq\label{eq:fdef}
X_k(\tau) =  \alpha_{k,j} f_k(\tau) + \beta_{k,j} f_k^*(\tau)\,,
\eeq
where $f_k(\tau)$ satisfies \eqref{eq:eomTau} in absence of $m^2(\tau)$: 
\Beq
\label{eq:eomf}
f_k''(\tau) + \left[k^2 -\frac{a''}{a}+a^2M^2 \right]\,f_k(\tau) = 0 \,.
\Eeq
The coefficients $\alpha_{k,j}$ and $\beta_{k,j}$ are the Bogoliubov coefficients after the $j$-th non-adiabatic event, with
\beq
\label{eq:abconst}
|\alpha_{k,j}|^2 - |\beta_{k,j}|^2 \;=\; 1\,.
\eeq
This constraint forces $f_k(\tau)$  to satisfy the same normalization condition as $X_k(\tau)$ above.  $f_k(\tau)$ is then completely specified (up to an irrelevant phase) provided it also satisfies the Bunch-Davies vacuum initial condition in the infinite past. Note that before any non-adiabatic interactions, $\alpha_{k,0}=1$ and $\beta_{k,0}=0$. The quantity $|\beta_{k,j}|^2$ can be interpreted as the {\it occupation number density} of particles of the field $\hat{X}$ with momentum $k$ after the $j$th non-adiabatic event.\footnote{The total number density would be $\int d^3k\, |\beta_{j,k}|^2$.} We caution that this interpretation of $|\beta_{k,j}|^2$ as an occupation number density, however, does not carry over easily on super-horizon scales \cite{Mukhanov:2007zz}. 
\subsubsection{Mode functions in de Sitter spacetime}
The focus of our discussion from Section~\ref{sec:DS} onwards will be the study of spectator fields in a de Sitter background where 
\Beq
H(t)=\dot{a}(t)/a(t)={\rm const.}\qquad\textrm{and}\qquad \tau =-1/aH<0\,.
\Eeq
General expressions for $f_k(\tau)$ are available in this case in terms of Hankel functions (see e.g.~\cite{Mukhanov:2007zz}). There are two cases where the form of $f_k(\tau)$ is even simpler. 
\\ \\
\noindent {\it Conformally Massive Fields} : When the mass of the field $M^2=2H^2$ in a de Sitter background, Eq.~(\ref{eq:eomf}) becomes the equation of motion for a free field in a non-expanding universe:
\beq
f_k''(\tau)+  k^2 f_k(\tau)=0\,,
\eeq
with a solution
\beq\label{eq:conf_fk}
f_k(\tau) \;=\; \frac{e^{-ik\tau}}{\sqrt{2k}}\,.
\eeq
Note that while $f_k(\tau)$ does not see the effects of expansion, the full solution $X_k(\tau)$ can depend on expansion through the non-adiabatic term $a^2m^2(\tau)$ in Eq.~\eqref{eq:eomTau}. %
\\ \\
\noindent {\it Massless Fields} : In this case $M^2=0$, and the mode functions have the form
\beq\label{eq:dsmass}
f_k(\tau)\;=\; \frac{e^{-ik\tau}}{\sqrt{2k}}\left(1-\frac{i}{k\tau }\right)\,.
\eeq
\subsection{The Transfer Matrix formalism}\label{sec:TM}
The assumption of localized interactions in the effective mass $m^2(\tau)$ allows for a transfer matrix approach for the determination of the coefficients $\alpha_k$ and $\beta_k$. We define the transfer matrix at the location of the $j$-th scattering, $\M_j$, to be such that
\beq\label{eq:bogo}
\begin{pmatrix}
\beta_j \\
\alpha_j
\end{pmatrix} = \M_j 
\begin{pmatrix}
\beta_{j-1} \\
\alpha_{j-1}
\end{pmatrix}\,.
\eeq
Here we have ignored the momentum dependence (ie. subscript $k$) for notational simplicity.  By chaining together $\M_j$ with all transfer matrices $\M_{i<j}$ we can  construct $\alpha_j,\beta_j$ from their initial values,
\beq
\begin{pmatrix}
\beta_j\\
\alpha_j
\end{pmatrix} = \M(j) 
\begin{pmatrix}
\beta_0\\
\alpha_0
\end{pmatrix}\,,\quad \text{where}\quad \M(j) = \M_j\M_{j-1}\cdots\M_1\,.
\eeq
Note that starting with fields in the Bunch-Davies vacuum is consistent with choosing $\{\beta_0,\alpha_0\}=\{0,1\}$. A general parametrization of the matrix $\M(j)$ can be written as~\cite{mello1988macroscopic}
\beq\label{eq:Mparg}
\M(j) =\begin{pmatrix} e^{i\phi} & 0 \\ 0 & e^{-i\phi} \end{pmatrix}
               \begin{pmatrix} \sqrt{1+n} &  \sqrt{n} \\ \sqrt{n} & \sqrt{1+n} \end{pmatrix}
               \begin{pmatrix} e^{i\psi} & 0 \\ 0 & e^{-i\psi} \end{pmatrix}\,,
\eeq
where the angular parameters can be identified as follows, 
\beq
\phi(j) = \frac{1}{2}\arg(\beta_j\alpha_j^*)\,,\qquad \psi(j) = -\frac{1}{2}\arg(\beta_j\alpha_j)\,,
\eeq
and $n(j)$ is the occupation number density:
\begin{align}
n(j) &= |\beta_j|^2\\
& = \frac{1}{4}{\rm Tr}\left[\M(j)\M^{\dagger}(j) - \I\right].
\end{align}
Note that two complex numbers $\alpha$ and $\beta$ keep track of the field evolution which we are interested in (see Eq.~\eqref{eq:fdef}). These numbers have one constraint \eqref{eq:abconst} which leaves 3 independent real numbers. These three numbers are conveniently parametrized by $n$, $\phi$ and $\psi$ in the transfer matrix $\M$.

Clearly, the functional dependence of $\M_j$ will be determined by the explicit form for $m^2(t)$. Following the analysis of~\cite{Amin:2015ftc,Amin:2017wvc}, in order to obtain analytically tractable expressions we will restrict ourselves to the assumption of Dirac-delta scatterers in cosmic time, 
\Beq
m^2(t)
&=\sum_j m_j\delta(t-t_j)\,,\\
&=\sum_j \frac{m_j}{a(\tau_j)}\,\delta(\tau-\tau_j)=m^2(\tau)\,.
\Eeq
Physically, we are assuming that the temporal width of the scatterers is much smaller than the characteristic period of $f_k(\tau)$. With such Dirac-Delta scatterers, (\ref{eq:eomTau}) takes the form
\beq\label{eq:KGf}
X_k''(\tau) + \left[k^2 -\frac{a''}{a}+a^2M^2 + \sum_{j}m_ja(\tau_j)\delta(\tau-\tau_j)\right]\,X_k(\tau) = 0 \,.
\eeq
The junction conditions at each $\tau_j$ correspond to 
\begin{align} \label{eq:junct1}
X_{k, j+1}(\tau_j) &= X_{k,j}(\tau_j)\,,\\ \label{eq:junct2}
X_{k,j+1}'(\tau_j)&= X_{k,j}'(\tau_j) - m_j a(\tau_j)X_{k,j}(\tau_j)\,.
\end{align}
Translated to (\ref{eq:bogo}), it implies the following general form for the transfer matrix:
\begin{align}\label{eq:Mjgen}
\M_j &= \I +  i{m}_j a(\tau_j) \begin{pmatrix}
|f(\tau_j)|^2 & f^{2}(\tau_j)\\
-f^{*2}(\tau_j) & -|f(\tau_j)|^2
\end{pmatrix} \,.
\end{align}
By multiplying transfer matrices with the form derived above, we can numerically compute the evolution of the occupation number density, phases and the field amplitude. 

Note that the addition of an extra scattering event to $\M(j)$ may be treated as a perturbation if the instantaneous scattering amplitude is such that
\beq\label{eq:weakdef}
{m}_j a(\tau_j) |f(\tau_j)|^2 \ll 1\,.
\eeq
For our general investigation (including our numerical simulations), we do not assume a small perturbation. However, the Fokker-Planck formalism described immediately below relies on this small perturbation assumption.
\subsection{The Fokker-Planck equation}\label{sec:FP}
The stochastic nature of the effective mass $m^2(t)$ implies that the non-adiabatic event amplitudes $m_j$ and locations $t_j$ are assumed to be drawn from some distribution. This in turn implies that the transfer matrices $\M_j$ and $\M(j)$ will also be stochastic in nature, and will take different values for different realizations of the $m_j,t_j$. We will therefore consider an ensemble of realizations for the amplitudes and locations of the scatterings, over which we can define a probability density $P_t(\M)$. Any physically meaningful quantity can then be obtained from expectation values with respect to this density.

The evolution equation for $P_t(\M)$ can be constructed by considering the addition of a small time interval $\delta t$ with a single weak scatterer (c.f.~\ref{eq:weakdef}) to an existing interval $t$ with $j$ scattering events. It can then be shown that the probability density of the enlarged time interval $P_{t+\delta t}(\M=\M_2\M_1)$, where $\M_1\equiv \M(j)$  and $\M_2\equiv \M_{j+1}$, corresponds to the convolution of the density for the transfer matrix of the extra strip of width $\delta t$: $P_{\delta t}(\M_2)$, with the density related to $j$ prior scatterings scatterings $P_t(\M_1=\M_2^{-1}\M)$\footnote{For a detailed derivation of the Smoluchowski and Fokker-Planck equations, see~\cite{mello2004quantum,Amin:2017wvc,Amin:2015ftc}}
\beq
P_{t + \delta t}(\M) = \int d\M_2\, P_{t}(\M_2^{-1}\M) P_{\delta t}(\M_2)\,.
\eeq
This integral equation (known as the {\em Smoluchowski} or {\em Chapman-Kolmogorov equation}) can be equivalently written as the Fokker-Planck equation
\beq\label{eq:FPgen}
\frac{\partial P}{\partial t} = - \sum_{b=1}^{2\Nf^2+\Nf} \frac{\partial}{\partial \lambda_b}\left[ \frac{\langle \delta\lambda_b\rangle_{\delta t}}{\delta t} P\right] + \frac{1}{2} \sum_{b,c=1}^{2\Nf^2+\Nf} \frac{\partial^2}{\partial \lambda_b \partial \lambda_c}\left[ \frac{\langle \delta\lambda_b \delta\lambda_c\rangle_{\delta t}}{\delta t} P\right]\,,
\eeq
where $\{\lambda_a\}$ denote the set of $2\Nf^2+\Nf$ parameters that characterize a general $\Nf$-field transfer matrix $\M$; in the single-field case $\{\lambda_a\}=\{n,\phi,\psi\}$. The $\delta\lambda_a$ denote the small increment in the parameters $\lambda_a$ due to the addition of an extra scattering. The expectation value is over the probability distribution describing the properties of the scatterer in the interval $\delta t$ (which includes location within this interval as well as strength/shape of the scatterers). 

In the present work we will focus on the evolution of the occupation number density and the magnitude of the scalar field mode functions. When this is the case, it is convenient to not track the full transfer matrix $\M$ but its square $\R(j) \equiv \M(j)\M^{\dagger}(j)$. This matrix is Hermitian and depends only on $\Nf^2+\Nf$ variables in the general case. For a single field, we can write
\beq\label{eq:Rparg}
\R(j) =\begin{pmatrix} e^{i\phi} & 0 \\ 0 & e^{-i\phi} \end{pmatrix}
               \begin{pmatrix} \lambda &  \tilde{\lambda} \\ \tilde{\lambda} & \lambda \end{pmatrix}
               \begin{pmatrix} e^{-i\phi} & 0 \\ 0 & e^{i\phi} \end{pmatrix}\,,
\eeq
where, somewhat abusing notation, we have defined $\lambda \equiv 2n+1$ and $\tilde{\lambda}\equiv \sqrt{\lambda^2-1}$. The Fokker-Planck equation (\ref{eq:FPgen}) would then be written as a three-variable PDE for the probability density $P(\lambda,\phi;t)$.

The re-parametrization in terms of $\R$ is particularly convenient for our purposes since the perturbations in the parameters $\lambda,\phi$ upon the addition of an extra scattering are known. They were derived in~\cite{Amin:2017wvc} under the assumption of Dirac-delta scatterers with zero-mean uncorrelated amplitudes,
\beq\label{eq:mstats}
\langle m_j\rangle = 0\,,\qquad \langle m_j m_i\rangle = \sigma^2 \delta_{ij}\,.
\eeq
Since $\R$ is quadratic in the $\M_j$ matrices, its instantaneous change contains first- and second-order corrections in the order parameter ${m}_j a(\tau_j) |f(\tau_j)|^2$. In terms of $\lambda$ and $\phi$, these are given by
\begin{subequations}  
\begin{align} \label{eq:deltalambda1}
\delta \lambda^{(1)} &= g^{(1)}\,,\\ \displaybreak[0]
\delta \lambda^{(2)} &= g^{(2)}\,, \\ \displaybreak[0]
\delta \phi^{(1)}  &= -\frac{i}{2\tilde{\lambda}^2}\left( \tilde{\lambda}\tilde{g}^{(1)} - \lambda g^{(1)} \right)\,,\\ \displaybreak[0]
\delta \phi^{(2)} & = -\frac{i}{2\tilde{\lambda}}\left[\tilde{g}^{(2)}-\frac{\lambda}{\tilde{\lambda}}\,g^{(2)} - \frac{(\tilde{g}^{(1)})^2}{2\tilde{\lambda}} + \frac{(\lambda^2+1)(g^{(1)})^2}{2\tilde{\lambda}^3} \right]\,,
\end{align}
\end{subequations}
where the $g,\tilde{g}$ are themselves functions of the parameters $\lambda,\phi$, 
\begin{subequations} 
\begin{align} \label{eq:g1g}
g^{(1)} & =  i\tilde{\lambda} {m}_j a(\tau_j) \left(e^{-2 i \phi}f^2(\tau_j) - e^{2 i \phi}f^{*2}(\tau_j) \right)\,,\\
\tilde{g}^{(1)} &= 2i {m}_j a(\tau_j) \left(\tilde{\lambda}|f(\tau_j)|^2 + \lambda e^{-2 i \phi}f^2(\tau_j) \right)\,,\\
g^{(2)} & =  2\lambda {m}^2_j a^2(\tau_j) |f(\tau_j)|^4 + \tilde{\lambda} {m}^2_j a^2(\tau_j) |f(\tau_j)|^2 \left(e^{-2 i \phi} f^2(\tau_j) + e^{2 i \phi}f^{*2}(\tau_j) \right)\,,\\ \label{eq:g4g}
\tilde{g}^{(2)} &= -\tilde{\lambda} {m}^2_j a^2(\tau_j) \left(|f(\tau_j)|^4+e^{-4 i \phi}f^4(\tau_j)\right) - 2\lambda {m}^2_j a^2(\tau_j) e^{-2 i \phi}|f(\tau_j)|^2f^2(\tau_j)\,.
\end{align}
\end{subequations}

To go further, the explicit expression for the mode-functions are necessary. We will show that, in spite of the apparent complexity of the equations, the equations predict certain universal results, which we turn to after we have discussed our numerical results first. For the moment, note that simplified, general expressions for expectation values of functions of $\lambda,\phi$ can be obtained by integration of the Fokker-Planck equation (as discussed in~\cite{Amin:2017wvc,Amin:2015ftc}).  

We will return back to this Fokker-Planck equation in Section~\ref{sec:analconf} and Section~\ref{sec:analnom} and use it to explain aspects of our numerical results for the conformal mass and massless fields.

\section{Conformally massive field in de Sitter background}\label{sec:DS}

\subsection{Numerical results}\label{sec:ssds}
In this section we focus on numerical results  for the evolution of the occupation number, the scalar field amplitude and its phase in the conformal mass case using the transfer matrix approach discussed in Section~\ref{sec:TM}. We  separate the discussion of our numerical results into four regimes, namely the sub- and super-horizon regimes with weak and strong scattering. 

The sub- and super-horizon regimes correspond to physical wavelengths smaller and larger than the horizon scale, $k/aH \gg 1$ and $k/aH\ll 1$, respectively. In an expanding de Sitter background $a(t)=e^{H(t-t_i)}$, any given comoving wavelength that starts inside the horizon will eventually cross outside the horizon at a time $t_k$ satisfying:
\beq
\frac{k}{a(t_k)H}\;=\;1\,.
\eeq
Numerically, this allows us to explore both the sub- and super-horizon regimes for a given Fourier mode by starting the computation for some time $t_i<t_k$ and finishing it at $t_f>t_k$; for definiteness, we have considered a total range of 40 Hubble times between the initial and final times, centered at horizon crossing. We will also set $a(t_i)=1$.

We will talk of weak or strong scattering depending on whether the parameter 
\begin{equation}\label{eq:paramdef}
\mathcal{N}_s \left(\frac{\sigma}{H}\right)^2 \equiv \frac{N_s}{H(t_f-t_i)} \left(\frac{\sigma}{H}\right)^2\,
\begin{cases*}
\ll 1& \textrm{weak scattering}\\
\sim \textrm{few} & \textrm{moderate scattering}\\
\gg 1 & \textrm{strong scattering}
\end{cases*}
\end{equation}
is much smaller or larger than unity. A more careful delineation will be provided later.
Here, $N_s$ is the number of scatterers in the interval $t_f-t_i$, $\mathcal{N}_s$ denotes the number of scatterers per Hubble time, and $\sigma^2=\langle m_i^2\rangle$ characterizes the strength of the scatterers. As we will demonstrate numerically and analytically below, it is the combination $\mathcal{N}_s (\sigma^2/H^2)$ that really determines the growth rate for the occupation number and the field amplitude. 

Note that our delineation of strong and weak scattering is different from our perturbativity condition (\ref{eq:weakdef}).
For the conformal de Sitter scenario ($M^2=2H^2$), the perturbativity condition can be rewritten as $(\sigma/k_{\rm phys})^2\ll 1$ where $k_{\rm phys}=k/a$. In the subhorizon regime, this condition is always satisfied if $(\sigma/H)^2\leq 1$, while outside the horizon the scattering amplitudes must satisfy the much more restrictive constraint $(\sigma/H)^2 \ll (k/aH)^2 \ll 1$. It is clear then that, for any given $(\sigma/H)^2<1$, perturbativity will be eventually lost outside the horizon. Because of this restriction, we naively expect the Fokker-Planck approach will properly account for the evolution inside the horizon for scattering amplitudes not greater than the Hubble scale, while outside the horizon it will fail unless $\mathcal{N}_s(\sigma/H)^2 \ll 1$. Our numerical approach does not require any such restriction. 

As we will discuss below, the universality of our results relies on the assumption that the number of scatterers per Hubble time $\mathcal{N}_s$ is large (see Fig.~\ref{fig:converg}). This implies that many thousands of operations over hundreds or even thousands of realizations are necessary to reach a stationary regime. This complexity, coupled with exponentially increasing or decreasing quantities, cries out for a numerical code capable of handling the extremely high precision required. To achieve this, we have built our (Fortran) code making extensive use of the thread-safe arbitrary precision package MPFUN-For written by David H. Bailey~\cite{mpfun}. We have confirmed that the precision used in our numerical simulations (500 digits) is adequate by ensuring that the constraint on the Bogoliubov coefficients, $|\alpha_j|^2-|\beta_j|^2=1$, holds up to the chosen precision for all realizations.
\subsubsection{Individual realizations}\label{sec:indrelcon}
\begin{figure}[t!]
\centering
    \includegraphics[width=\textwidth]{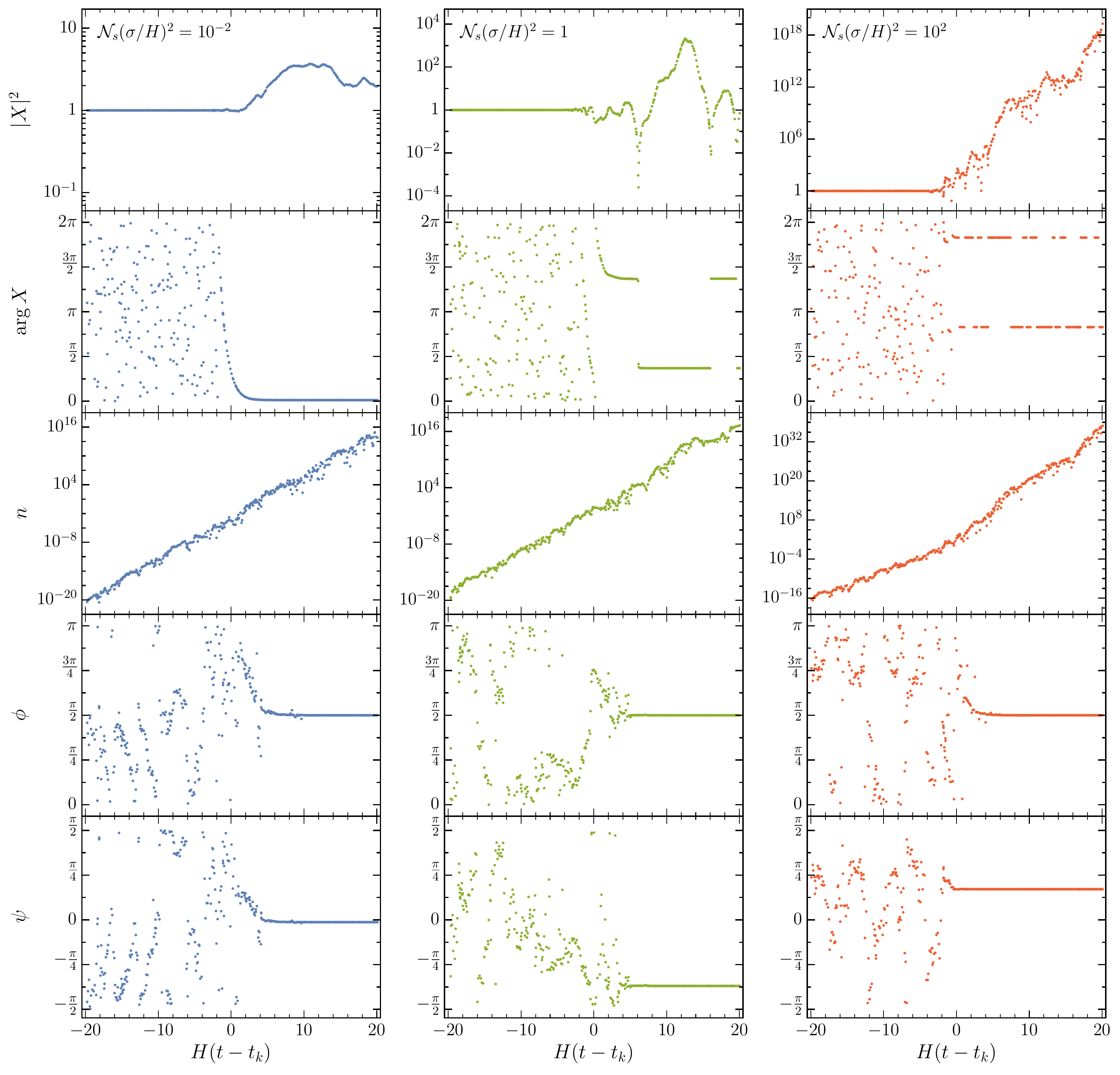}
    \caption{Evolution of the occupation number, the field squared magnitude and its phase, and the transfer matrix phases $\phi$ and $\psi$, as functions of cosmic time, in the conformal mass case with uniformly distributed amplitudes and locations of the non-adiabatic events. Three representative values of the parameter $\mathcal{N}_s(\sigma/H)^2$ have been chosen, corresponding to weak, moderate or strong scattering. The plots are generated for $k=e^{20}H$, and a subscript $k$ of the $y$-axis quantities is suppressed to reduced clutter. Note the re-scaling (\ref{eq:rescx}).}
    \label{fig:all_grid}
\end{figure}

Fig.~\ref{fig:all_grid} shows the evolution of the the field amplitude and its phase, as well as transfer matrix parameters $\{n,\phi,\psi\}$  as functions of time. Here we have assumed that both the amplitudes and the locations of the non-adiabatic events are uniformly distributed in the intervals $m_j\in(-\sqrt{3}\sigma,\sqrt{3}\sigma)$ and $\delta t_j\in(0,1/H\mathcal{N}_s)$, where $\delta t_j$ denotes the interval between event locations, $\delta t_j\equiv t_j - t_{j-1}$. For definiteness we have taken $N_s=300$ and $k=e^{20}H$. The factor of $e^{20}$ allows for $20$ e-folds before horizon crossing.\footnote{In order to allow a simpler reading of our numerical results, in this figure and all other figures that follow, we present the canonically normalized field re-scaled by its magnitude in the Bunch-Davies vacuum. That is, in all figures the physical value of $X$ can be recovered by taking
\beq\label{eq:rescx}
X \;\longrightarrow\; \sqrt{2k}\,X\,.
\eeq
}\textsuperscript{,}\footnote{Our choice of 20 e-folds after horizon crossing runs afoul of backreaction constraints for strong scattering (see Appendix~\ref{app:back}). Nevertheless, we display our results for ease of comparison with weak scattering.}

\begin{figure}[t!]
\centering
    \includegraphics[width=\textwidth]{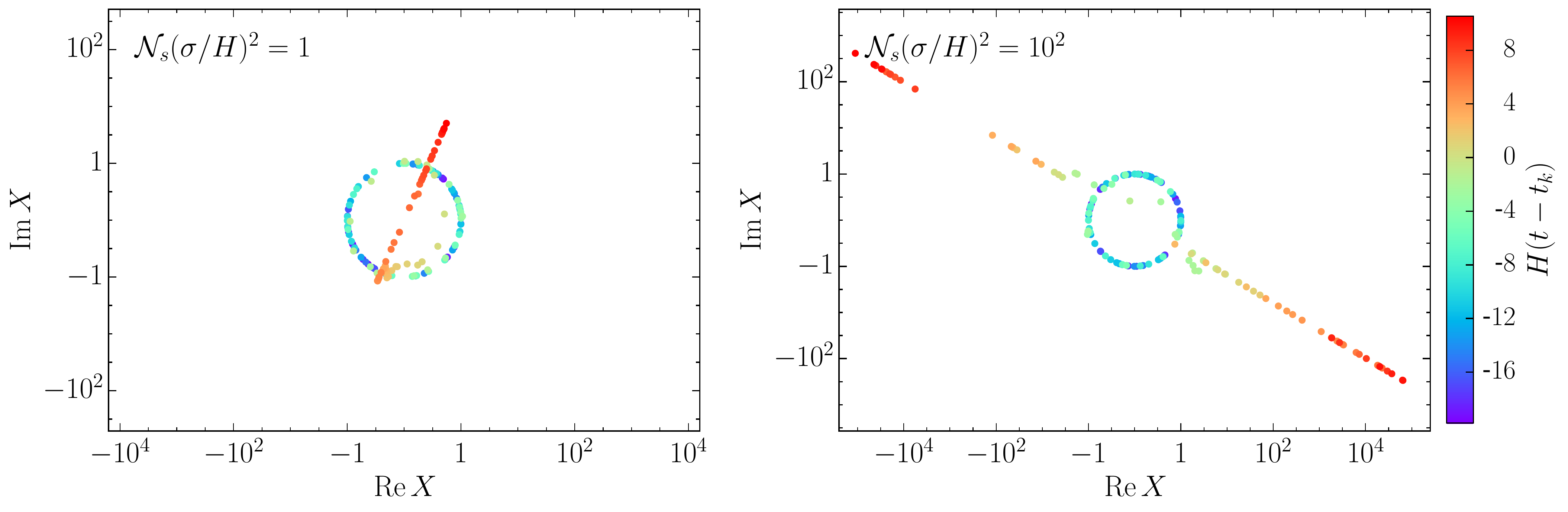}
    \caption{Evolution of the (re-scaled) real and imaginary parts of the canonically normalized conformally massive field $X$ as functions of time in the strong scattering regime, for two values of $\mathcal{N}_s(\sigma/H)^2$. The numerical results shown here correspond to those of the center and right panels of Fig.~\ref{fig:all_grid}, sampled every other point for clarity. The real and imaginary parts of the mode amplitude evolve on a circle when the mode is sub-horizon; after leaving the horizon they converging to line at a random angle in this plane. For sufficiently weak scattering (not shown in this figure), this line will be along the real axis, with $|{\rm Re}\, X|\lesssim \mathcal{O}[{\rm few}]$.}
    \label{fig:reimchi}
\end{figure}

Each column in Fig.~\ref{fig:all_grid} corresponds to a single realization of the disorder $\{(m_1,t_1),(m_2,t_2),\hdots\}$. The different columns correspond to disorder realizations drawn from distributions which  correspond to different values of the parameter $\mathcal{N}_s(\sigma/H)^2$, one in the weak regime (left), one for a ``moderate'' value (center), and one in the strong scattering regime (right). Despite the fact that the results in Fig.~\ref{fig:all_grid} correspond to a single realization of the amplitudes and locations of the non-adiabatic events, we can still read off the main features of the evolution of the parameters of interest, namely
\begin{enumerate}
\item The magnitude of the (re-scaled) canonically normalized field, $|X|$, remains very close to one, with virtually no influence from scattering on subhorizon scales. In other words, $|\chi|=|X|/a$ decreases exponentially with its decay rate determined by the scale factor. Outside the horizon, the magnitude of $X$ remains $\mathcal{O}(1)$ for weak scattering, which implies that the decaying trend for $|\chi|$ is continued after horizon crossing. For moderate scattering, the effect of the stochastic non-adiabaticity is capable of exciting $|X|$ by a couple of orders of magnitude away from its vacuum value, although no clear increasing or decreasing trend is noticeable. Finally, in the case of strong scattering, $|X|$ grows exponentially outside the horizon, with a rate dependent on $\mathcal{N}_s(\sigma/H)^2$. In the case shown in Fig.~\ref{fig:all_grid}, this growth is sufficiently large to overcome the decay of $|\chi|$ and to make it grow for $|k \tau|\ll 1$.
\item The scalar field phase, $\arg X$, is uniformly distributed in the interval $(0,2\pi)$ before horizon crossing. After horizon crossing, if scattering is weak, this phase freezes asymptotically to a small value, $\arg X\ll 1$. If scattering is moderate or strong, the phase becomes almost frozen along a random direction, with $X$ evolving along a ray in the complex plane. Also see Fig.~\ref{fig:reimchi} and the discussion after this list of observations.
\item The occupation number density, $n=|\beta|^2$,  grows exponentially (for the given Fourier mode). For weak scattering the exponential growth rate is constant throughout the evolution (with $n\propto a^2$), while for strong scattering, the rate increases shortly before horizon crossing. The numerical value of these rates depends on $\mathcal{N}_s(\sigma/H)^2$.  Note that for very weak scattering (or when $n\ll 1$ more generally) $n\propto a^2$ seems counter-intuitive at first glance. However, this result follows from the observation that each scatterer (as seen in Eq. \eqref{eq:KGf}) comes with an increasing strength $a(\tau_j)m_j$ as a function of time. 
\item The transfer matrix phase $\phi$ (c.f.~\eqref{eq:Mparg}) is naturally defined on the domain $(0,\pi)$. The phase varies randomly over this domain inside the horizon, and freezes asymptotically to $\phi\sim \pi/2$ far outside the horizon.
\item From the statistical point of view, as we will discuss below, the natural range for the second transfer matrix phase $\psi$ (c.f.~\eqref{eq:Mparg}) corresponds to $(-\pi/2,\pi/2)$, where it varies randomly for $|k\tau|\gg 1$. For weak scattering, it freezes to $|\psi|\ll 1$ in super-horizon scales, while for strong scattering it freezes to a seemingly random value.
\end{enumerate}

\begin{figure}[t!]
\centering
    \includegraphics[width=\textwidth]{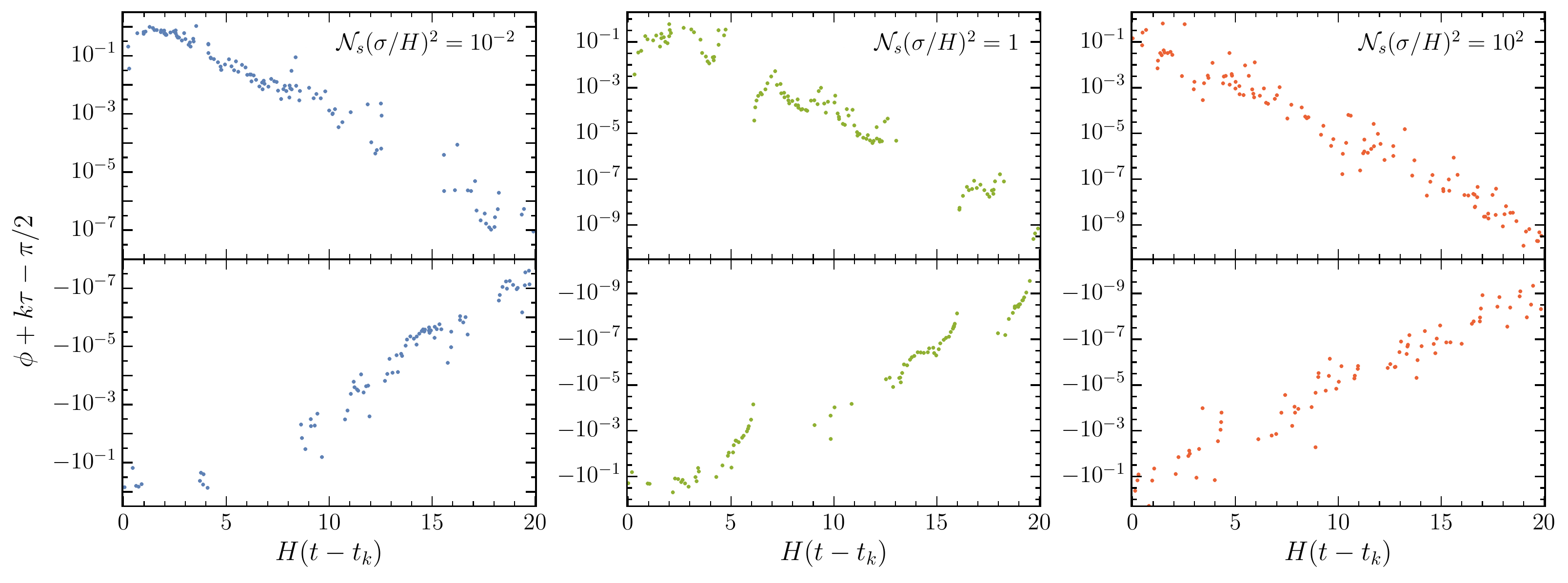}
    \caption{Super-horizon evolution of the transfer matrix phase $\phi$ in the weak and strong scattering regimes, for a conformally massive spectator field. We have shifted its value by $k\tau-\pi/2$ to demonstrate the asymptotic behavior. Notice that the sign of $\phi+k\tau-\pi/2$ continues to flip randomly even as $|\phi+k\tau-\pi/2|\rightarrow 0$. }
    \label{fig:phigrid}
\end{figure}

The curious behavior of $\arg X$ for moderate and strong scattering is displayed in a clearer fashion in Fig.~\ref{fig:reimchi}. There, the evolution of $X$ is shown in the complex plane for $\mathcal{N}_s(\sigma/H)^2=1$ and $10^2$. It is immediately clear that, as discussed above, the field amplitude is constant in time in subhorizon scales, and the random, but uniformly distributed phase, results in a random walk of the (re-scaled) field $X$ on the unit circle. As $k\sim aH$, the phase of $X$ locks along a random line in the complex plane, and $X$ evolves along this ray; for strong scattering it grows exponentially, jumping between diametrically opposite directions. These diametrical jumps can be understood as follows. In terms of the transfer matrix $\M$-parameters $\{n,\phi,\psi\}$ defined in \eqref{eq:Mparg}, the spectator field can be written in general as
\beq\label{eq:genchi}
X \;=\; \frac{1}{\sqrt{2k}} \left[ (1+n)^{1/2}e^{-i(\phi+\psi+k\tau)} + n^{1/2}e^{i(\phi-\psi+k\tau)} \right]\,.
\eeq
Inside the horizon, $n\ll 1$, and $\arg X \simeq -(\phi+\psi+k\tau) \pmod {2\pi}$, which is randomly distributed, not only because $\phi$ and $\psi$ themselves are, but most importantly because $|k\tau|\gg 1$, completely scrambling the phase. Outside the horizon, however, $n\gg 1$, and 
\begin{flalign}\label{eq:chisuph}
& \text{($|k\tau|\ll 1$)} & \Cen{3}{X \;\simeq\; \sqrt{\frac{2n}{k}}\cos(\phi+k\tau)e^{-i\psi} \,,}      &&  
\end{flalign}
which implies that $\arg \chi$ is mostly determined by $\psi$. However, a curious behavior regarding the sign of $X$ arises due to the asymptotic behavior of $\phi$. As we discussed above, Fig.~\ref{fig:all_grid} shows that $\phi\rightarrow \pi/2$ as $|k\tau|\rightarrow 0$. Moreover, Fig.~\ref{fig:phigrid} demonstrates that the argument of the cosine in (\ref{eq:chisuph}) is driven exponentially fast in cosmic time towards $\pi/2$, alternating signs randomly. Straightforward expansion implies then that
\begin{flalign}\label{eq:chisuph2}
& \text{($|k\tau|\ll 1$)} & \Cen{3}{X \;\simeq\; -\sqrt{\frac{2n}{k}}(\phi+k\tau-\pi/2)\,e^{-i\psi} \;=\; \sqrt{\frac{2n}{k}}|\phi+k\tau-\pi/2|\, e^{i(\zeta\pi-\psi)}\,,}      &&  
\end{flalign}
with $\zeta=0\,\,{\rm or}\,\,1$ after each scattering.
\subsubsection{Means and variances}\label{sec:meanvarconf}
In the previous subsection we discussed the evolution of the transfer matrix parameters $\{n,\phi,\psi\}$ and the scalar field amplitude $|X|^2$ and its argument for particular realizations of the locations and amplitudes of the scattering events. We now turn to the description of the dynamics of the system given an ensemble of realizations of the scatterers. In this section we will discuss the evolution of the lowest moments of the angles $\phi$ and $\psi$, as well as those for $\ln(1+n)$ and $\ln|X|^2$; in Section~\ref{sec:pdfs} we study the form of their probability distributions. Note the focus on logarithms of $n$ and $|X|$ is related to the observation that both $n$ and $|X|$ show an exponential behavior with cosmic time. We will also find that $\ln |X|^2$ is normally distributed both inside and outside the horizon (for any strength of scattering), making it a simpler variable to work with.

\begin{figure}[t!]
\centering
    \includegraphics[width=0.95\textwidth]{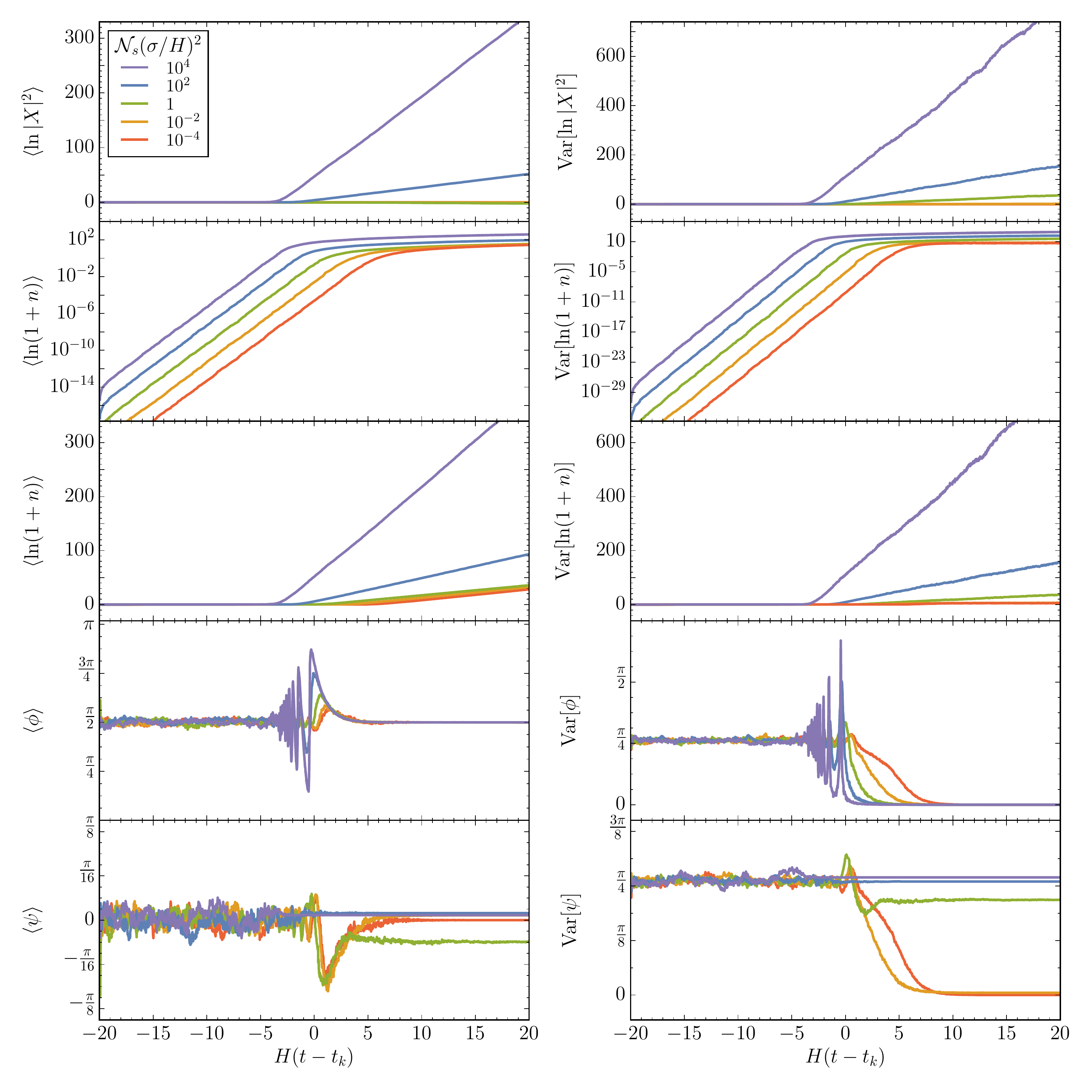}
    \caption{Sub- and super-horizon evolution of the mean and variance of the transfer matrix parameters $\{n,\phi,\psi\}$ and the re-scaled scalar field amplitude $|X|$, for different values of $\mathcal{N}_s(\sigma/H)^2$ in the conformal case. On super-horizon scales, the means and variances of $\ln |X|^2$ and $\ln (1+n)$ evolve linearly with cosmic time with growth rates depending on $\mathcal{N}_s(\sigma/H)^2$. On sufficiently sub-horizon scales, the growth rates are independent of $\mathcal{N}_s(\sigma/H)^2$. Note that the second and third rows are the same data points displayed with and without a log scale to demonstrate exponential and linear behavior with cosmic time on super- and sub-horizon scales respectively. The behavior of the means and variances of $\phi$ and $\psi$ are discussed in the text. Here we have taken $k=e^{20}H$, $N_s=2000$ and the averages and variances are taken over 2000 different realizations of the amplitudes and locations of the non-adiabatic interactions. }
    \label{fig:sigmeanvar}\vspace{10pt}
\end{figure}

\begin{figure}[t!]
\centering
    \includegraphics[width=\textwidth]{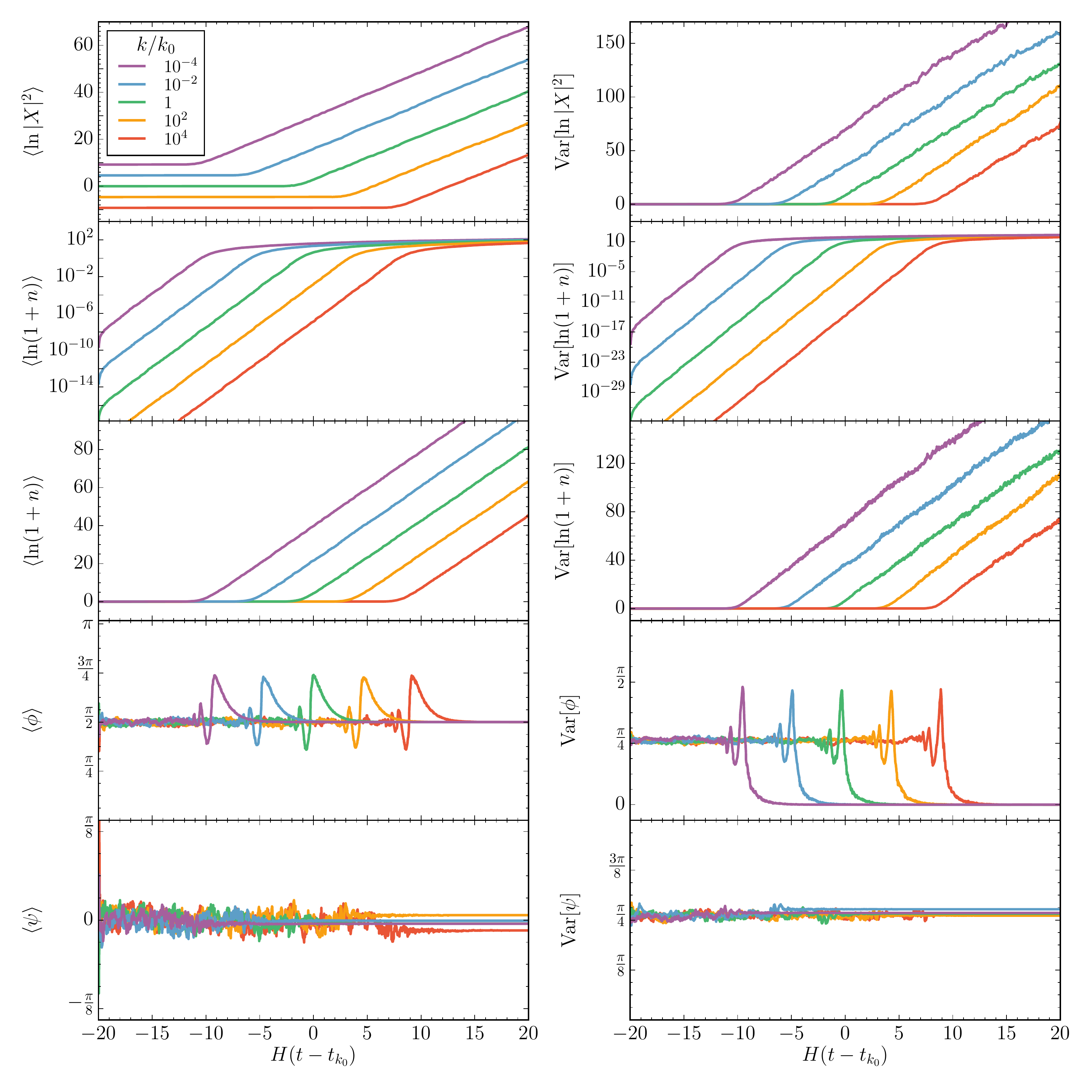}
    \caption{Sub- and super-horizon evolution of the mean and variance of the transfer matrix parameters $\{n,\phi,\psi\}$ and the scalar field amplitude, for different values of $k/H$ in the conformal case. For $\ln(1+n)$ and $\ln |X|^2$, while horizon crossing is determined by the $k$, the growth rates inside and outside the horizon become independent of $k$ as is to be expected. The behavior for the phases is discussed further in the text. Here $\mathcal{N}_s(\sigma/H)^2=60$, $N_s=2000$ and we consider 2000 different realizations of the amplitudes and locations of the non-adiabatic interactions. The time interval is chosen to be symmetric with respect to the time $t_{k_0}$ when the mode with momentum $k_0=e^{20}H$ crosses the horizon. Here all re-scalings of $X$ are to be taken with respect to $k_0$, $X\rightarrow \sqrt{2k_0}\,X$.}
    \label{fig:kmeanvar}
\end{figure}

Figs.~\ref{fig:sigmeanvar} and \ref{fig:kmeanvar} show the dependence of the mean and variance of $\ln(1+n)$, $\ln|X|^2$, $\phi$ and $\psi$ on the scattering-strength parameter $\mathcal{N}_s(\sigma/H)^2$ and the wavenumber $k$, respectively. For Fig.~\ref{fig:sigmeanvar}, we have set $k/H=e^{20}$ for definiteness, while for Fig.~\ref{fig:kmeanvar} we have fixed $\mathcal{N}_s(\sigma/H)^2 =60$. In both cases we consider $N_s=2000$ non-adiabatic events, drawn from an ensemble of 2000 members for the scattering locations and strengths $m_j$ and $\delta t_j$. For simplicity we have assumed that the amplitudes and locations of the scatterings are drawn from uniform distributions, as in Fig.~\ref{fig:all_grid}. Nevertheless, we will show in Appendix~\ref{ap:conv} that the results discussed here are not sensitive to the ensemble distributions, provided that both the $m_j$ and $\delta t_j$ are random. \par\bigskip
\par

\noindent{\bf Moments of $\ln |X|^2$ on sub-horizon scales}: 
In the first row of Figs.~\ref{fig:sigmeanvar} and \ref{fig:kmeanvar}, the panels show the evolution of the moments for the (logarithmic) scalar field amplitude. The mean $\langle \ln|X|^2\rangle$ is approximately constant on sub-horizon scales, and independent of the value of $\mathcal{N}_s(\sigma/H)^2$. In light of (\ref{eq:rescx}), we can then simply write
\begin{flalign}\label{eq:subhlnch}
& \text{($|k\tau|\gg 1$)} & \Cen{3}{\langle \ln|X|^2\rangle \;\simeq\; - \ln(2k)\,.}      &&  
\end{flalign}
Although it is not obvious from the figure, unlike the mean, the variance of $\ln|X|^2$ increases exponentially with cosmic time, with a rate that is independent of $\mathcal{N}_s(\sigma/H)^2$ and $k$. Its functional dependence can be approximated as
\begin{flalign}\label{eq:subhvarlnch}
& \text{($|k\tau|\gg 1$)} & \Cen{3}{ {\rm Var}\left[ \ln|X|^2\right] \;\simeq\; \frac{\mathcal{N}_s}{4}\left(\frac{\sigma}{k_{\rm phys}}\right)^2\;=\; \frac{1}{4}\mathcal{N}_s\frac{\sigma^2}{H^2}\left(k\tau\right)^{-2}\,,}      &&  
\end{flalign}
where $k_{\rm phys}=k/a$. Note that while the mean of $\ln |X|^2$ remains similar to its value in the vacuum, the variance is growing $\propto a^2$. \par\bigskip 

\begin{figure}[t!]
\centering
    \includegraphics[width=0.87\textwidth]{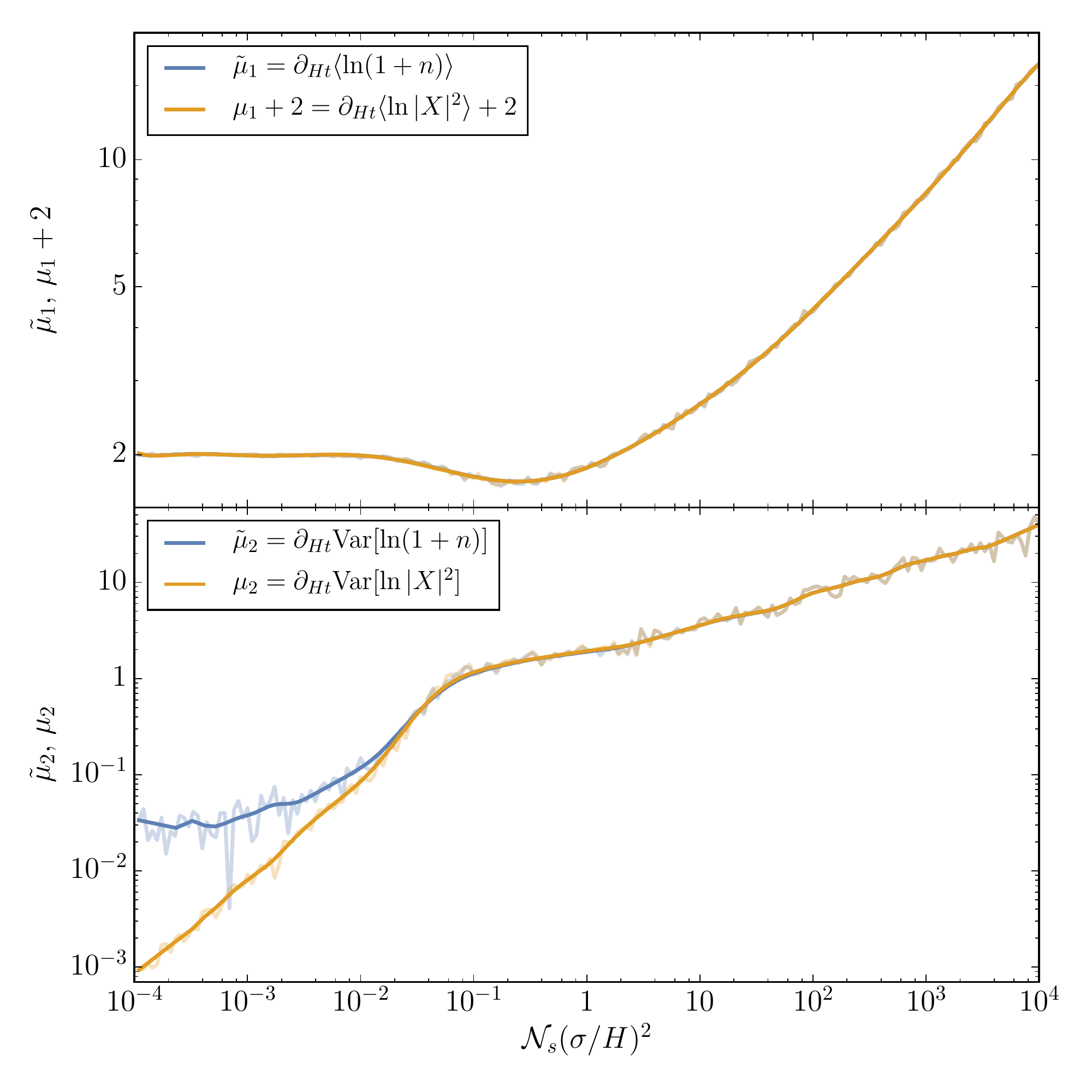}
    \caption{Numerically evaluated rates of growth on super-horizon scales for the mean and variance of the  log of occupation number and the scalar field mode amplitude as functions of the parameter $\mathcal{N}_s(\sigma/H)^2$. Note $\partial_{Ht}\langle\ln |X|^2\rangle$, $\partial_{Ht}\textrm{Var}[\ln |X|^2]\ll 1$ for sufficiently weak scattering, as expected.   For very strong scattering, the rates are approximate power laws in $\mathcal{N}_s(\sigma/H)^2$. We do not yet understand the curious dip in the rates near $\mathcal{N}_s(\sigma/H)^2\sim 1$ (see top panel), which indicates the fields decay faster than the case where there is no scattering. For the above plots $M^2=2H^2$ for the spectator field, and we have chosen $k/H=e^{20}$, $N_s=3000$ over a total of 40 Hubble times and we have checked that these rates are independent of $k$ on super-horizon scales. The plotted values correspond to the average of 400 realizations per value of $\mathcal{N}_s(\sigma/H)^2$ (transparent), further smoothed with a polynomial fit (solid). Amplitudes and locations of scatterers are drawn from uniform distributions.
    }
    \label{fig:nfrates}
\end{figure}

\noindent{\bf Moments of $\ln |X|^2$ on super-horizon scales}: 
In the case of very weak scattering, $\ln|X|^2$ continues to be approximately constant as it is inside the horizon; it is only for strong scattering that the field can overcome the expansion and either decay or grow. Moreover, this rate is also independent of $k$. The variance of $\ln|X|^2$ is also a linear function of cosmic time for $|k\tau|\ll 1$, with a rate that is independent of the wavenumber $k$. We can then write 
\begin{flalign} \label{eq:lnchirateconf}
& \text{($|k\tau|\ll 1$)} & \Cen{3}{
\begin{aligned}
\partial_{Ht} \langle \ln |X|^2\rangle \;&=\; \mu_1 \left(\mathcal{N}_s(\sigma/H)^2\right)\,,\\
\partial_{Ht} {\rm Var}\left[ \ln |X|^2\right] \;&=\; \mu_2\left(\mathcal{N}_s(\sigma/H)^2\right)\,,
\end{aligned}}      &&  
\end{flalign}
where $\mu_1$ and $\mu_2$ are functions of $\mathcal{N}_s(\sigma/H)^2$ and are shown in Fig.~\ref{fig:nfrates}. The curves displayed in the figure are the result of a linear fit to the averaged moments over 400 realizations in the super-horizon regime.

For $\mathcal{N}_s(\sigma/H)^2\ll 10^{-2}$, $\partial_{Ht}\langle \ln|X|^2\rangle\sim 0$. Along with Fig.~\ref{fig:nfrates}, this feature is also evident in the top left panel of Fig.~\ref{fig:sigmeanvar}. For $10^{-2}\lesssim \mathcal{N}_s(\sigma/H)^2\lesssim 2$, the rate of change of $\langle \ln|X|^2\rangle$ is negative (notice the dip in the top panel of  Fig.~\ref{fig:nfrates}). This peculiar behavior is also demonstrated by the green curve in the third panel on the right of Fig.~\ref{fig:sigmeanvar}. In other words, $\langle \ln|\chi|^2\rangle$ decays faster than in the vacuum (recall that $X=a\chi$) in this regime of $\mathcal{N}_s(\sigma/H)^2$. If scattering is stronger, this rate of decay is smaller, until it vanishes for $\mathcal{N}_s(\sigma/H)^2\simeq 60$. When this is the case, 
$\langle \ln|\chi|^2\rangle \sim {\rm const.}$, or equivalently $\langle \ln|X|^2\rangle \sim a^2$ while scatterings continue taking place. For $\mathcal{N}_s(\sigma/H)^2\gg 1$, the field grows exponentially, with rate $\mu_1\sim \left[ \mathcal{N}_s(\sigma/H)^2 \right]^{0.3}-2$ (top panel of see Fig. \ref{fig:nfrates}).

The dependence of ${\mu}_2=\partial_{Ht}{\rm Var}[\ln|X|^2]$ on $\mathcal{N}_s (\sigma^2/H^2)$ is shown in the lower panel of Fig.~\ref{fig:nfrates}. A remarkable feature of this variation is the sharp distinction in the evolution for weak and strong scattering. In the former case, the variance grows with $\mu_2 \sim  \mathcal{N}_s(\sigma/H)^2 $, while in the later 
it grows as 
 $\mu_2\sim \left[ \mathcal{N}_s(\sigma/H)^2 \right]^{0.33}$.\par\bigskip

\noindent{\bf Moments of $\ln(1+n)$ on sub-horizon scales}: 
The second row of Fig.~\ref{fig:sigmeanvar} shows the evolution of $\langle \ln(1+n)\rangle$ and Var$[\ln(1+n)]$ in a log-scale, both showing an exponential growth ($\propto a^2$) in the sub-horizon regime. The growth rates are independent of $\mathcal{N}_s(\sigma/H)^2$, but the absolute magnitude of the moments depends on it. Notice that this trend is maintained until $\langle \ln(1+n)\rangle \sim \mathcal{O}(1)$, which occurs shortly after horizon crossing for weak scattering and before for strong scattering. The second row of Fig.~\ref{fig:kmeanvar} further shows that while the mean and the variance on sub-horizon scales depend on the value of $k$, their growth rates are independent of $k$. These observations lead us to write:
\begin{flalign}\label{eq:subhnres}
& \text{($|k\tau|\gg 1,\,n\ll 1$)} & \Cen{3}{\langle \ln(1+n)\rangle \;\simeq\; \left( {\rm Var}\left[ \ln(1+n)\right]\right)^{1/2} \;\simeq\; \frac{\mathcal{N}_s}{8}\left(\frac{\sigma}{k_{\rm phys}}\right)^2\,,}&&
\end{flalign}
which we derive analytically in Section~\ref{sec:analconf1}. Note that the exponential growth of $\langle \ln(1+n)\rangle$ and Var$[\ln(1+n)]$ comes from $k_{\rm phys}=k/a$ where $a=e^{H(t-t_i)}$. At first sight this seems inconsistent with the results from earlier papers by some of us \cite{Amin:2015ftc,Amin:2017wvc} which calculated particle production in a non-expanding universe and found that $\ln (1+n)$ grows linearly with time. On sufficiently sub-horizon scales, one could expect the above result and the non-expanding case to agree. A closer look reveals that there is no inconsistency. The linear growth of $\ln(1+n)$ was really true for $n\gg 1$, whereas above we have $n\ll 1$. In detail, for $n\ll 1$, since $\ln(1+n)\sim n$, both $\ln(1+n)$ and $n$ grow exponentially with cosmic time. When $n\gg 1$, find that $\ln (1+n)\sim \ln (n)$ will grow linearly with cosmic time, whereas $n$ will grow exponentially.
\\ \\
\noindent{\bf Moments of $\ln(1+n)$ on super-horizon scales}: 
Third row from the top in Figs.~\ref{fig:sigmeanvar} and \ref{fig:kmeanvar}, the left and right panels  show the evolution of the moments of the occupation number, but now in a linear scale to demonstrate the linear increase of $\langle \ln(1+n)\rangle \simeq \langle \ln(n)\rangle$ for $|k\tau|\ll 1$ (ie. on super-horizon scales). In this regime the growth rate is clearly dependent on $\mathcal{N}_s(\sigma/H)^2$, being steeper for strong scattering, and it is seemingly independent of the wavenumber $k$. In analogy with $\ln |X|^2$, we define these super-horizon growth rates of $\langle \ln(1+n)\rangle$ and Var$[\ln(1+n)]$ as follows:
\begin{flalign}
& \text{($|k\tau|\ll 1$)} & \Cen{3}{
\begin{aligned}
\partial_{Ht} \langle \ln(1+n)\rangle \;&=\; \tilde{\mu}_1\left(\mathcal{N}_s(\sigma/H)^2\right)\,,\\
\partial_{Ht} {\rm Var}\left[ \ln(1+n)\right] \;&=\; \tilde{\mu}_2\left(\mathcal{N}_s(\sigma/H)^2\right)\,.
\end{aligned}}      &&  
\end{flalign}
The functions $\tilde{\mu}_1$ and $\tilde{\mu}_2$ are also shown in Fig.~\ref{fig:nfrates}. As seen in this figure, these rates are closely connected with the corresponding rates for $\ln |X|^2$. Explicitly, $\tilde{\mu}_1\simeq \mu_1+2$ everywhere, whereas $\tilde{\mu}_2\simeq\mu_2$ for $\mathcal{N}_s(\sigma/H)^2\gtrsim 10^{-1}$.

For $\mathcal{N}_s(\sigma/H)^2\leq 10^{-2}$, the growth rate of the mean is constant, $\partial_{Ht}\langle \ln(1+n)\rangle \simeq 2$. For the occupation number, this implies that $n\sim a^2$ at all times when scattering is weak. We will revisit this result with the Fokker-Planck formalism in Section~\ref{sec:analconf}. For $10^{-2}\lesssim \mathcal{N}_s(\sigma/H)^2\lesssim 2$, the typical occupation number grows at a slower rate. In the strong scattering regime, the mean grows in a power-like fashion with the scattering strength parameter, $\tilde{\mu}_1\sim \left[ \mathcal{N}_s(\sigma/H)^2 \right]^{0.3}$.

The rate of growth with time for the variance of $\ln(1+n)$ is shown in the lower panel of Fig.~\ref{fig:nfrates}. For $\mathcal{N}_s(\sigma/H)^2\ll 1$, it is approximately constant, $\tilde{\mu}_2\sim 0.025$, but it rises sharply as the scattering parameter increases. For $\mathcal{N}_s(\sigma/H)^2\gtrsim 10^{-1}$ the rate follows a power-law dependence, $\tilde{\mu}_2\sim \left[ \mathcal{N}_s(\sigma/H)^2 \right]^{0.33}$, slightly steeper than that of the mean. \\

It is important to note that the rapid growth of the variance in sub- and super-horizon scales (for both $\ln (1+n)$ and $\ln |X|^2$) sheds some doubt on our characterization of their means corresponding to the ``most probable'' member of the ensemble of realizations. We dismiss these concerns in detail in Appendix~\ref{ap:typ}, by constructing ratios of means and standard deviations of these quantities and showing that while the standard deviations grow, the means grow even faster.\par\bigskip

\noindent{\bf Moments of $\phi$ on sub-horizon scales}:
The evolution of the angular parameter $\phi$ is shown in the fourth row in Figs.~\ref{fig:sigmeanvar} and \ref{fig:kmeanvar}. Deep inside the horizon we find
\begin{flalign}
& \text{($|k\tau|\gg 1$)} & \Cen{3}{\langle \phi\rangle \simeq \frac{\pi}{2}\,,\qquad {\rm Var}\left[ \phi \right] \simeq \frac{\pi^2}{12}\,,}      &&  
\end{flalign}
values which are consistent with a uniformly distributed random variable in $(0,\pi)$.\\
\\
\noindent{\bf Moments of $\phi$ on super-horizon scales}: 
As the mode leaves the horizon, both the mean and variance oscillate about these values, with the amplitude of these oscillations being dependent on the scattering strength parameter. Once the mode is far outside the horizon, the oscillations stop, and the moments settle down to
\begin{flalign}\label{eq:phimvcn}
& \text{($|k\tau|\ll 1$)} & \Cen{3}{\langle \phi\rangle \simeq \frac{\pi}{2}\,,\qquad {\rm Var}\left[ \phi \right] \rightarrow 0\,,}      &&  
\end{flalign}
with an exponentially decreasing variance. Numerically we find that the final value (at $H(t-t_k)=20$) of the mean of $\phi$ is equal to $\pi/2$ for all $\mathcal{N}_s(\sigma/H)^2$ up to a numerical error smaller than one part in $10^{-6}$. We also find for the time rate of the log of the variance that
\begin{flalign}\label{eq:phivrcn}
& \text{($|k\tau|\ll 1$)} & \Cen{3}{\partial_{Ht}\ln\left({\rm Var}\left[ \phi \right] \right) \;\simeq\; -1\,,}      &&  
\end{flalign}
for any scattering strength, up to a $\lesssim 8\%$ deviation that lacks a simple dependence on $\mathcal{N}_s(\sigma/H)^2$.\par\bigskip

\noindent{\bf Moments of $\psi$ on sub-horizon scales}: 
Finally, the time-dependence of the moments of $\psi$ are shown in the bottom left and right panels of Figs.~\ref{fig:sigmeanvar} and \ref{fig:kmeanvar}. Similarly to $\phi$, these results are consistent with a uniformly distributed random variable inside the horizon (on the interval $(-\pi/2,\pi/2)$):
\begin{flalign}\label{eq:psmvsub}
& \text{($|k\tau|\gg 1$)} & \Cen{3}{\langle \psi\rangle \simeq 0\,,\qquad {\rm Var}\left[ \psi \right] \simeq \frac{\pi^2}{12}\,.}      &&  
\end{flalign}

Also similar to the $\phi$ case, the mean and variance of $\psi$ appears to be perturbed away from these values during horizon crossing.\par\bigskip

\noindent{\bf Moments of $\psi$ on super-horizon scales}: In the super-horizon regime, both moments of $\psi$ appear to asymptote to values dependent on the scattering strength parameter. From Fig.~\ref{fig:sigmeanvar} it is apparent that $\psi$ retains its uniform distribution for strong scattering, while for weak scattering the moments are consistent with a narrow probability density centered at $\psi\simeq 0$. Numerically we find that a good approximation to the lowest moments of $\psi$ is given by
\begin{flalign}\label{eq:psimvsup}
& \text{($|k\tau|\ll 1$)} & \Cen{3}{
\begin{aligned}
\langle \psi \rangle &\;\simeq\; 0\,,\\
{\rm Var}\left[ \psi \right] &\;\simeq\;  \frac{\pi}{2}\times\begin{cases}
\mathcal{N}_s(\sigma/H)^2\,, & \mathcal{N}_s(\sigma/H)^2 \lesssim \pi/6\\[5pt]
\dfrac{\pi}{6}\,, & \mathcal{N}_s(\sigma/H)^2 \gtrsim \pi/6
\end{cases}\,.
\end{aligned}}      &&  
\end{flalign}
To arrive to the previous expressions we have ignored a mild but complicated dependence on the scattering parameter for $\langle\psi\rangle$. The maximum deviation is found at $\mathcal{N}_s(\sigma/H)^2\sim 1$, for which $|\langle \psi\rangle| \lesssim 0.1$. Note that the variance freezes at small values in the case of weak scattering, while for strong scattering it freezes with the same value as in (\ref{eq:psmvsub}), consistent with a non-evolving probability distribution.\\

In our previous discussion we have mostly focused on the super-horizon behavior of the transfer matrix parameters under the assumption that this late-time evolution is controlled only by the scattering strength parameter $\mathcal{N}_s(\sigma/H)^2$. Furthermore, all the results previously presented assume an underlying uniform distribution for the strength and location of the non-adiabatic events that drive particle creation. Nevertheless, it can be shown that these results are unchanged if the previously mentioned assumptions are broken, as long as the density of scatterers $\mathcal{N}_s$ is sufficiently high. See Appendix~\ref{ap:conv} for details.

\subsubsection{Probability densities}\label{sec:pdfs}
\begin{figure}[t!]
\centering
    \includegraphics[angle=270,origin=c,width=0.97\textwidth]{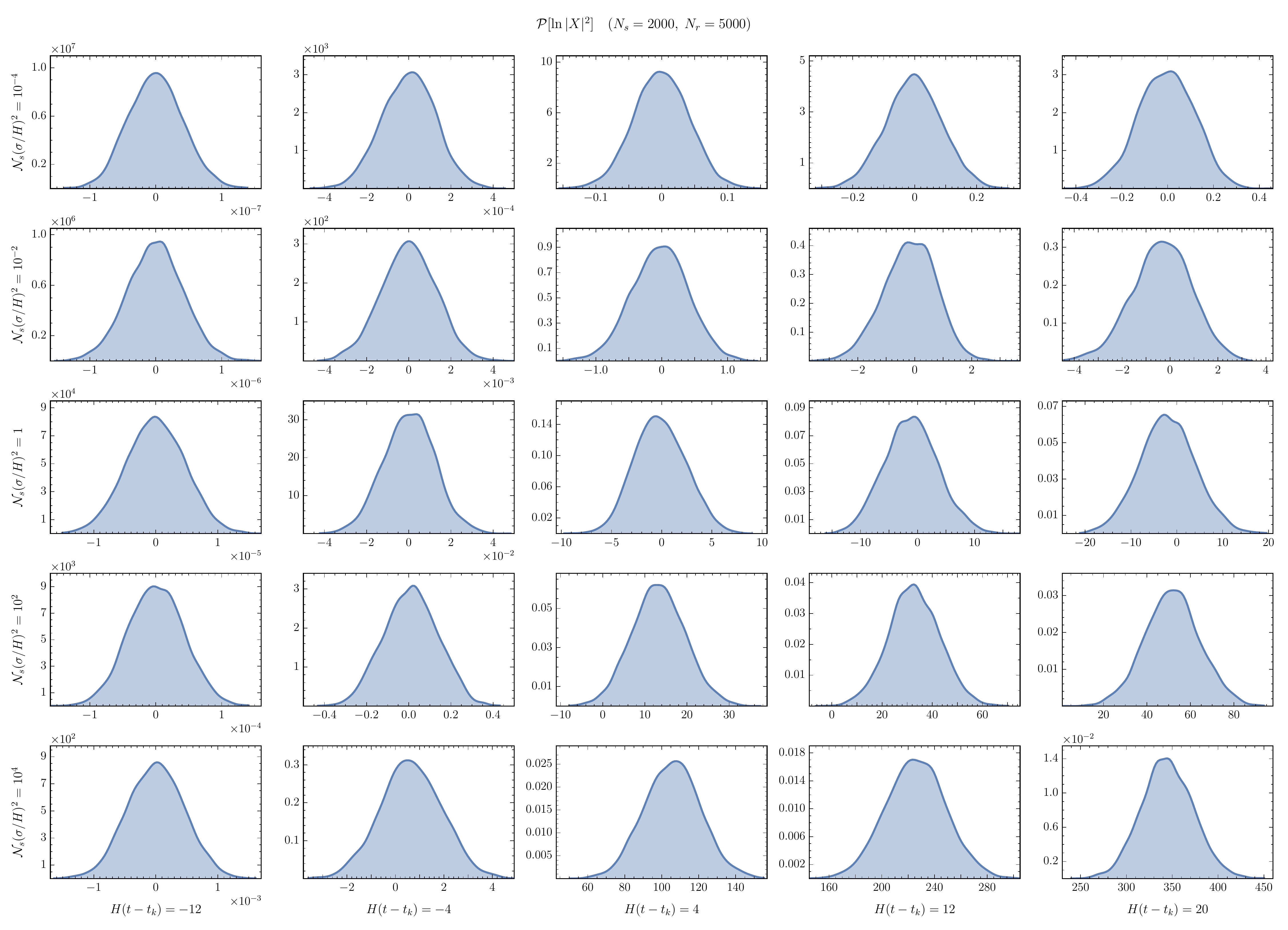}
    \caption{Pdf for $\ln|X|^2$ as a function of time and scattering strength. The distribution is always lognormal, on sub- and super-horizon scales, for weak, moderate strong scattering.}
    \label{fig:fdist}\vspace{-50pt}
\end{figure}
\begin{figure}[t!]
\centering
    \includegraphics[angle=270,origin=c,width=\textwidth]{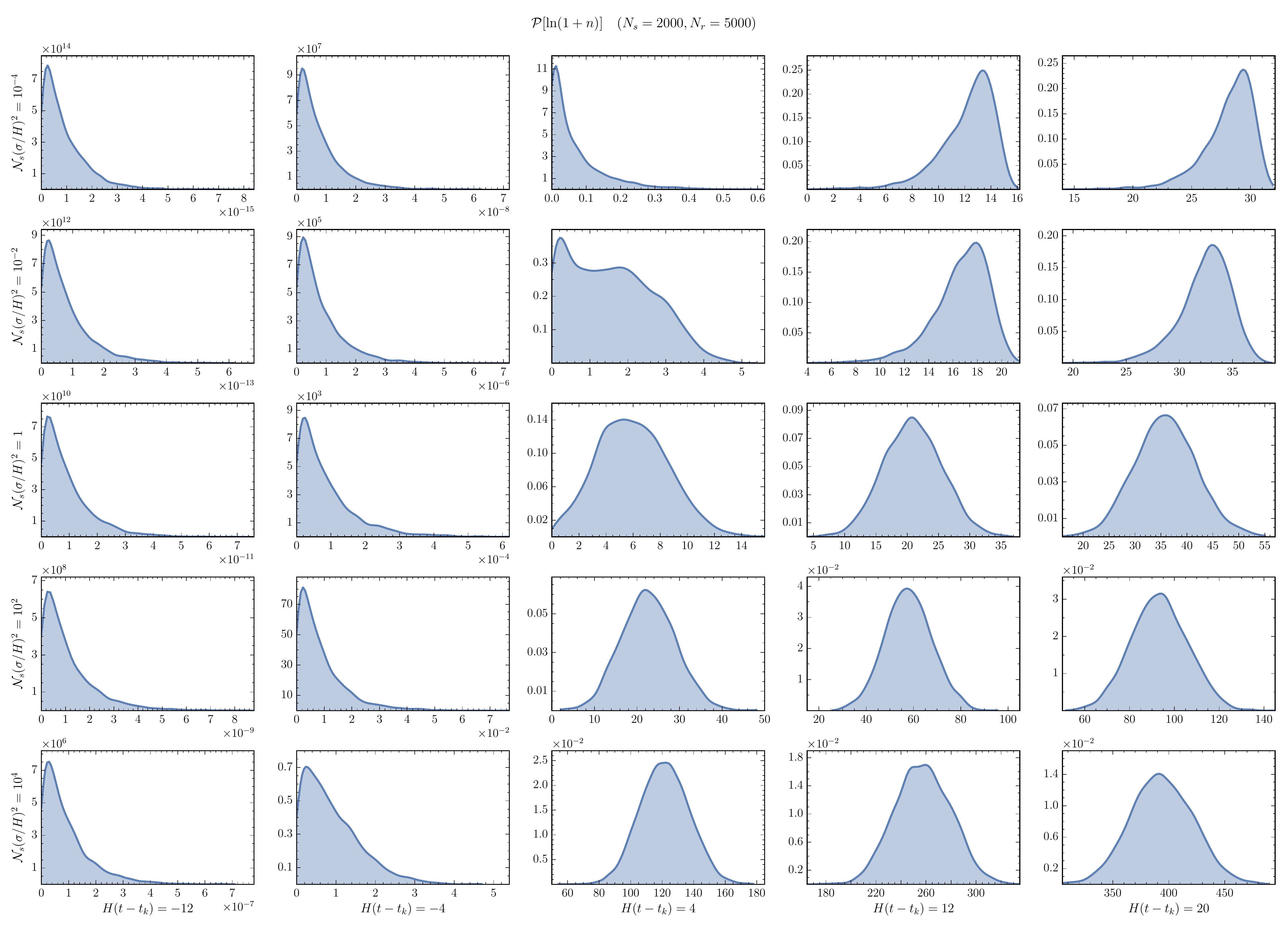}
    \caption{Pdf for $\ln(1+n)$ as a function of time and scattering strength.}
    \label{fig:ndist}\vspace{-50pt}
\end{figure}
\begin{figure}[t!]
\centering
    \includegraphics[angle=270,origin=c,width=\textwidth]{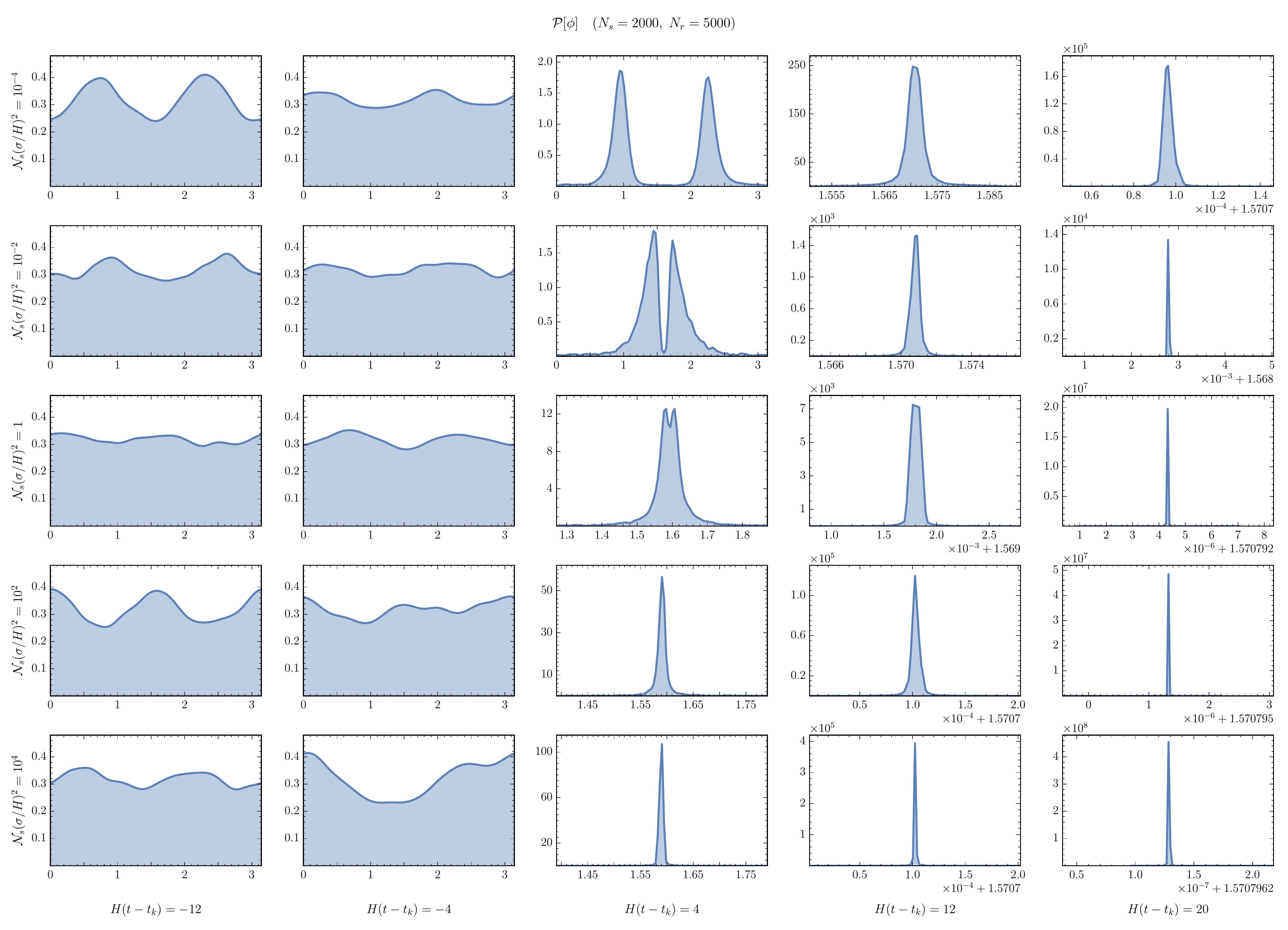}
    \caption{Pdf for $\phi$ as a function of time and scattering strength.}
    \label{fig:phdist}\vspace{-50pt}
\end{figure}
\begin{figure}[t!]
\centering
    \includegraphics[angle=270,origin=c,width=\textwidth]{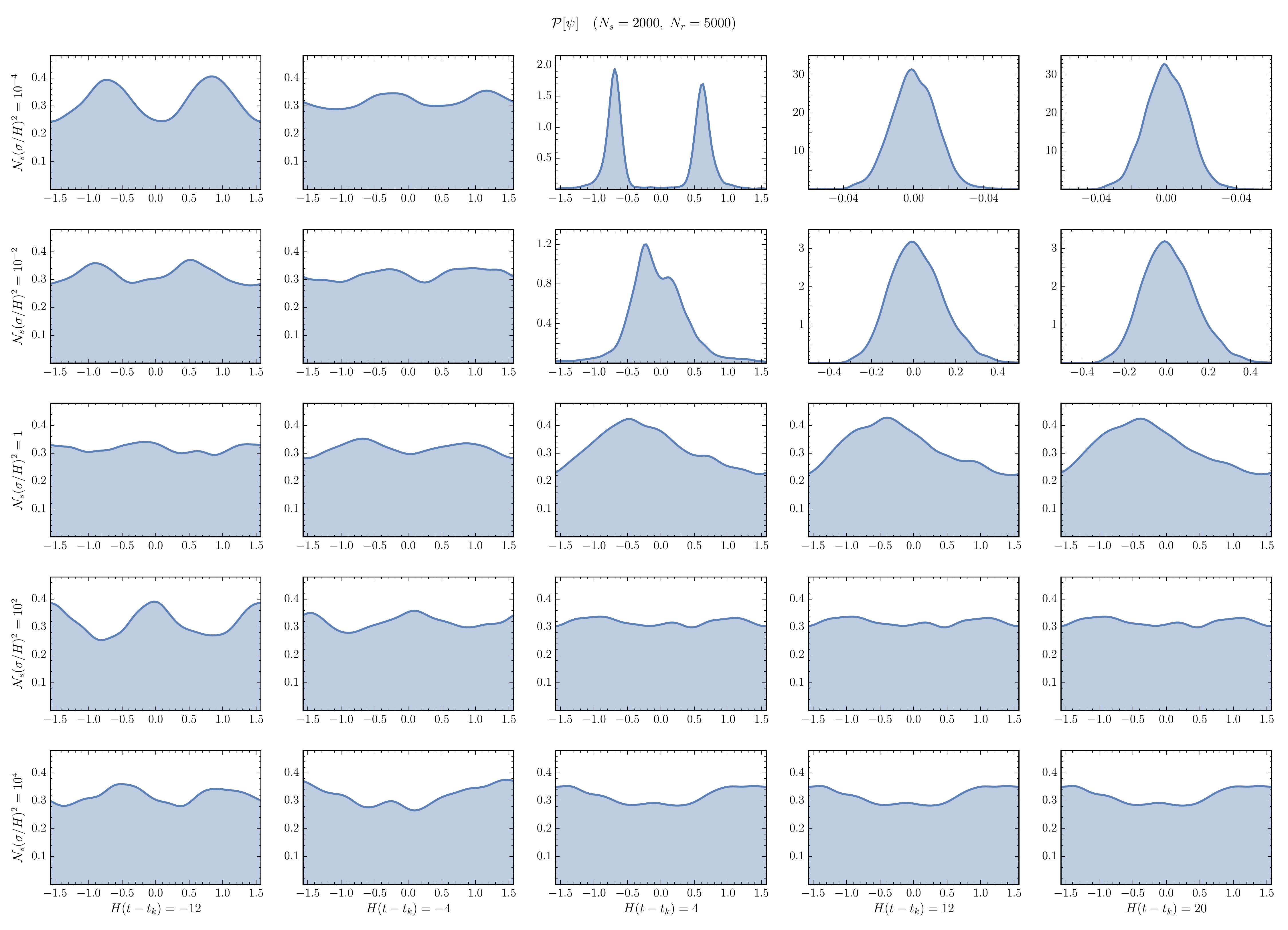}
    \caption{Pdf for $\psi$ as a function of time and scattering strength.}
    \label{fig:psdist}\vspace{-50pt}
\end{figure}
In the previous section we have described the sub- and super-horizon evolution of the lowest moments of $\ln |X|^2$ and  the transfer matrix parameters. In this section we now study in a mostly qualitative fashion the form and dynamics of the full probability density functions (pdf) for these random variables.

\begin{figure}[t!]
\centering
    \includegraphics[width=0.94\textwidth]{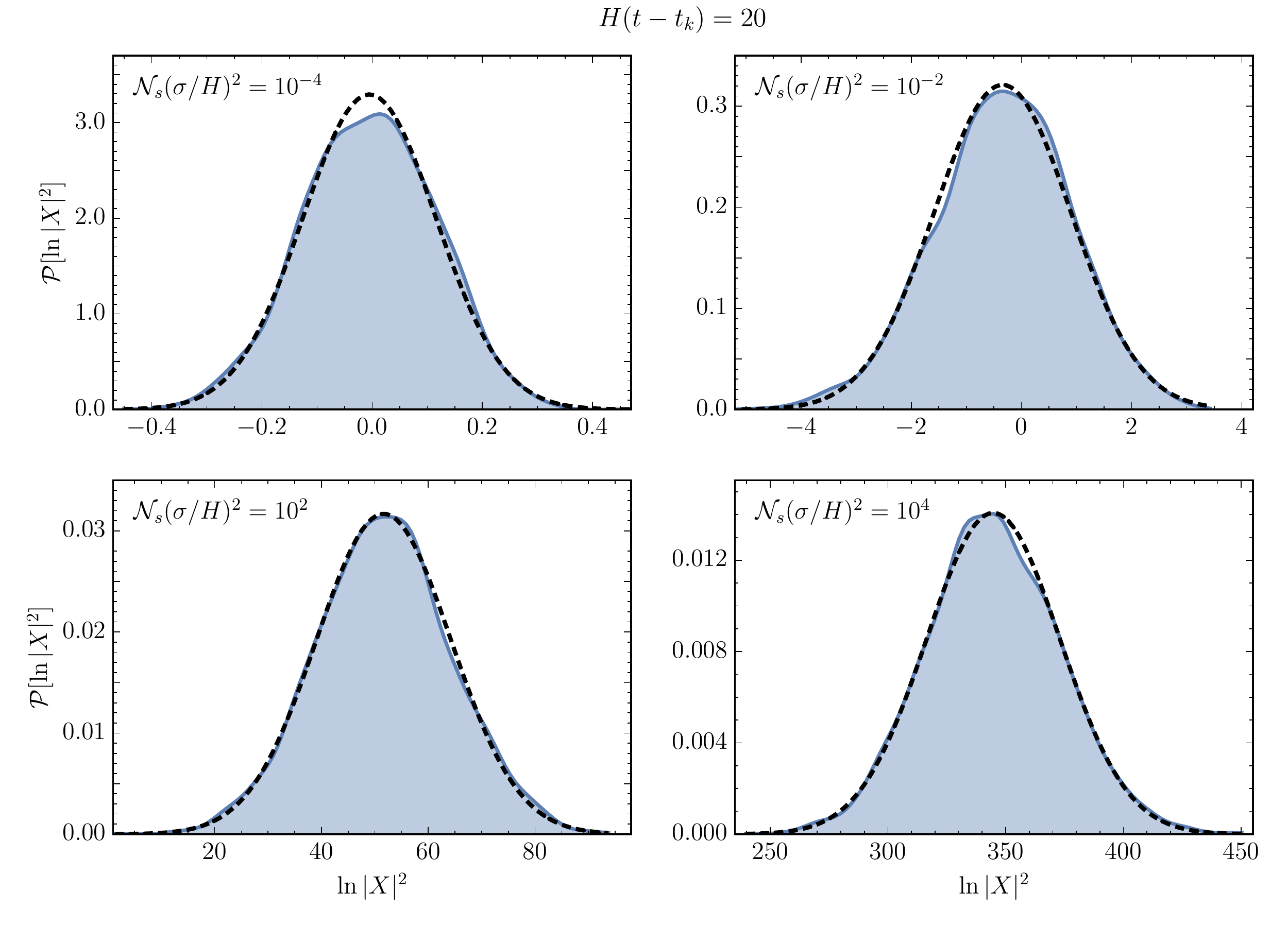}
    \caption{Pdf for $\ln|X|^2$ for selected values of $\mathcal{N}_s(\sigma/H)^2$ at $H(t-t_k)=20$ (conformal case). Blue, continuous: numerical result. Black, dashed: Gaussian fit.}
    \label{fig:ffit}
\end{figure}

Figs.~\ref{fig:ndist}-\ref{fig:psdist} display snapshots of the time evolution of the instantaneous normalized pdfs for the field and $\M$-parameters for selected values of $\mathcal{N}_s(\sigma/H)^2$. In all cases we have considered $k/H=e^{20}$ and $N_s=2000$ with $N_r=5000$ realizations. As before, the total evaluation time interval corresponds to $H(t_i-t_f)=40$. The pdfs are built using a Gaussian kernel density estimator of variable bin size. For the angular variables $\phi$ and $\psi$, the data has been extended periodically in the cases where the pdf support is of size $\pi$, to minimize edge effects. \\
\\
\noindent{\bf Pdf for $\ln|X|^2$}: 
Fig.~\ref{fig:fdist} shows the pdfs for the logarithmic field amplitude $\ln|X|^2$. The description of the distribution is exceptionally simple: for all times and values of $\mathcal{N}_s(\sigma/H)^2$ a normal pdf is a good fit for the data.  We prove this fact analytically for sub-horizon modes in Section~\ref{sec:analconf1} (see Fig.~\ref{fig:ffitg}). Far outside the horizon, at $H(t-t_k)=20$, the normal form of the pdf is demonstrated in Fig.~\ref{fig:ffit}, where a Gaussian fit is superimposed for small and large values of the scattering strength. We can therefore conclude that:\par\medskip

\noindent\rule{\textwidth}{1pt}\\
The squared-field amplitude, $|X|^2$, follows a log-normal distribution both inside and outside the horizon, for weak and strong scattering.\\
\noindent\rule{\textwidth}{1pt}
\par

\noindent{\bf Pdf for $\ln(1+n)$}: 
The instantaneous pdfs for $\ln(1+n)$ are shown in Fig.~\ref{fig:ndist}. In the first two columns from the left, turning a blind eye to the axis tick values, it is clear that this pdf ``flows'' in a way almost independent of the strength of the scattering. At very early times it starts with a highly right-skewed, almost exponential shape.\footnote{This is consistent with a coefficient of variation $\tau_{\ln(1+n)} \simeq 1$ (see Appendix~\ref{ap:typ}).} In Section~\ref{sec:analconf1} we will confirm this fact analytically (see Fig.~\ref{fig:nfit}). As time increases, the position of the maximum increases, together with the width of the distribution, maintaining its shape in all cases save for the  strongest scattering case, where the shape is now slightly distorted. It is not until the mode is stretched to super-horizon scales that  the difference between weak and strong scattering is evident. 

The middle column of Fig.~\ref{fig:ndist} shows the transition regime, and demonstrates the delay in evolution of the weak scattering case compared to the strong scattering ones. As it is clearly exhibited by the $\mathcal{N}_s(\sigma/H)^2=10^{-2}$ case, the pdf shifts from the left-lobed exponential-like distribution to a right- or center-lobed normal-like distribution. The last two columns of the figure show finally the pdf outside the horizon. In all cases the distribution has a marked peak, with the weak cases retaining a significant tail of realizations with low occupation numbers, while the strong cases show symmetry with respect to the peak of the distribution. 

\begin{figure}[t!]
\centering
    \includegraphics[width=0.94\textwidth]{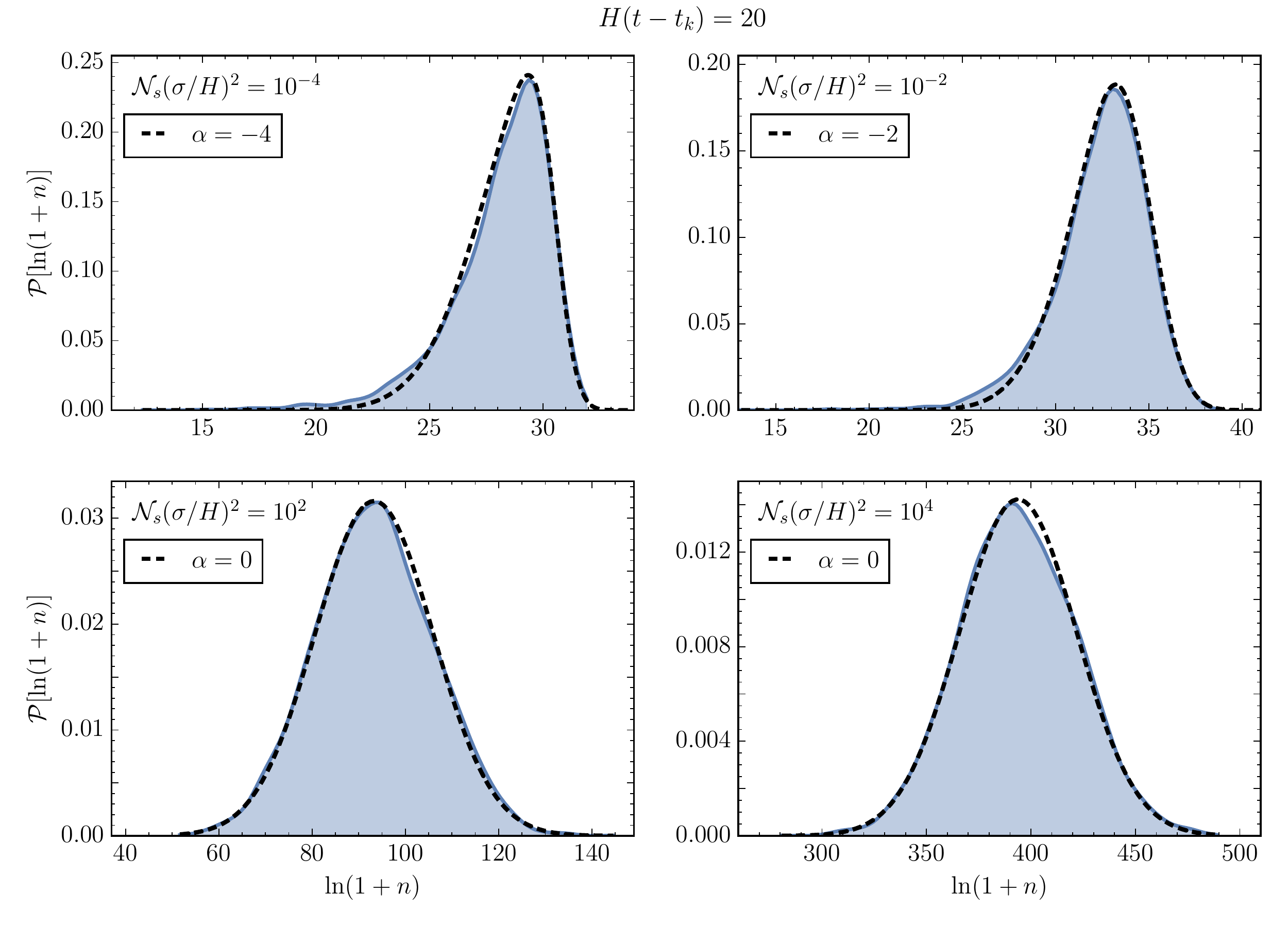}
    \caption{Pdf for $\ln(1+n)$ for selected values of $\mathcal{N}_s(\sigma/H)^2$ at $H(t-t_k)=20$ (conformal case). Blue, continuous: numerical result. Black, dashed: skew-normal fit with shape parameter $\alpha$.}
    \label{fig:nfit_skw}
\end{figure}

In Fig.~\ref{fig:nfit_skw} we show four of the five panels of the last column of Fig.~\ref{fig:ndist}, with a skew-normal fit to the data shown as a black dashed curve. As a reminder, a random variable $x$ is skew-normal distributed if its pdf is given by
\beq\label{eq:skew}
\mathcal{P}(x) = \frac{2}{\sqrt{2\pi\omega^2}}e^{-\frac{(x-x_0)^2}{2\omega^2}}\int_{-\infty}^{\alpha\left(\frac{x-x_0}{\omega}\right)}e^{-\frac{t^2}{2}}\,dt\,,
\eeq
where $x_0$, $\omega$ and $\alpha$ denote the location, scale and shape parameters, respectively~\cite{skew1,skew2}. As it is clear in the top panels of the figure, the distributions for low values of $\mathcal{N}_s(\sigma/H)^2$ show significant skewness, deviating noticeably from normality. In contrast, large values of the scattering strength parameter exhibit a Gaussian shape, with $\alpha=0$ and $x_0$ and $\omega$ given by the mean and variance of the discrete data for the fit. Hence, the occupation number $n$ will be log-skew-normally distributed in general, with a decreasing skewness for increasing $\mathcal{N}_s(\sigma/H)^2$.\par\bigskip

\noindent{\bf Pdf for $\phi$}: 
The $\phi$-distribution is shown in Fig.~\ref{fig:phdist}. This pdf presents a non-trivial evolution, strongly dependent on time and $\mathcal{N}_s(\sigma/H)^2$. As anticipated in Sections \ref{sec:indrelcon} and \ref{sec:meanvarconf}, the distribution of $\phi$ is approximately uniform in sub-horizon scales. Although some structure is visible for some pdfs (at $H(t-t_k) = -12$ with $\mathcal{N}_s(\sigma/H)^2=10^{-2}$, for example), we believe that upon increasing the number of realizations any features will be mostly smoothed out. Note that a uniform distribution is obtained independent of the strength of scattering.

The central column of Fig.~\ref{fig:phdist} shows the transition forms of $\mathcal{P}[\phi]$.  As the mode leaves the horizon the uniformity of the pdf is lost, and a two-lobed distribution arises, with the lobes being of approximately the same size and located  symmetrically with respect to $\phi=\pi/2$. As time increases, these lobes approach, but never fully merge. This is expected from the ``jumping'' behavior of $\phi$ with respect to $\pi/2$ shown earlier in Fig.~\ref{fig:phigrid}.

The super-horizon form of the $\phi$ pdf is shown in the two rightmost columns of Fig.~\ref{fig:phdist}. In these cases, the maxima of the two lobes approach $\pi/2$ exponentially fast, while their widths also decrease  exponentially. Clearly the rates of approach and narrowing are dependent on $\mathcal{N}_s(\sigma/H)^2$; for strong scattering the rates are so high that our pdf estimator is not capable of showing clearly the structure of the distribution.  In Section \ref{sec:analconf2} we discuss an analytical approximation to the super-horizon evolution of $\mathcal{P}[\phi]$ in terms of a Fokker-Planck equation (see Fig.~\ref{fig:phfit}) which captures this behaviour.\par\bigskip

\begin{figure}[t!]
\centering
    \includegraphics[width=0.95\textwidth]{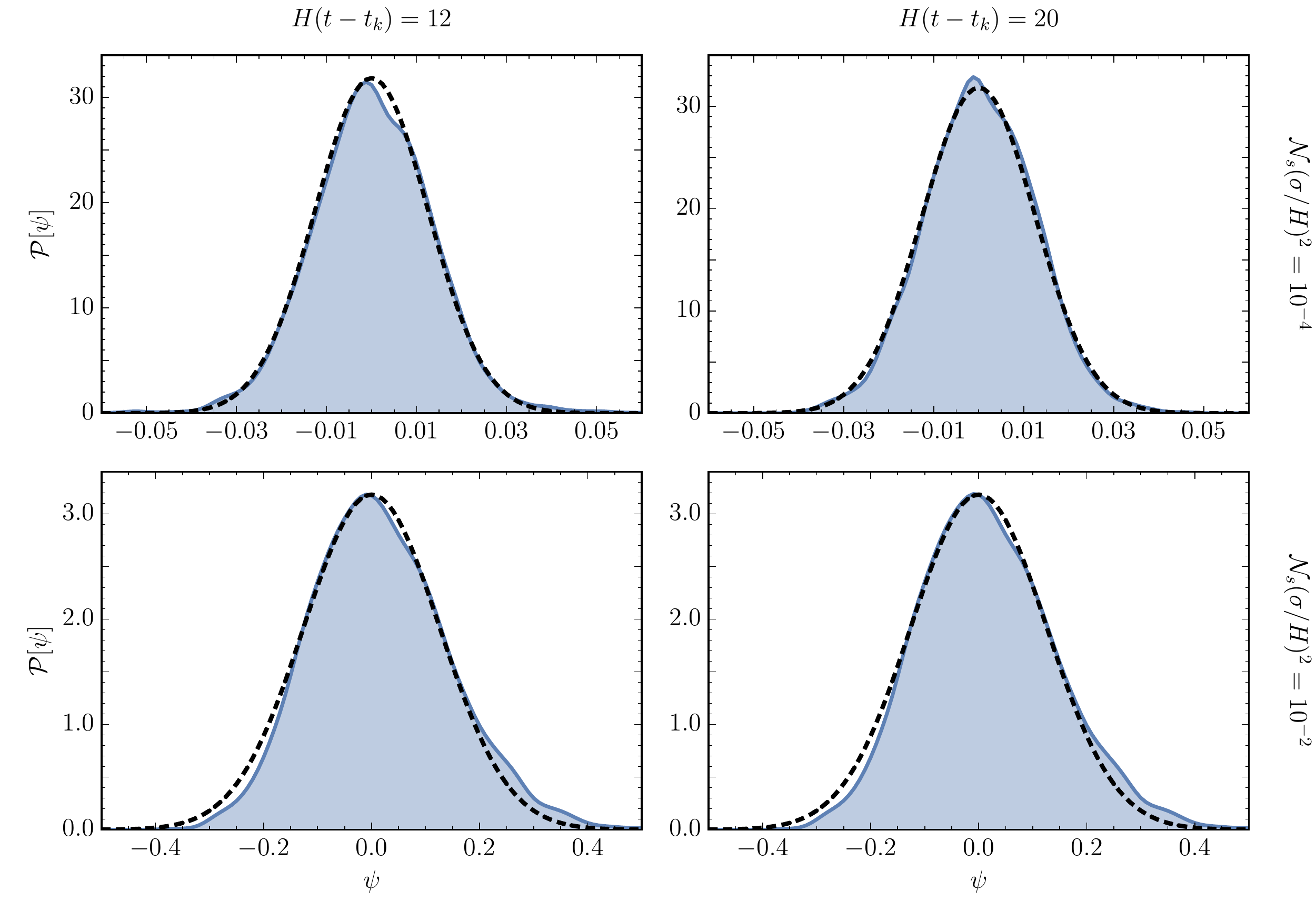}
    \caption{Pdf for $\psi$ for selected values of $\mathcal{N}_s(\sigma/H)^2\ll 1$ and time (conformal case). Blue, continuous: numerical result. Black, dashed: normal distribution with mean and variance (\ref{eq:psimvsup}).}
    \label{fig:psfit}
\end{figure}

\noindent{\bf Pdf for $\psi$}: 
Fig.~\ref{fig:psdist} shows the evolution of the probability distribution for the transfer matrix phase $\psi$. In analogy with the pdf for $\phi$, for $|k\tau|\gg 1$ the distribution of $\psi$ is uniform for any scattering strength, albeit over the interval $(-\pi/2,\pi/2)$. We also observe some features on the pdfs, but we believe that they are mostly an artifact of our finite ensemble of realizations. For strong scattering, the uniformity of the distribution is preserved into super-horizon scales, where a frozen pdf is evident in the last two rows of the figure in question. This is consistent with (\ref{eq:psimvsup}). For weak scattering, the distribution becomes two-lobed as the mode leaves the horizon. However, unlike the $\phi$ case, these two lobes merge in a finite time around $\psi\simeq 0$, and lead to a peaked, frozen distribution for $|k\tau|\ll 1$. Fig.~\ref{fig:psfit} shows the four upper right panels of Fig.~\ref{fig:psdist} compared to a normal distribution of zero mean and variance $(\pi/2)\mathcal{N}_s(\sigma/H)^2$, as per (\ref{eq:psimvsup}). This clearly shows that for weak scattering, $\psi$ is normally distributed outside the horizon. Finally, the central row of Fig.~\ref{fig:psdist} shows a super-horizon pdf intermediate between a uniform and a normal distribution, clearly dependent on the value $\S$.

\par

\subsection{The field two-point function}\label{sec:twopconf}
Arguably, the most remarkable result from our previous numerical explorations consists in the fact that the spectator field amplitude $|X|^2$ is lognormally distributed at all times for any scattering strength. Moreover, outside the horizon, the one-point pdf of $\ln|X|^2$ possesses a mean and a variance that increases linearly with cosmic time; eqs.~(\ref{eq:lnchirateconf}) may be rewritten as
\begin{flalign}
& \text{($|k\tau|\ll 1$)} & \Cen{3}{
\begin{aligned}
\langle \ln|X_{k}(t)|^2\rangle \;&\simeq\; \mu_1 H(t-t_k) - \ln\left(2k\right)\,,\\
{\rm Var}\left[ \ln |X_{k}(t)|^2\right] \;&\simeq\; \mu_2 H (t-t_k) \,,
\end{aligned}}      &&  
\end{flalign}
where $t_k$ is the time of horizon-crossing for the given mode, and we have restored for convenience the momentum-dependence of the mode function. A normal one-point pdf with mean and variance that linearly grow with time are characteristic features of {\em Brownian motion (random walk, Wiener processes) with drift}~\cite{stark2002probability}. Also characteristic of Wiener processes is the property that the unequal time two point correlation function is linearly proportional to the smaller of the  two times. We check this property below.
\begin{figure}[t!]
\centering
    \includegraphics[width=\textwidth]{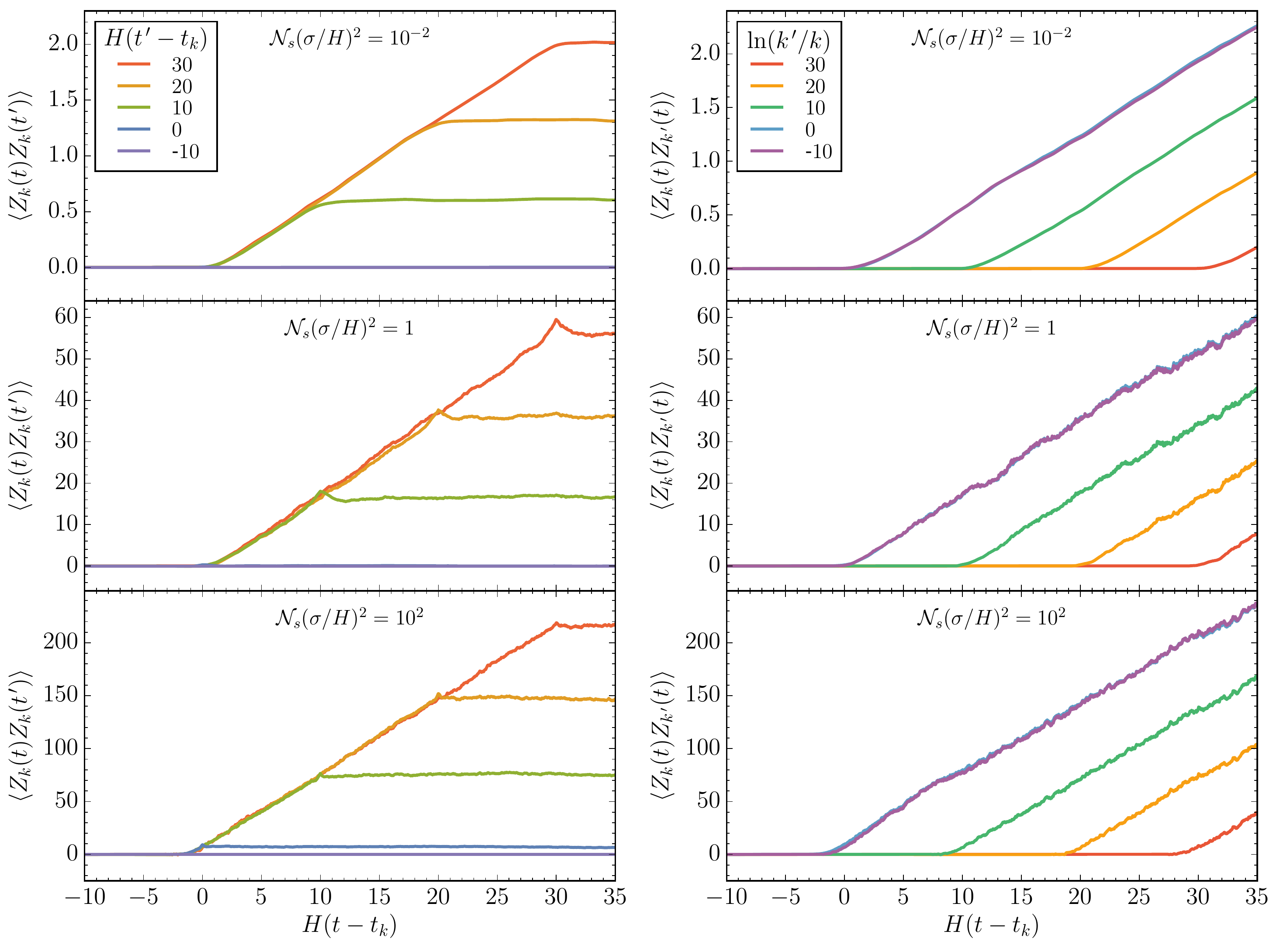}
    \caption{Sub- and super-horizon evolution of the field two-point function for equal momenta and unequal time (left), and unequal momenta and equal time (right), for different values of $\mathcal{N}_s(\sigma/H)^2$ in the conformal case. $t_k, t_{k'}$ indicates the time of horizon crossing for $k$ and $k'$ modes. For the equal momenta and unequal time case, $\langle Z_{k}(t) Z_{k}(t') \rangle=\mu_2 H\,{\rm min}[t-t_k,t'-t_k]$ on super-horizon scales, and zero otherwise. For the unequal momenta, equal time case, $\langle Z_{k}(t) Z_{k'}(t) \rangle=\mu_2 H\,{\rm min}[t-t_k,t-t_{k'}]$ with both $k$ and $k'$ being super-horizon, and zero otherwise. Here we have taken $k=e^{20}H$ and the averages and variances are taken over 2000 different realizations of the amplitudes and locations of the non-adiabatic interactions.}
    \label{fig:2point_conf}
\end{figure}

In order to compute the two-point function of the logarithm of the field amplitude, let us define the driftless (zero mean) variable
\beq\label{eq:zdef}
Z_{k}(t) \;\equiv\; \ln|X_{k}|^2 - \langle  \ln|X_{k}|^2\rangle\,.
\eeq
We then define the expectation value 
\beq\label{eq:twopdef}
\langle Z_{k}(t) Z_{k'}(t') \rangle\equiv \textrm{unequal momentum and unequal time two-point function}\,.
\eeq
Fig.~\ref{fig:2point_conf} shows the time-dependence ($t$) of the two-point function in two cases.\par\bigskip

\noindent{\bf Unequal time}: The left column corresponds to considering the equal momenta $(k'=k)$, unequal time ($t\ne t'$) scenario for discrete values of $t'$ and $\mathcal{N}_s(\sigma/H)^2$. For all three values of $\S$, the qualitative behavior of the curves shown in similar. For $t<t_k$, the magnitude of the two-point function is negligible for any $t'$. If $t' \lesssim t_k$, this non-growing trend is preserved after the mode in question crosses outside the horizon, as demonstrated by the purple and blue curves. Assuming now that $t'>t_k$ (green, orange and red curves), we observe that the two point function grows at the same rate as the variance of $\ln|X_{k}|^2$ does (c.f.~Section~\ref{sec:meanvarconf}) for $t_k<t<t'$, and is frozen at its value at $t=t'$ for $t>t'$. In summary, 
\begin{equation}\label{eq:twoptime}
\langle Z_{k}(t) Z_{k}(t') \rangle \;\simeq\;
\mu_2H\,\textrm{min}[t-t_k,t'-t_k]\, \theta(t-t_k)\theta(t'-t_k)
\end{equation}
indicating that we have an approximately Wiener process on super-horizon scales.

It is worth noting that for fixed $k$, $Z_k(t)$ describes a Gaussian process and therefore all its higher point correlation functions may be computed in terms of its two-point function (\ref{eq:twoptime}),
\beq
\langle Z_k(t_1)Z_k(t_2)\cdots Z_k(t_n)\rangle \;=\; \sum_{\rm pairings}\ \prod_{\rm pairs} \langle Z_k(t_a)Z_k(t_b)\rangle\,.
\eeq\par\bigskip

\noindent{\bf Unequal Momenta}:
The right column of Fig.~\ref{fig:2point_conf} corresponds to the equal time but unequal momenta case. Note here that for the purple and blue curves, for which $k'\leq k$, the two-point function grows linearly with time for $t>t_k$. For the green, orange and red curves, which correspond to $k'>k$, the two-point function grows only for $t>t_{k'}$. We therefore conclude  that 
\begin{equation}\label{eq:twopk}
\langle Z_{k}(t) Z_{k'}(t) \rangle \;\simeq\;
\mu_2H\,\textrm{min}[t-t_k,t-t_{k'}]\, \theta(t-t_k)\theta(t-t_{k'})
\end{equation}
which again confirms our expectation for a Wiener process when both modes are super-horizon.
\\ \\
\noindent{\bf Unequal Momenta and Time}: 
For the general case of unequal time and unequal momenta correlators, we have thus found that the equation
\beq\label{eq:twopointconf}
\langle Z_{k}(t) Z_{k'}(t') \rangle \;\simeq \; \mu_2 H \min\left[(t-t_k),(t'-t_k),(t-t_{k'}),(t'-t_{k'})\right]\theta(t-t_k)\theta(t-t_{k'})\theta(t'-t_k)\theta(t'-t_{k'}) \,,
\eeq
is a good approximation. 

The above results show that $\ln|X_{k}|^2$ satisfies the auto-correlation properties of a Brownian motion with drift for super-horizon $k$. Hence,  the field amplitude $|X_{k}|^2$ (and consequently $|\chi_{k}|^2$) describes a {\em geometric Brownian motion with drift} in cosmic time as long as modes are outside the horizon. For unequal momenta, when either of the modes is inside the horizon, the correlation is vanishing for unequal times.

It is worth noting that, in terms of the two-point function (\ref{eq:twopointconf}), the $n$-point correlation function for the squared field magnitude can be written in general as follows,
\beq\label{eq:npointX}
\langle |X_{k_1}(t_1)|^2\cdots |X_{k_n}(t_n)|^2 \rangle \;=\; \exp\left[\sum_{i=1}^n \langle \ln|X_{k_i}(t_i)|^2\rangle + \frac{1}{2}\sum_{i,j=1}^n \langle Z_{k_i}(t_i)Z_{k_j}(t_j) \rangle \right]\,.
\eeq
This result follows trivially from the lognormality of $|X_k|^2$.

\subsection{Analytical results}\label{sec:analconf}
Up to this point we have only discussed the numerically-obtained trends and values for the transfer matrix parameters and the scalar field magnitude, without referring to analytical expectations. We have decided to follow this ``inverted'' program because the numerical results compose an almost complete picture that will not be attainable with the analytical tools at our disposal. In particular, our discussion following Eq.~(\ref{eq:paramdef}) suggests that only the very weak scattering regime $\mathcal{N}_s(\sigma/H)^2\ll 1$ can be reliably probed analytically using the formalism laid out in Section~\ref{sec:FP}. As an example, obtaining the precise functional form for the functions $\mu_{1,2}$ and $\tilde{\mu}_{1,2}$ is beyond our present study. Nevertheless, as we will show, it is possible to confirm the functional dependence of {\em some} moments on $\mathcal{N}_s(\sigma/H)^2$ and $k$ for weak scattering, as well as that of the probability densities. In the next two sections we derive the form of the Fokker-Planck equation (\ref{eq:FPgen}) which corresponds to a conformally-massive scalar field in a de Sitter expanding background in the sub- and super-horizon regimes and we use it to derive analytical and semi-analytical expressions for moments and pdfs. The main results in this section include: (i) the full pdf of $\phi$, $\ln(1+n)$ and $\ln |X|^2$ on sub-horizon scales, and (ii) the rather non-trivial pdf of $\phi$ and the time-evolution rates for $\langle \ln(1+n)\rangle $ and $\langle \ln|X|^2\rangle $ on super-horizon scales. These results are consistent with our numerical investigations.
\subsubsection{Sufficiently sub-horizon}\label{sec:analconf1}
The general form for the coefficients for the two-parameter FP equation was derived in Section~\ref{sec:FP}, under the assumption of Dirac-delta scatterers with uncorrelated amplitudes of vanishing mean. 
For a conformally massive scalar field in an expanding de Sitter background, the coefficient functions (\ref{eq:g1g})-(\ref{eq:g4g}) can be written as follows, 
\begin{subequations} 
\begin{align} \label{eq:confgs1}
g^{(1)} & =  -\frac{i\tilde{\lambda} m_j}{ 2  H k \tau_j} \left(e^{-2 i (\phi+k \tau_j)} - e^{2 i (\phi+k \tau_j)} \right)\,,\\
\tilde{g}^{(1)} &= -\frac{i m_j}{H k \tau_j} \left(\tilde{\lambda} + \lambda e^{-2 i (\phi + k \tau_j)} \right)\,,\\
g^{(2)} & =  \frac{m_j^2}{4H^2 (k \tau_j)^2} \left[2\lambda +  \tilde{\lambda} \left(e^{-2 i (\phi+k \tau_j)} + e^{2 i (\phi + k \tau_j)} \right) \right]\,,\\  \label{eq:confgs4}
\tilde{g}^{(2)} &= -\frac{m_j^2}{4H^2 (k \tau_j)^2} \left[ \tilde{\lambda} \left(1 + e^{-4 i (\phi + k \tau_j)}\right)  + 2\lambda  e^{-2 i (\phi + k \tau_j)} \right]\,.
\end{align}
\end{subequations}
where $\lambda=2n+1$, $\tilde{\lambda}=\sqrt{\lambda^2-1}$. With these expressions at hand we can compute the coefficients of the FP equation (\ref{eq:FPgen}). By means of an example we will be able to find a pattern that will allow us to bypass the need to compute the disorder averages for all coefficients. From (\ref{eq:deltalambda1}) it follows that
\beq
\langle \delta\lambda^{(1)}\delta\lambda^{(1)}\rangle_{\delta t} = -\frac{\tilde{\lambda}^2}{4 H^2} \left\langle \frac{m_j^2}{(k \tau_j)^2}  \left(e^{-2 i (\phi+k \tau_j)} - e^{2 i (\phi+k \tau_j)} \right)^2 \right\rangle_{\delta t}\,.
\eeq
Let us evaluate the first term of the previous expression in full detail,
\begin{align} \notag
-\frac{\tilde{\lambda}^2e^{-4i\phi}}{4 H^2} \left\langle \frac{m_j^2}{(k \tau_j)^2}  e^{-4 i k \tau_j}  \right\rangle_{\delta t} &\equiv -\frac{\tilde{\lambda}^2e^{-4i\phi}}{4 H^2\delta t} \int_{t}^{t+\delta t} \frac{\langle m_j^2\rangle }{(k \tau_j)^2}  e^{-4 i k \tau_j}\,  dt_j \\ \notag
& = -\frac{(k \tau)\tilde{\lambda}^2 \sigma^2 e^{-4i\phi}}{4 H^2(k \delta \tau)} \int_{k \tau}^{k \tau+k \delta \tau} y^{-3}  e^{-4 i y}\,  dy\\ \label{eq:partcal}
& = -\frac{(k \tau)\tilde{\lambda}^2 \sigma^2 e^{-4i\phi}}{4 H^2} \times
\begin{cases}
(k \tau)^{-3} e^{-4ik \tau}  + \mathcal{O}(k \delta \tau)\,, & |k \delta \tau| \ll 1\\
\mathcal{O}\left((k \delta \tau)^{-1}\right)\,, & |k\delta \tau|\gg 1
\end{cases}\,.
\end{align}
Here in the second line we have made the variable change $t_j\rightarrow y=k \tau_j$, we have recalled the definition for $\sigma^2=\langle m_j^2\rangle$ in (\ref{eq:mstats}), and we used $\delta t=a(\tau)\delta \tau$. Note that the time interval over which the disorder average is taken should be at most of the order of the separation between the non-adiabatic events, $\delta t \lesssim \delta t_i$. Assuming for simplicity that the scattering locations are uniformly distributed in cosmic time, we can identify 
\beq\label{eq:dtdef}
\delta t \;=\; \frac{t_f-t_i}{N_s} \;=\; \frac{1}{H\mathcal{N}_s}\,.
\eeq
where we again recall that $\mathcal{N}_s$ is the number of scatterers per Hubble time. This implies that the parameter delineating different regimes for the coefficients of the FP equation in the previous calculation corresponds roughly to the ratio of the physical wavenumber to the Hubble scale, weighed by the density of scatterers per Hubble time, 
\beq\label{eq:nsktau}
|k \delta \tau| = k_{\rm phys}\delta t \;\sim\; \frac{|k\tau|}{\mathcal{N}_s}\,.
\eeq
In the super-horizon regime in (\ref{eq:partcal}), $|k \delta \tau| \ll 1$, the disorder average is equivalent to simply evaluating the coefficient function at $\tau=\tau_j$; this is to be expected as the time period of oscillations of the mode function is larger than the mean free path determined by the separation between scattering events. On the other side, deep inside the horizon, the mode function oscillates a large number of times in between events, resulting in a vanishing expectation value. We can generalize this result for any non-oscillatory (e.g.~polynomial) function $g(\tau)$ as follows:
\begin{flalign}
& (\mathcal{N}_s^{-1}|k\tau|\gg1 ) & &\langle g(\tau_j)\rangle_{\delta t} \simeq g(\tau_j)\,, \quad \langle g(\tau_j)e^{\pm i n k \tau_j}\rangle_{\delta t} \simeq 0\,,      &&  \\ \label{eq:avrule2}
& (\mathcal{N}_s^{-1}|k\tau|\ll1 ) & &\langle g(\tau_j)\rangle_{\delta t} \simeq g(\tau_j)\,, \quad \langle g(\tau_j)e^{\pm i n k \tau_j}\rangle_{\delta t} \simeq g(\tau_j)e^{\pm i n k \tau_j}\,.      &&  
\end{flalign}

With the previous result at hand, we can immediately write the full set of correlators for the FP equation in the deep sub-horizon regime,
\begin{subequations}  
\begin{align}\label{eq:confMEAa} \displaybreak[0]
\langle \delta \lambda^{(1)} \delta \lambda^{(1)}\rangle_{\delta t}  &\;\simeq\; \frac{\tilde{\lambda}^2\sigma^2}{2H^2(k \tau)^2}  \;=\; \frac{\tilde{\lambda}^2\sigma^2}{2k_{\rm phys}^2}\,,\\ \displaybreak[0]
\langle \delta \lambda^{(1)} \delta \phi^{(1)}\rangle_{\delta t} &\;\simeq\; 0\,, \\ \displaybreak[0]
\langle \delta \phi^{(1)} \delta \phi^{(1)}\rangle_{\delta t}  &\;\simeq\; \frac{(\tilde{\lambda}^2+\lambda^2) \sigma^2}{4\tilde{\lambda}^2H^2(k \tau)^2}  \;=\; \frac{(\tilde{\lambda}^2+\lambda^2) \sigma^2}{4\tilde{\lambda}^2k_{\rm phys}^2}\,,\\ \displaybreak[0]
\langle \delta \lambda^{(2)}\rangle_{\delta t} &\;\simeq\; \frac{ \lambda \sigma^2}{2H^2(k \tau)^2}  \;=\; \frac{ \lambda \sigma^2}{2k_{\rm phys}^2}\,,\\ \label{eq:confMEAe}
\langle \delta \phi^{(2)} \rangle_{\delta t} & \;\simeq\; 0\,.
\end{align}
\end{subequations}
In a completely analogous manner to the non-expanding scenario, all the expectation values are independent of the angular variable~\cite{Amin:2015ftc,Amin:2017wvc}. Therefore, we can immediately conclude that the probability density $P$ which is a solution to the FP equation \eqref{eq:FPgen} is independent of $\phi$, or equivalently, 
\begin{center}
$\phi$ is uniformly distributed deep inside the horizon.\vspace{10pt}
\end{center}
\noindent {\bf Moments of $\ln(1+n)$}: Before attempting to solve the FP equation, let us consider the expectation value $\langle \ln(1+n)\rangle$. The multiplication of (\ref{eq:FPgen}) by $\ln(1+n)$ and integration with respect to both $\lambda$ and $\phi$ leads to the expression
\beq\label{eq:lngen}
\frac{\partial}{\partial t} \langle \ln(1+n)\rangle = \left\langle \frac{1}{2(1+n)}  \frac{\langle \delta\lambda \rangle_{\delta t}}{\delta t} - \frac{1}{8(1+n)^2} \frac{\langle (\delta\lambda)^2\rangle_{\delta t}}{\delta t} \right\rangle \,,
\eeq
Using (\ref{eq:paramdef}),  (\ref{eq:dtdef}) and (\ref{eq:confMEAa})-(\ref{eq:confMEAe}) we can immediately rewrite the above equation as follows:\footnote{Note that this expression is consistent with the expectation from a non-expanding universe \cite{Amin:2015ftc,Amin:2017wvc} where the right hand side of the above equation was $\sigma^2/k^2$). The extra factor of $4$ is explained by a slight change in the definition of $\sigma^2$, whereas the appearance of $\mathcal{N}_s$ is related to the choice of time variable.}
\beq\label{eq:meannsub}
\frac{\partial}{\partial Ht}\langle \ln(1+n)\rangle \;=\; \mathcal{N}_s \left(\frac{\sigma}{2k_{\rm phys}}\right)^2 \left\langle \frac{\lambda}{1+n}-\frac{\tilde{\lambda}^2}{4(1+n)^2}\right\rangle \;=\; \mathcal{N}_s \left(\frac{\sigma}{2k_{\rm phys}}\right)^2  \,.
\eeq
Integration with respect to time leads to $\langle \ln(1+n)\rangle \simeq \mathcal{N}_s(\sigma/k_{\rm phys})^2/8$ (c.f.~\ref{eq:subhnres}), as verified via numerical simulations. Note that no assumptions regarding the magnitude of $n$ have been made to derive this result. 

\begin{figure}[t!]
\centering
    \includegraphics[width=0.94\textwidth]{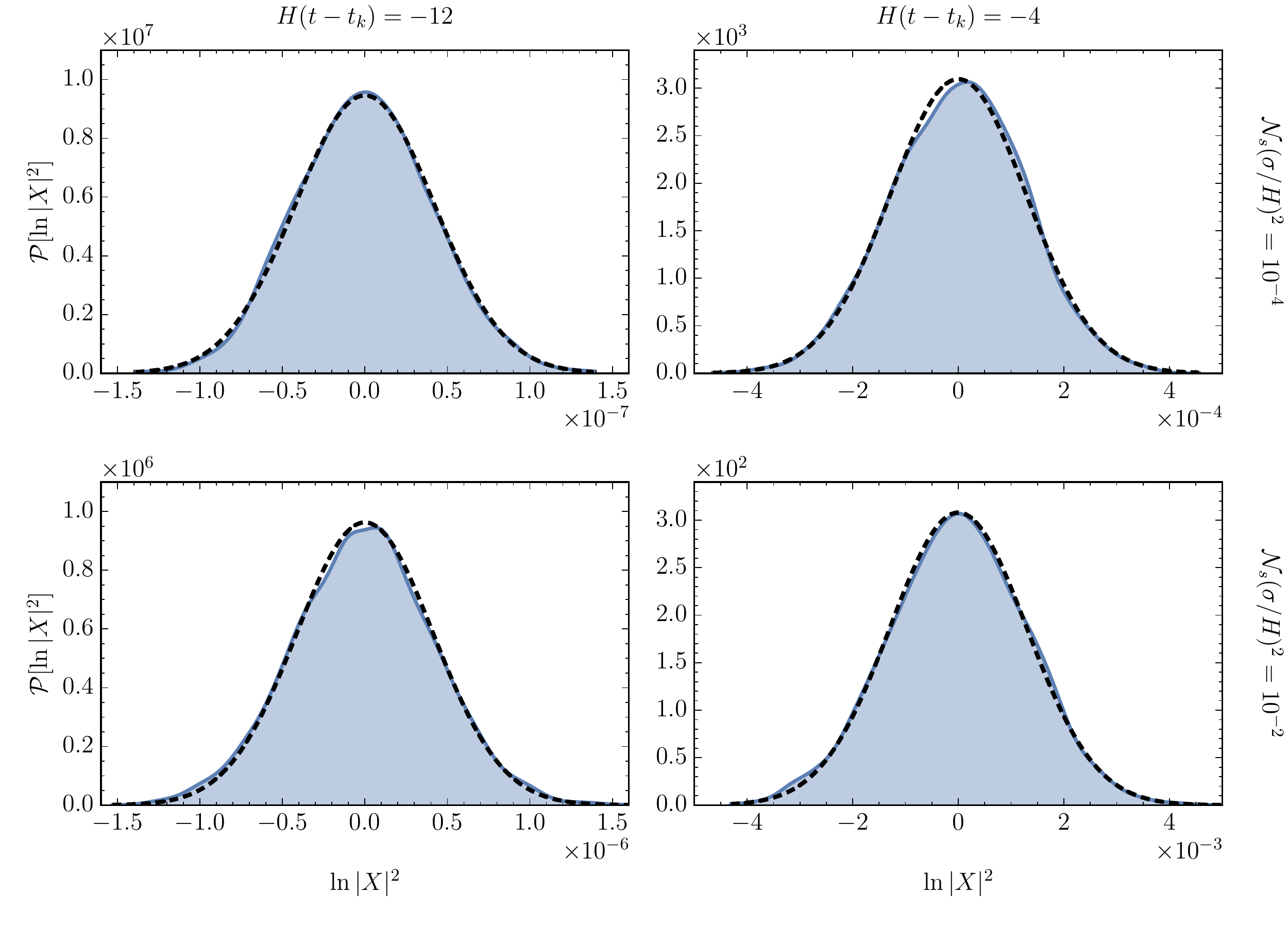}
    \caption{Pdf for $\ln|X|^2$ for selected values of $\mathcal{N}_s(\sigma/H)^2$ and time in sub-horizon scales (conformal case). Blue, continuous: numerical result. Black, dashed: the approximation (\ref{eq:pchiin}). Note the re-scaling (\ref{eq:rescx}).}
    \label{fig:ffitg}
\end{figure}

We can repeat this exercise for the variance if we multiply the general FP Eq.~(\ref{eq:FPgen}) with $[\ln(1+n)]^2$ and integrate over $\phi$, and $\lambda$, obtaining
\begin{align} \notag
\frac{\partial}{\partial Ht} \langle [\ln(1+n)]^2\rangle \;&=\;\mathcal{N}_s\left\langle \frac{1}{2}\frac{\partial}{\partial n} [\ln(1+n)]^2  \langle \delta\lambda \rangle_{\delta t}  + \frac{1}{8}\frac{\partial^2}{\partial n^2} [\ln(1+n)]^2  \langle (\delta\lambda)^2\rangle_{\delta t}  \right\rangle\\ \notag
&=\; \mathcal{N}_s\left(\frac{\sigma}{2 k_{\rm phys}}\right)^2 \left[ 2\langle \ln(1+n)\rangle + 2\langle n(1+n)^{-1}\rangle \right]\\
&\simeq\; 2\mathcal{N}_s^2\left(\frac{\sigma}{2 k_{\rm phys}}\right)^4\,,
\end{align}
where in the third line we have approximated $\langle \ln(1+n)\rangle \simeq \langle n\rangle $ deep inside the horizon, and we have used (\ref{eq:meannsub}) for its value. The previous expression can be integrated to give
\beq
 {\rm Var}[\ln(1+n)] \;=\; \langle [\ln(1+n)]^2\rangle - \langle \ln(1+n)\rangle^2 \;=\; \frac{\mathcal{N}_s^2}{4}\left(\frac{\sigma}{2 k_{\rm phys}}\right)^4\,,
\eeq
which reproduces the numerically obtained result $\langle \ln(1+n)\rangle \simeq {\rm Var}[\ln(1+n)]^{1/2}$ (c.f. \ref{eq:subhnres}).\\
\\
\noindent{\bf Pdf for $\ln(1+n)$}: Let us now consider the $\phi$-independent FP equation. Upon substitution of (\ref{eq:paramdef}), (\ref{eq:dtdef}) and (\ref{eq:confMEAa})-(\ref{eq:confMEAe}) in the general FP Eq.~(\ref{eq:FPgen}), we have 
\beq
\frac{1}{\mathcal{N}_s}\left(\frac{2k_{\rm phys}}{\sigma}\right)^2\frac{\partial P}{\partial Ht}\; =\; -2 \frac{\partial}{\partial \lambda} \left(\lambda P\right) + \frac{\partial^2}{\partial \lambda^2}\left( \tilde{\lambda}^2 P\right)\,.
\eeq
In terms of the new variables
\beq \label{eq:xidef}
\xi\;\equiv\; \frac{\mathcal{N}_s}{2}\left(\frac{\sigma}{2k_{\rm phys}}\right)^2\propto a^2\,,\qquad \rho \;\equiv\; \frac{1}{2}(\lambda + 1) \;=\; 1+n\,,
\eeq
it can be rewritten as
\beq
\frac{\partial P}{\partial \xi} \;=\; -\frac{\partial}{\partial \rho}\Big[(2\rho-1)P\Big] + \frac{\partial^2}{\partial\rho^2}\Big[ \rho(\rho-1) P\Big]\,,
\eeq
which has the integral-form solution~\cite{Abrikosov1981997}
\beq
P(\rho,\xi) \;=\; \frac{2}{(\pi\xi^3)^{1/2}}\int_{{\rm acosh}\sqrt{\rho}}^{\infty} \frac{x \exp\left[-(x^2/\xi + \xi/4)\right]}{(\cosh^2 x - \rho)^{1/2}}\,dx\,.
\eeq
We can find an approximate expression for this probability density in the deep sub-horizon regime ($\xi\ll 1)$, if we re-write it in terms of $\ln\rho = \ln(1+n) \ll 1$,
\begin{align} \notag
P(\ln(1+n),\xi) \;&\simeq\; \frac{2}{(\pi\xi^3)^{1/2}} \int_{\sqrt{\ln(1+n)}}^{\infty} \frac{x e^{-x^2/\xi}}{(x^2-\ln(1+n))^{1/2}}\,dx\\ \label{eq:distappn}
&\simeq\; \frac{1}{\xi}e^{-\ln(1+n)/\xi}\,.
\end{align}
where $\xi=\mathcal{N}_s(\sigma^2/k_{\rm phys})^2/8\propto a^2$. From this distribution it is straightforward to verify that $\langle \ln(1+n)\rangle = \left({\rm Var}[\ln(1+n)] \right)^{1/2} = \xi$, consistent with the previously derived and numerically verified results. Moreover, the exponential form of the pdf is compatible with the correlation coefficient $\tau_{\ln(1+n)}\simeq 1$, computed numerically in Appendix~\ref{ap:typ}. Fig.~\ref{fig:nfit} shows the agreement between the expression (\ref{eq:distappn}) and four selected panels from Fig.~\ref{fig:ndist}.
\begin{figure}[t!]
\centering
    \includegraphics[width=0.94\textwidth]{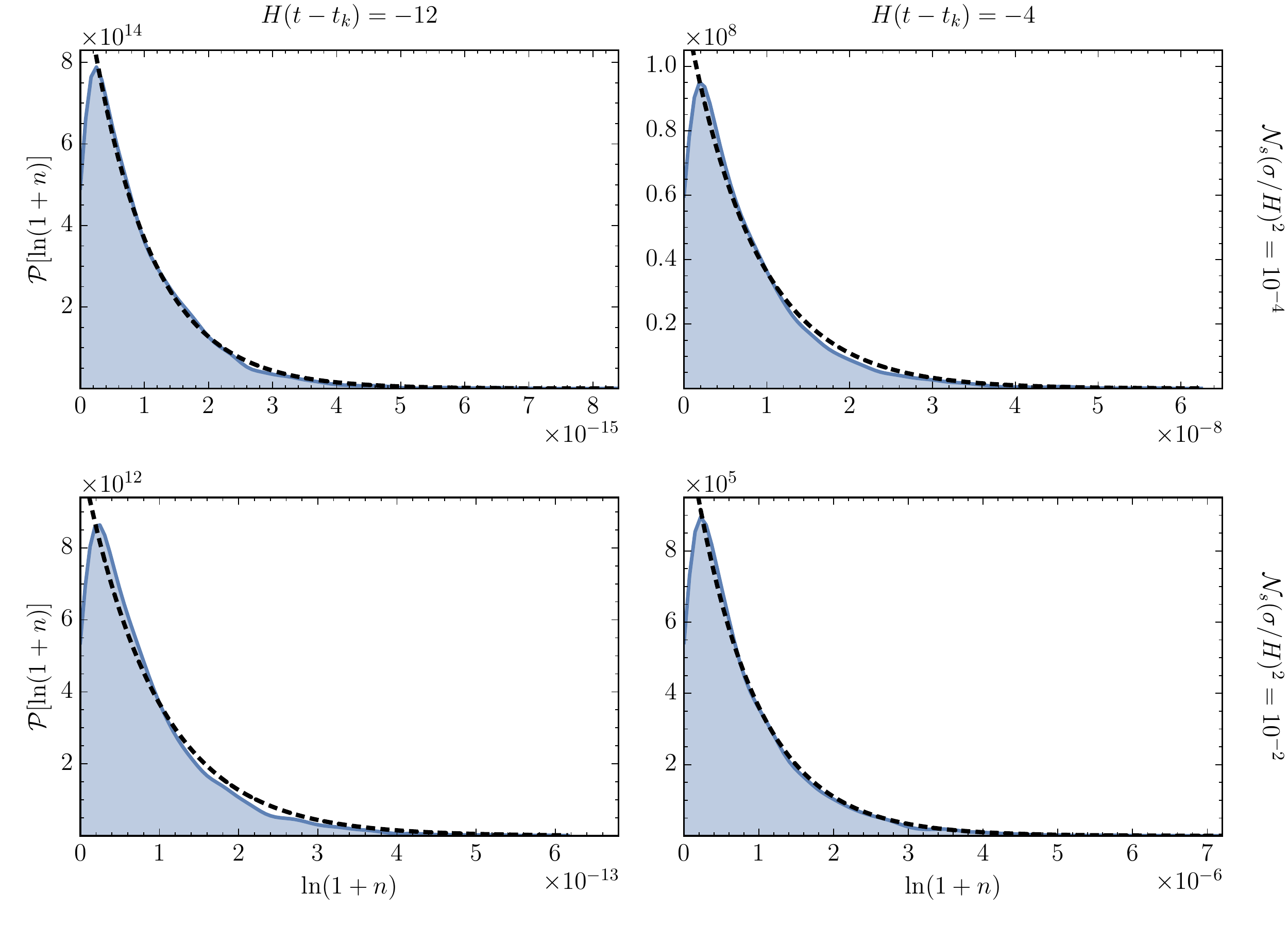}
    \caption{Pdf for $\ln(1+n)$ for selected values of $\mathcal{N}_s(\sigma/H)^2$ and time in sub-horizon scales (conformal case). Blue, continuous: numerical result. Black, dashed: the approximation (\ref{eq:distappn}).}
    \label{fig:nfit}
\end{figure}
\\ \\
\noindent{\bf Pdf for $\ln|X|^2$}: With the pdf for $\phi$ and $n$ at hand, we can now compute the corresponding pdf for the scalar field amplitude inside the horizon. Starting from (\ref{eq:genchi}) we can write
\beq
|X|^2 = \frac{1}{2k}\left[1+2n+2\sqrt{n(1+n)}\cos\left(2(\phi+k\tau)\right)\right]\,.
\eeq
As $n\ll 1$, we can approximate the logarithm of the amplitude as
\begin{align}\notag
\ln|X|^2 \;&\simeq\; -\ln(2k) + 2\sqrt{n}\cos\left(2(\phi+k\tau)\right)\,, &&P(n)\simeq \frac{1}{\xi}e^{-n/\xi},\ \ P(\phi) = \frac{1}{\pi}\\[5pt] \notag
&\simeq\; -\ln(2k) + uv\,, &&P(u)= \frac{2 u }{\xi }e^{-\frac{u^2}{\xi }},\ \ P(v) = \frac{1}{\pi  \sqrt{4-v^2}}\\[5pt]
&\simeq\; -\ln(2k) + y\,, &&P(y)= \frac{1}{2 \sqrt{\pi  \xi }}e^{-\frac{y^2}{4 \xi }}
\end{align}
and therefore,
\beq\label{eq:pchiin}
P(\ln|X|^2)\;\simeq\; \frac{1}{2 \sqrt{\pi  \xi }}\exp\left[-\frac{\left(\ln|X|^2 +\ln(2k) \right)^2}{4 \xi } \right]\,.
\eeq
where recall that $\xi=\mathcal{N}_s(\sigma^2/k_{\rm phys})^2/8\propto a^2$. Consistent with our numerical exploration of section~\ref{sec:pdfs}, we have found that $\ln|X|^2$ is normally distributed, or equivalently, $|X|^2$ is log-normally distributed inside the horizon. The pdf (\ref{eq:pchiin}) immediately implies that $\langle \ln|X|^2\rangle \simeq -\ln(2k)$ and ${\rm Var}\,[\ln|\chi|^2]\simeq 2\xi$, in agreement with the numerical fits (\ref{eq:subhlnch}) and (\ref{eq:subhvarlnch}). Fig.~\ref{fig:ffitg} further shows the agreement between (\ref{eq:pchiin}) and four selected weak scattering panels from Fig.~\ref{fig:fdist}.

\subsubsection{Outside the horizon}\label{sec:analconf2} 
\begin{figure}[t!]
\centering
    \includegraphics[width=0.87\textwidth]{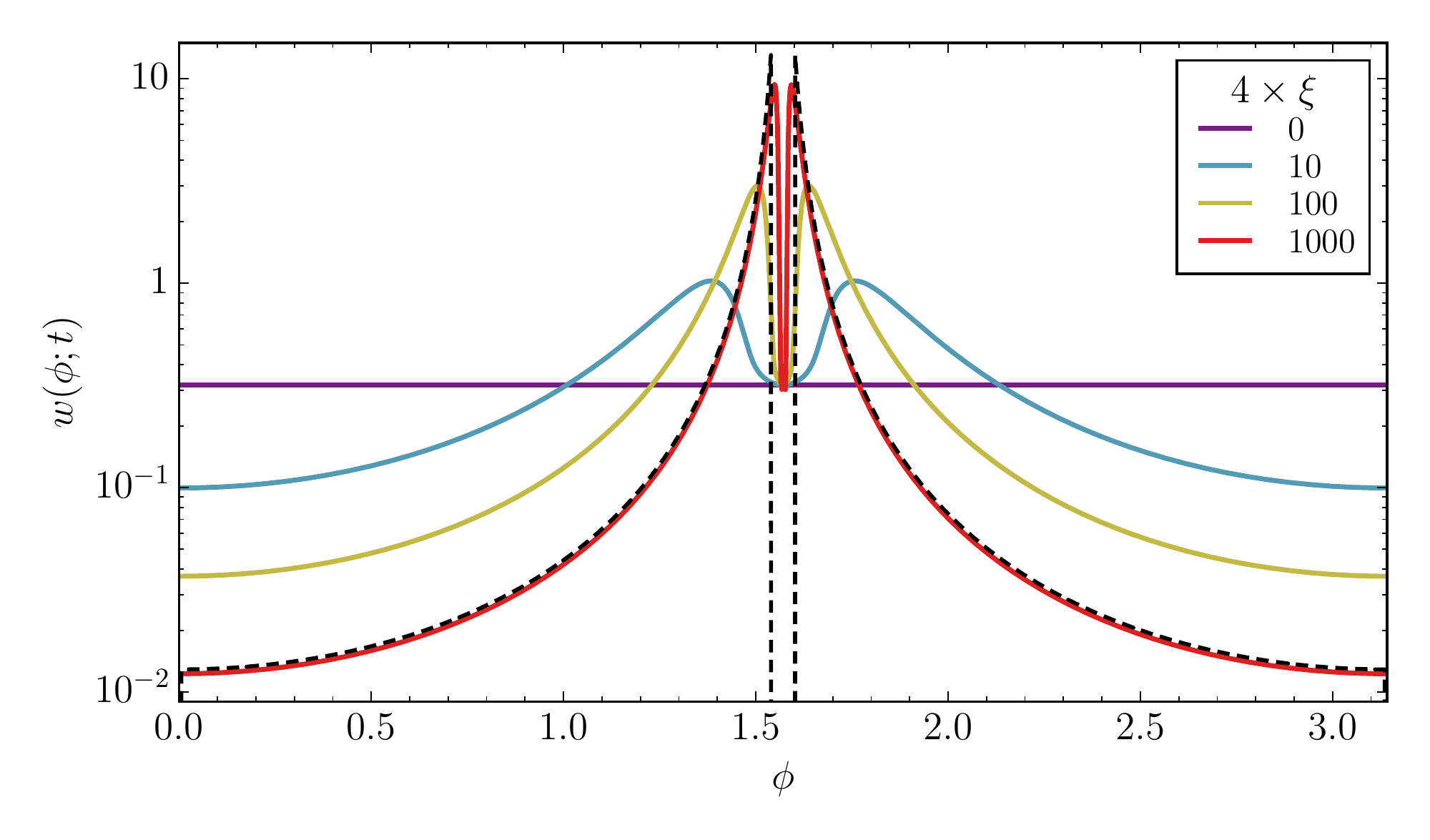}
    \caption{Numerical solution of the FP equation (\ref{eq:wmarc}) at different times, with initial condition $w(\phi;0)=1/\pi$. The late-time approximation (\ref{eq:latewc}) is shown as the dashed black curve.}
    \label{fig:phfit2}
\end{figure}

\begin{figure}[t!]
\centering
    \includegraphics[width=0.94\textwidth]{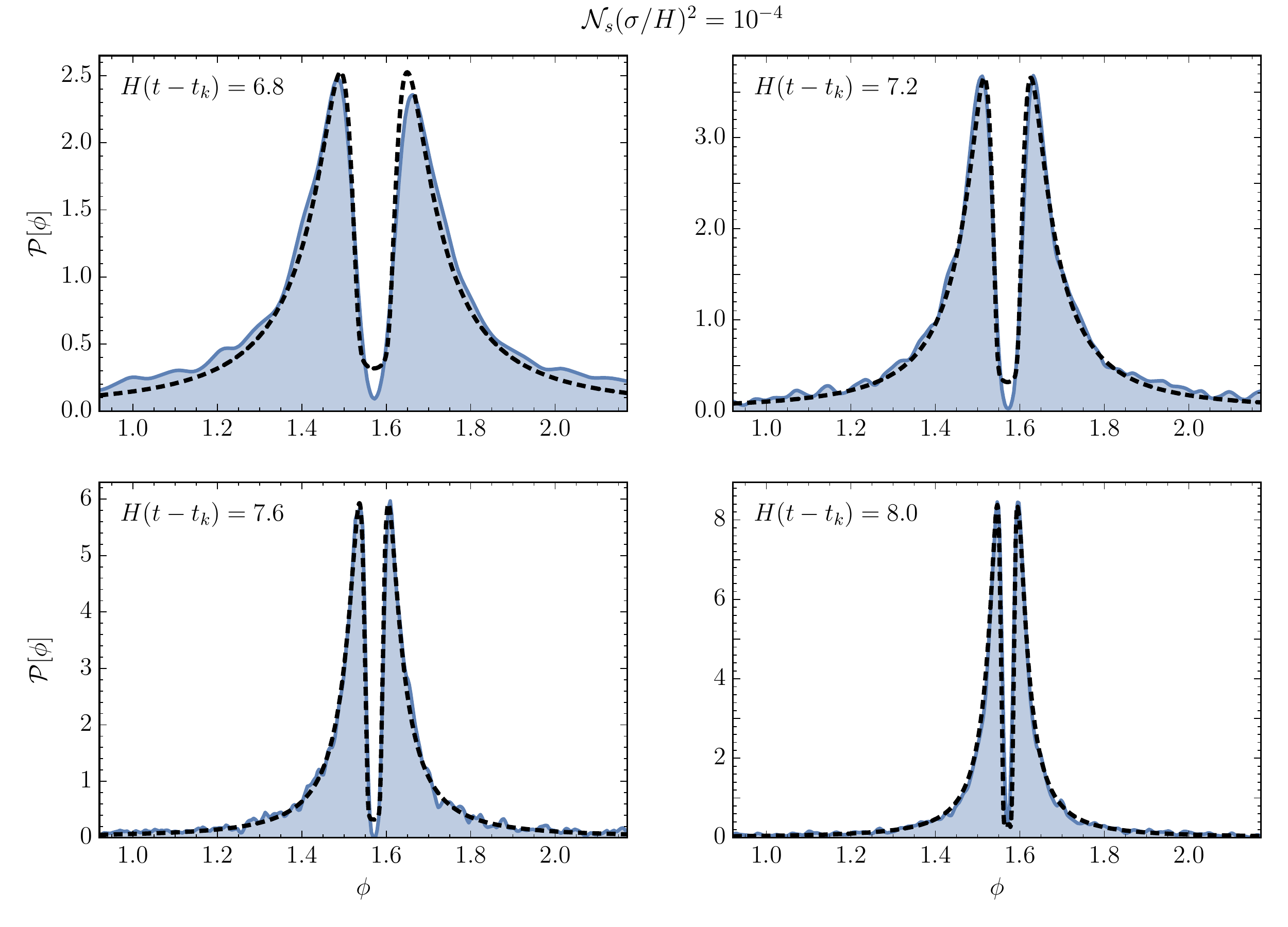}
    \caption{Pdf for $\phi$ for selected values of time in super-horizon scales; here $\mathcal{N}_s(\sigma/H)^2=10^{-4}$ (conformal case). Blue, continuous: fully numerical result. Black, dashed: $w(\phi)$ from the numerical integration of (\ref{eq:wmarc}).}
    \label{fig:phfit}
\end{figure}

Let us now find the form of the FP equation far outside the horizon. In this case, the disorder average for the FP coefficients must be taken as in (\ref{eq:avrule2}), where averages are replaced essentially by their instantaneous values. In this late-time limit, we will assume for simplicity that the occupation number has grown sufficiently so that the approximation $\tilde{\lambda}\approx \lambda \approx 2n$ is valid (recall that $\lambda\equiv2n+1$ and $\tilde{\lambda}\equiv\sqrt{\lambda^2-1}$). After some algebra we obtain the following set of correlators,
\begin{subequations}  
\begin{align}\label{eq:confSHa}
\langle \delta \lambda^{(1)} \delta \lambda^{(1)}\rangle_{\delta t}  &\;\simeq\; \frac{\lambda^2\sigma^2}{k_{\rm phys}^2}\,    \sin^2\left(2 (\phi + k \tau)\right)  \,,\\
\langle \delta \lambda^{(1)} \delta \phi^{(1)}\rangle_{\delta t} &\;\simeq\; 
\frac{2 \lambda  \sigma^2}{k_{\rm phys}^2}\,  \cos ^3(\phi + k \tau ) \sin (\phi + k \tau ) \,, \\
\langle \delta \phi^{(1)} \delta \phi^{(1)}\rangle_{\delta t}  &\;\simeq\; \frac{\sigma^2}{k_{\rm phys}^2}\,  \cos ^4(\phi + k \tau ) \,,\\
\langle \delta \lambda^{(2)}\rangle_{\delta t} &\;\simeq\; \frac{\lambda  \sigma^2}{k_{\rm phys}^2}\,  \cos ^2(\phi+ k \tau) \,,\\ \label{eq:confSHe}
\langle \delta \phi^{(2)} \rangle_{\delta t}  & \;\simeq\; -\frac{\sigma^2}{k_{\rm phys}^2}\, \cos ^3(\phi + k \tau) \sin(\phi + k \tau)\,.
\end{align}
\end{subequations}
\\
\noindent{\bf Pdf for $\phi$}: In terms of the shifted variable
\beq\label{eq:phishift}
\varphi \;\equiv\; \phi + k\tau\,,
\eeq 
the FP equation (\ref{eq:FPgen}) takes then the form
\begin{align} \notag
\frac{1}{\mathcal{N}_s}\left(\frac{k_{\rm phys}}{\sigma}\right)^2\frac{\partial P}{\partial Ht} \;&= \; -\frac{\partial}{\partial \lambda}\Big[ \lambda \cos^2\varphi\,P \Big] + \frac{\partial}{\partial\phi}\Big[ \cos^3\varphi \sin\varphi\,P\Big] + \frac{1}{2} \frac{\partial^2}{\partial\lambda^2} \Big[\lambda^2\sin^2 2\varphi\,P\Big]\\ \label{eq:FPsuphc}
&\qquad + 2\frac{\partial^2}{\partial\lambda\partial\phi} \Big[\lambda \cos^3\varphi \sin\varphi\,P\Big] + \frac{1}{2}\frac{\partial^2}{\partial\phi^2}\Big[ \cos^4\varphi\,P\Big]\,.
\end{align}
We do not attempt to find a closed-form solution to this equation. Nevertheless, we can find an approximate expression for the time-dependent marginal probability distribution
\beq\label{eq:defmarg}
w(\phi;t) \equiv \int d\lambda\, P(\lambda,\phi;t)\,,
\eeq
which in turn will allow us to calculate the mean particle production rate. Integrating both sides of (\ref{eq:FPsuphc}) with respect to $\lambda$, and re-parametrizing the time-dependence in terms of $\xi$ defined in (\ref{eq:xidef}), we obtain the following expression for the equation of motion of $w$, 
\beq\label{eq:wmarc}
\frac{\partial w}{\partial (4\xi)} \;=\; \frac{1}{2}\frac{\partial}{\partial\phi} \left[\cos^4\varphi \left( \frac{\partial w}{\partial\phi} - 2 w\,\tan\varphi \right)\right]\,.
\eeq
In the very-late time limit one could naively expect that the temporal dependence is negligible, and the marginal distribution tends to a limiting pdf. Were this the case, the FP equation for this limit distribution would have the form
\beq
w'(\phi)-2w(\phi) \tan\phi \;=\; 0\,,
\eeq
which has the solution $w(\phi)\propto \sec^2\phi$. However, this function is divergent at $\phi=\pi/2$ and it is not normalizable, implying that the time dependence in (\ref{eq:wmarc}) cannot be outright disregarded. Nevertheless, as it turns out, this solution correctly describes the qualitative behavior of the marginal distribution at late times, save for a time-dependent cutoff of the divergence at $\pi/2$. Fig.~\ref{fig:phfit2} shows the numerical solution of the FP equation (\ref{eq:wmarc}), in solid curves, for different values of the temporal parameter $\xi$, assuming the initial condition $w(\phi,0)=1/\pi$, i.e.~the sub-horizon uniform distribution. The dashed black curve is given by the approximation
\beq\label{eq:latewc}
w(\phi;t) \;\simeq\; \begin{cases}
\dfrac{\sec^2\phi}{2\cot\delta}\,, & \phi\in(0,\frac{\pi}{2}-\delta)\cup(\frac{\pi}{2}+\delta,\pi)\,,\\[5pt]
0\,,& \phi\in(\frac{\pi}{2}-\delta,\frac{\pi}{2}+\delta)\,,
\end{cases}
\eeq
where we find the cutoff $\delta$ to be approximately given by
\beq
\delta \simeq \frac{\xi^{-1/2}}{2}\propto a^{-1}\,.
\eeq
Fig.~\ref{fig:phfit} shows a comparison between the numerical solution of the marginal FP equation (\ref{eq:wmarc}) and the fully numerically calculated pdf for $\phi$ for selected time slices, with weak scattering $\mathcal{N}_s(\sigma/H)^2=10^{-4}$ (c.f.~Section~\ref{sec:pdfs}). The agreement between both results is clear, and it improves as we move farther outside the horizon. From the approximation (\ref{eq:latewc}) we also obtain
\beq
\langle \phi\rangle = \frac{\pi}{2}\,,\qquad {\rm Var}\,\phi \;\simeq\; \pi\ln(2)\,\delta \;=\; \pi\ln(2) \left(\frac{2}{\mathcal{N}_s(\sigma/H)^2}\right)^{1/2} |k\tau|\,,
\eeq
which agree with the numerical results (\ref{eq:phimvcn}), (\ref{eq:phivrcn}).\\
\\
\noindent{\bf Rates $\langle \ln(1+n)\rangle$ and $\langle \ln|X|^2\rangle$}: Given the list of correlators (\ref{eq:confSHa})-(\ref{eq:confSHe}), we can calculate the expectation value for the particle production rate for $|k\tau|\ll 1$. Substitution into (\ref{eq:lngen}) gives
\begin{align} \notag
\frac{1}{\mathcal{N}_s}\frac{\partial}{\partial Ht}\langle \ln(1+n)\rangle \;&= \; \left( \frac{\sigma}{k_{\rm phys}} \right)^2 \left\langle \frac{\lambda}{2(1+n)}  \,  \cos ^2\varphi - \frac{\lambda^2}{8(1+n)^2} \,    \sin^2\left(2 \varphi \right) \right\rangle\\ \notag
&\simeq \; \left( \frac{\sigma}{k_{\rm phys}} \right)^2 \left\langle \cos ^2\varphi  \cos\left(2 \varphi \right) \right\rangle \\
&\simeq \; \frac{2}{\mathcal{N}_s}\frac{\sin^2\delta}{\delta^2}\,.
\label{eq:nrateA}
\end{align}
At very late times $\delta\ll1$, hence we recover the result $\partial_{Ht}\langle \ln(1+n)\rangle\simeq 2$ outside the horizon for weak scattering (c.f.~Fig. \ref{fig:nfrates}). Moreover, from $|X|=\sqrt{2n/k}|\varphi-\pi/2|$ (see Eq.~(\ref{eq:chisuph2})), we can write
\beq
\label{eq:nXrateA}
\partial_{Ht}\langle\ln|X|^2\rangle  \simeq \partial_{Ht}\langle \ln(n)\rangle + \partial_{Ht}\langle \ln|\varphi-\pi/2|^2\rangle \,.
\eeq
Integration using the approximation (\ref{eq:latewc}) yields $\langle \ln|\varphi-\pi/2|^2\rangle\simeq 2\ln\delta + 2$. Therefore, in the super-horizon regime (where $\delta\ll 1$) from Eq.~\eqref{eq:nrateA} and \eqref{eq:nXrateA}, we have
\begin{align}
\partial_{Ht} \langle \ln(1+n)\rangle \;&\simeq\; 2\,,\\
\partial_{Ht} \langle \ln|X|^2\rangle \;&\simeq\; 0\,,
\end{align}
in agreement with the numerical result shown in Fig.~\ref{fig:nfrates} in the weak scattering limit.

\section{Massless field in de Sitter background}\label{sec:MDS}
\subsection{Numerical results}\label{sec:mssds}
We now turn to the discussion of the numerical results for the massless case. This analysis will mirror our previous study for the conformal case: we consider the sub- and super-horizon regimes with weak and strong scattering, where the former are explored by computing the evolution of a Fourier mode over a range of 40 Hubble times centered at horizon crossing, while the later are defined depending on the magnitude of the scattering strength parameter $\mathcal{N}_s(\sigma/H)^2$ defined in (\ref{eq:paramdef}).  A straightforward substitution of the mode functions for the free massless field (\ref{eq:dsmass}) into (\ref{eq:Mjgen}) provides the instantaneous transfer matrix used to the derive the results discussed below.
\subsubsection{Individual realizations}
\begin{figure}[t!]
\centering
    \includegraphics[width=\textwidth]{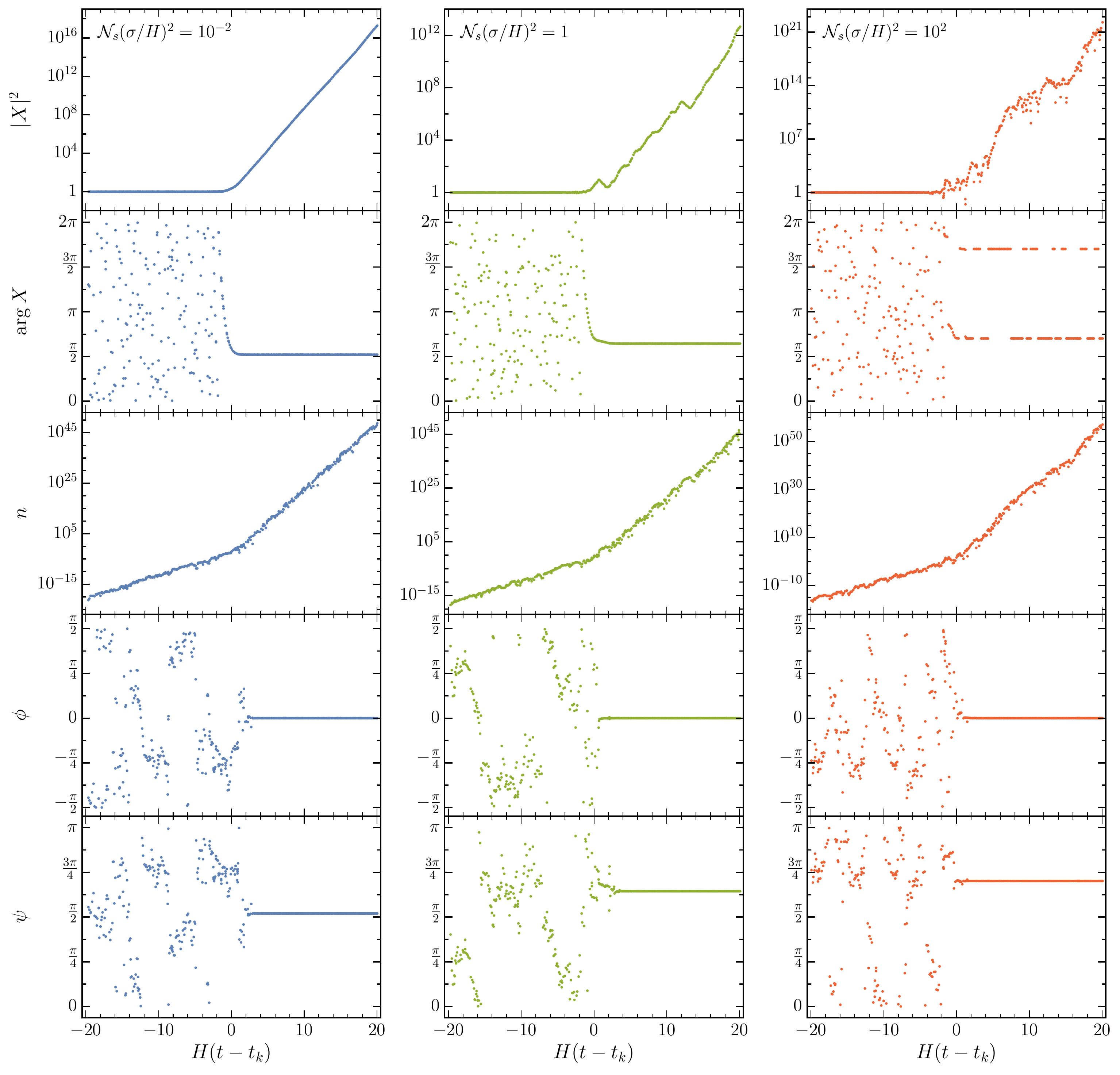}
    \caption{Evolution of the field squared magnitude and its phase, the occupation number, and the transfer matrix phases $\phi$ and $\psi$, as functions of cosmic time, in the massless case with uniformly distributed amplitudes and locations of the non-adiabatic events. The same three representative values of the parameter $\mathcal{N}_s(\sigma/H)^2$ in Fig.~\ref{fig:all_grid} have been chosen.}
    \label{fig:all_grid_m}
\end{figure}

Fig.~\ref{fig:all_grid_m} shows the evolution of the field amplitude and its phase, as well as transfer matrix parameters $\{n,\phi,\psi\}$,  as functions of time. All assumptions on the scattering parameters coincide with those for the conformal mass case discussed in Section~\ref{sec:indrelcon}: the amplitudes and the locations of the non-adiabatic events are uniformly distributed in the intervals $m_j\in(-\sqrt{3}\sigma,\sqrt{3}\sigma)$ and $\delta t_j\in(0,1/H\mathcal{N}_s)$, with $N_s=300$ and $k=e^{20}H$, and to avoid cumbersome notation we have also adopted the re-scaling convention (\ref{eq:rescx}). Each plot corresponds to a single realization of the disorder for the same three scattering strength parameters as in Fig.~\ref{fig:all_grid}. The features that one can read from these results are similar to those for the conformal case, except for a few key differences,
\begin{enumerate}

\item The magnitude of the canonically normalized field $X$ is constant with virtually no spread in subhorizon scales. For $|\chi|$, this implies an exponential decrease with the rate determined by the inverse of the scale factor. Outside the horizon, $|X|$ grows exponentially. In the case of weak scattering, the growth rate is to a good approximation exactly that given by the scale factor; equivalently, $|\chi|$ is frozen to a constant value after horizon crossing, as expected from the mode function (\ref{eq:dsmass}). For moderate scattering, $|X|$ clearly grows at a slightly slower rate than $a$, signifying an exponential decrease for $|\chi|$. In turn, for strong scattering, the rate of growth of $|X|$ is significantly larger than that for the scale factor, leading to the exponential increase of $|\chi|$. 
\item The behavior of the field phase is analogous to that of a conformally massive field, it fluctuates uniformly in $(0,2\pi)$ for $|k\tau|>1$, and it is frozen at $\pi/2$ with $|k\tau|<1$ for weak scattering, or along a random direction for strong scattering.
\item The occupation number grows exponentially. In all cases the growth rate increases approximately at horizon crossing, with a value dependent on  $\mathcal{N}_s(\sigma/H)^2$.
\item Unlike the conformal case, the angular parameter $\phi$ is defined here on the domain $(-\pi/2,\pi/2)$, over which it fluctuates randomly inside the horizon. It freezes asymptotically to $\phi\simeq 0$ far outside the horizon.
\item In the massless scenario, the natural domain for the phase $\psi$ is $(0,\pi)$. This angular parameter varies randomly over its whole range for $|k\tau|\gg 1$. When scattering is weak, $\psi\rightarrow \pi/2$ outside the horizon, while for strong scattering $\psi$ freezes to a random value.
\end{enumerate}

\begin{figure}[t!]
\centering
    \includegraphics[width=\textwidth]{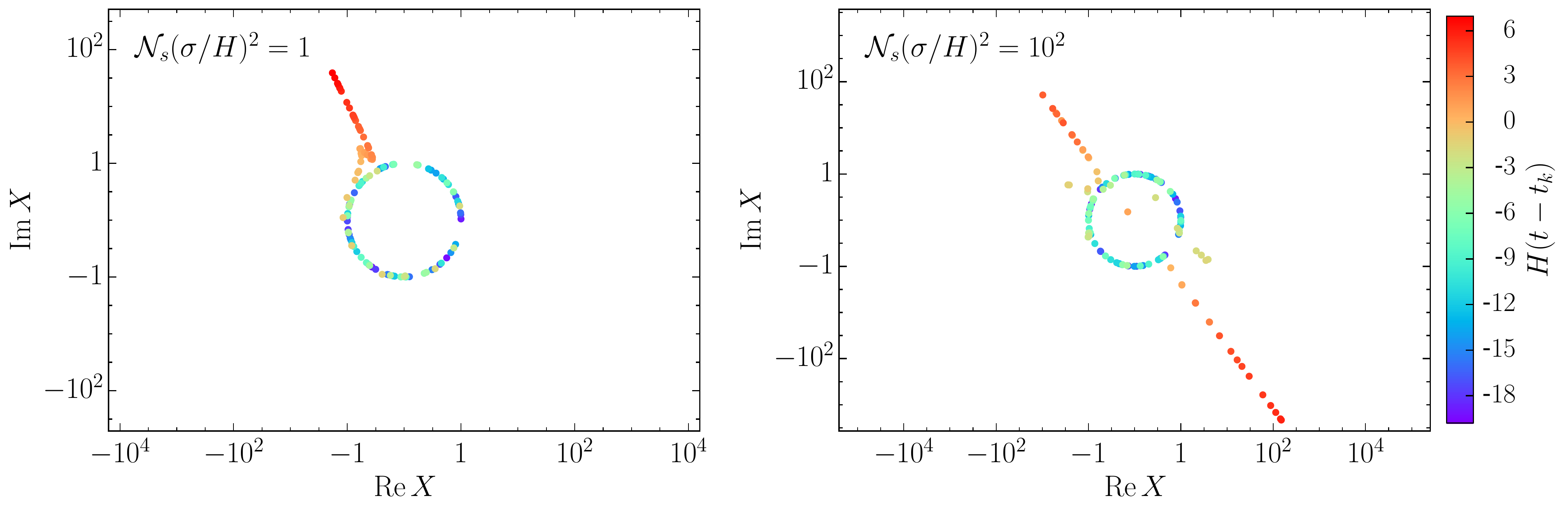}
    \caption{Evolution of the (re-scaled) real and imaginary parts of the massless field $X$ as functions of time in the strong scattering regime, for two values of $\mathcal{N}_s(\sigma/H)^2$. The numerical results shown here correspond to those of the center and right panels of Fig.~\ref{fig:all_grid_m}, sampled every other point for clarity.}
    \label{fig:reimchi_m}
\end{figure}

\begin{figure}[t!]
\centering
    \includegraphics[width=\textwidth]{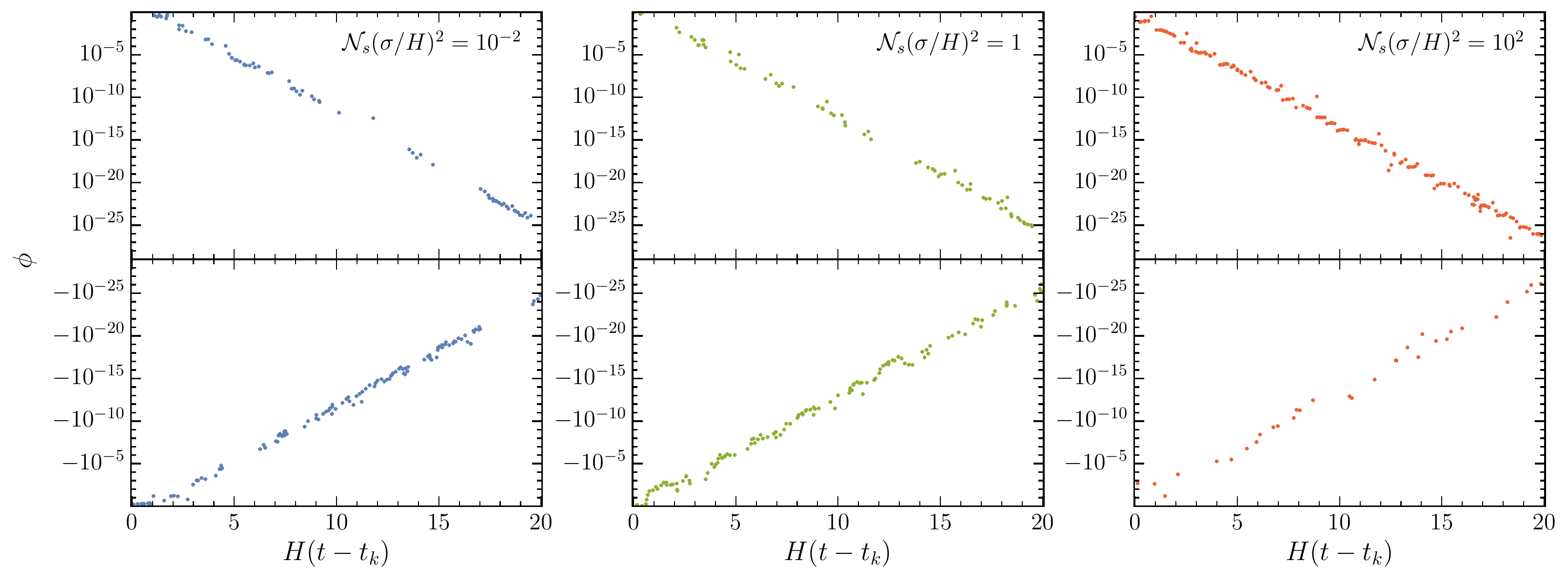}
    \caption{Super-horizon evolution of the transfer matrix phase $\phi$ in the weak and strong scattering regimes, for a massless spectator field.}
    \label{fig:phigrid_m}
\end{figure}

Note that, qualitatively, in the strong scattering limit $\arg X$ behaves in the same manner as in the conformal case, being locked along a ray in the complex plane and jumping between diametrically opposite directions after the mode in question has left the horizon. This result is more clearly seen in Fig.~\ref{fig:reimchi_m}, which shows the evolution of $X$ in the complex plane with a color coded time dependence. Analytically, in terms of the transfer matrix parameters, the massless scalar can be written as
\beq\label{eq:genchi_m}
X \;=\; \frac{1}{\sqrt{2k}} \left[ (1+n)^{1/2}e^{-i(\phi+\psi+k\tau)}\left(1-\frac{i}{k\tau}\right) + n^{1/2}e^{i(\phi-\psi+k\tau)}\left(1+\frac{i}{k\tau}\right) \right]\,.
\eeq
Deep inside the horizon, with $|k\tau|\gg 1$ and $n\ll 1$, $\arg X \simeq -(\phi+\psi+k\tau) \pmod {2\pi}$, implying a random uniform distribution for the field phase. Conversely, outside the horizon,
\begin{flalign}\label{eq:chisuph_m}
& \text{($|k\tau|\ll 1$)} & \Cen{3}{X \;\simeq\; -\frac{1}{k\tau}\sqrt{\frac{2n}{k}}\sin(\phi+k\tau)e^{-i\psi} \,,}      &&  
\end{flalign}
showing that $\arg X$ is determined by $\psi$ up to a sign, determined by the asymptotic value of $\phi$. Fig.~\ref{fig:phigrid_m} shows an enhancement around $\phi=0$ of the $\phi$-panels in Fig.~\ref{fig:all_grid_m}. Is is clear that, with $|k\tau|\rightarrow 0$, the argument of the sine in (\ref{eq:chisuph_m}) will change signs as it is driven to zero. Expanding around this value we can then write
\begin{flalign}\label{eq:chisuph2_m}
& \text{($|k\tau|\ll 1$)} & \Cen{3}{X \;\simeq\; -\frac{1}{k\tau}\sqrt{\frac{2n}{k}}(\phi+k\tau)\,e^{-i\psi} \;=\; \frac{1}{k\tau}\sqrt{\frac{2n}{k}}|\phi+k\tau|\, e^{i(\zeta\pi-\psi)}\,,}      &&  
\end{flalign}
with $\zeta=\{0,1\}$ randomly, leading to the diametrical flip of the field phase.
\subsubsection{Means and variances}\label{sec:meanvarconf_m}
Let us now discuss the dynamics of the moments of the field and transfer matrix parameters given an ensemble of realizations. As we did in Section~\ref{sec:meanvarconf} in the conformal case, we will focus on the lowest moments for $\ln|X|^2$, $\ln(1+n)$, $\phi$ and $\psi$, leaving the discussion of their probability distributions to the next section. 

\begin{figure}[t!]
\centering
    \includegraphics[width=\textwidth]{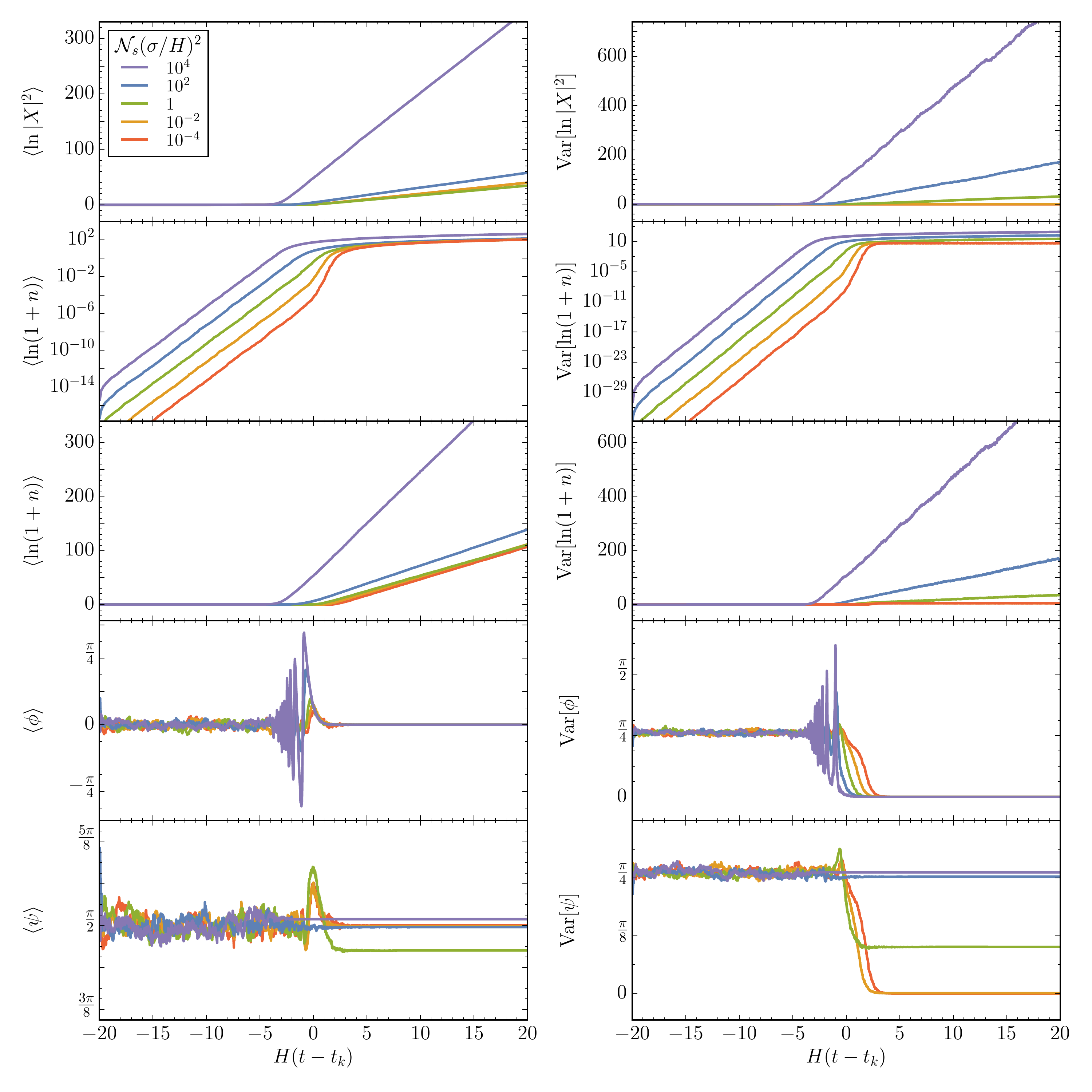}
    \caption{Sub- and super-horizon evolution of the mean and variance of the transfer matrix parameters $\{n,\phi,\psi\}$ and the re-scaled scalar field amplitude, for different values of $\mathcal{N}_s(\sigma/H)^2$ in the massless case. All parameters are chosen as in Fig.~\ref{fig:sigmeanvar}.}
    \label{fig:sigmeanvar_m}
\end{figure}

\begin{figure}[t!]
\centering
    \includegraphics[width=\textwidth]{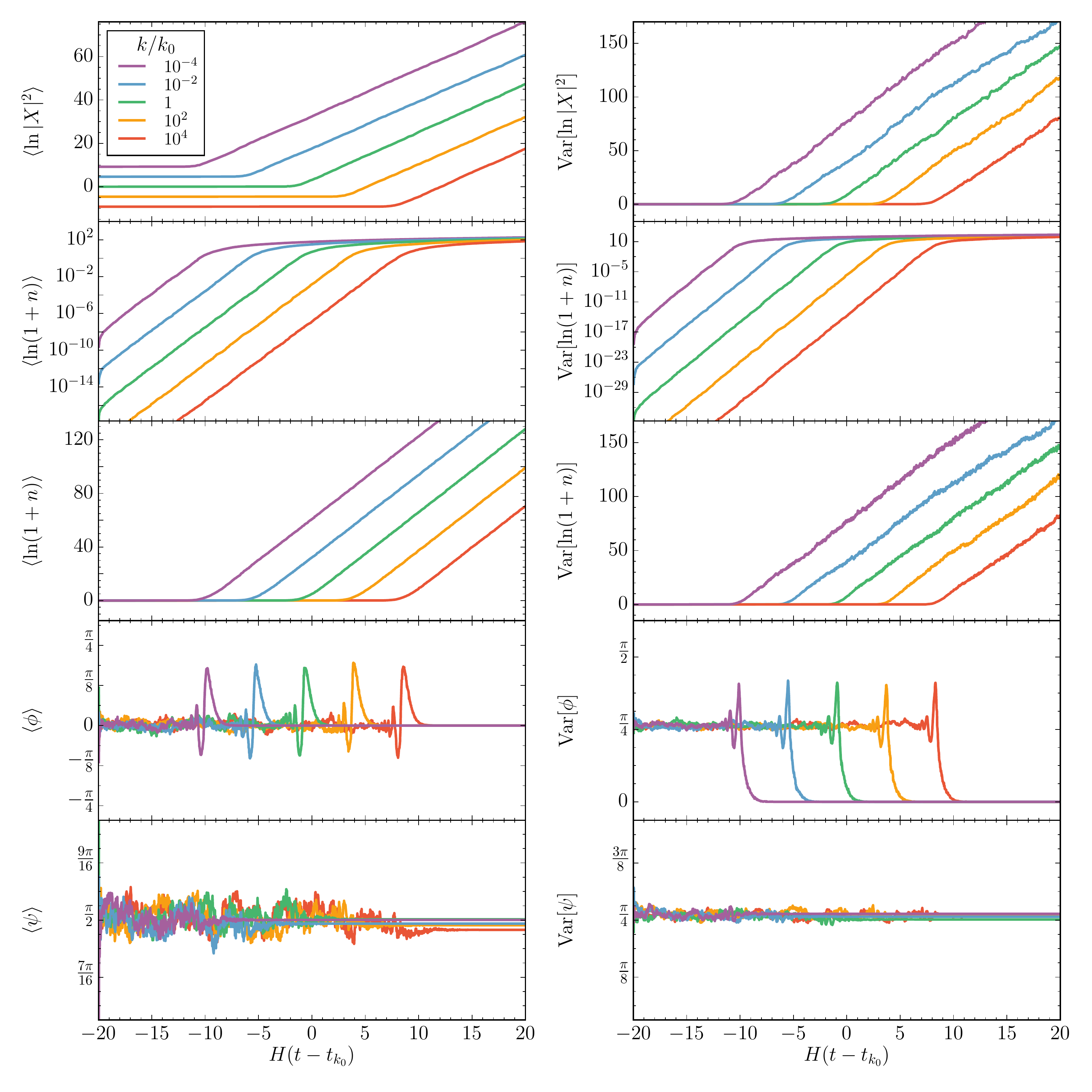}
    \caption{Sub- and super-horizon evolution of the mean and variance of the transfer matrix parameters $\{n,\phi,\psi\}$ and the scalar field amplitude, for different values of $k/H$ in the massless case. All parameters and normalizations are chosen as in Fig.~\ref{fig:kmeanvar}.}
    \label{fig:kmeanvar_m}
\end{figure}

The dependence of the means and variances of $\ln|X|^2$, $\ln(1+n)$, $\phi$ and $\psi$ on $\mathcal{N}_s(\sigma/H)^2$ and $k$ are shown in Figs.~\ref{fig:sigmeanvar_m} and \ref{fig:kmeanvar_m}, respectively. To allow a straightforward comparison, all parameters have been chosen as their conformal counterparts shown in Figs.~\ref{fig:sigmeanvar} and \ref{fig:kmeanvar}. The same is true for the distribution of the local disorder parameters, namely a uniform distribution for amplitudes and locations of scatterings. This assumption may be broken to allow for different disorder distributions, but it always leads to the same results provided that both the $m_j$ and $\delta t_j$ are random. We state this fact without explicitly showing our checks, as they are qualitatively indistinguishable from those discussed in Appendix~\ref{ap:conv}. A similar argument follows for the convergence test for the particle production rate at large $N_s$, also addressed in Appendix~\ref{ap:conv} for the conformal case.
\par
\bigskip

\noindent{\bf Moments of $\ln |X|^2$ on sub-horizon scales}: 
The time-evolution of the massless scalar field amplitude is shown in the top of Figs.~\ref{fig:sigmeanvar_m} and \ref{fig:kmeanvar_m}. Inside the horizon,  $\langle \ln|X|^2\rangle$ is constant, independently of both the scattering strength and the wavenumber of the corresponding mode. From (\ref{eq:genchi_m}), for $n\ll1$, we can parametrize this behavior using the conformal expression (\ref{eq:subhlnch}). Conversely, ${\rm Var}\left[ \ln|X|^2\right]$ grows exponentially, with a rate that is independent of $\mathcal{N}_s(\sigma/H)^2$ and $k$, and approximately equal to (\ref{eq:subhvarlnch}). Note that this implies that the variance grows $\propto a^2$.\par\bigskip

\begin{figure}[t!]
\centering
    \includegraphics[width=0.87\textwidth]{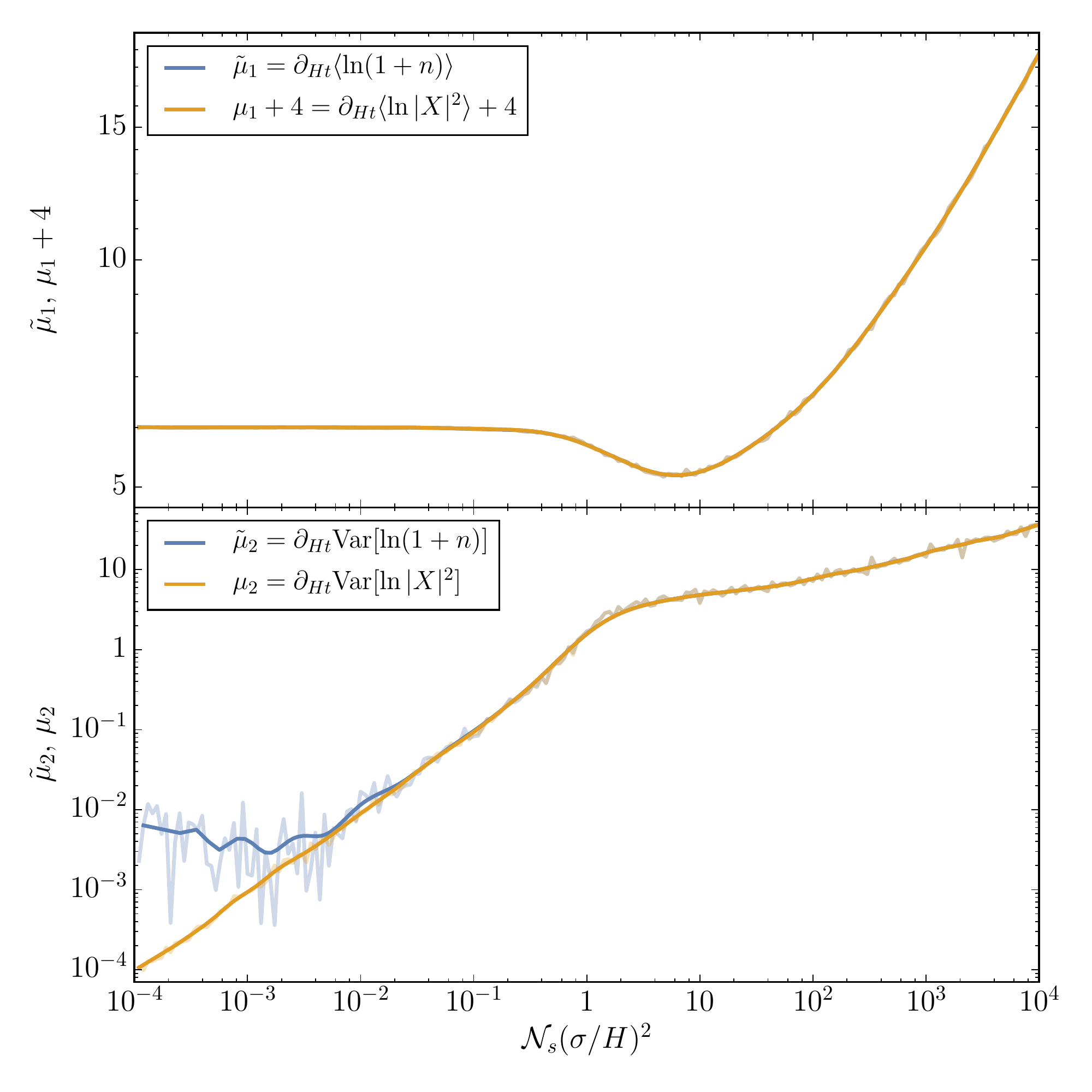}
    \caption{Numerically evaluated super-horizon cosmic time rates for the occupation number and the scalar field magnitude as functions of the parameter $\mathcal{N}_s(\sigma/H)^2$, in the massless case. Parameters are chosen as in Fig.~\ref{fig:nfrates}.}
    \label{fig:nfrates_m}
\end{figure}

\noindent{\bf Moments of $\ln |X|^2$ on super-horizon scales}: 
Outside the horizon, the difference between the massless and the conformal scenarios becomes evident. Unlike the conformal case, in which the magnitude of $X$ remains approximately constant for weak scattering, for a massless spectator field the magnitude grows with the scale factor in super-horizon scales if $\mathcal{N}_s(\sigma/H)^2\ll 1$; this is of course the expected behavior for a massless adiabatic mode, for which $|\chi|\sim {\rm const}$. In order to break from the adiabatic limit, the strong scattering regime must be considered. Fig.~\ref{fig:kmeanvar_m} suggests that the rate of this growth is independent of the mode wavenumber. In analogy with the conformal case we therefore write,
\begin{flalign}
& \text{($|k\tau|\ll 1$)} & \Cen{3}{
\begin{aligned}
\partial_{Ht} \langle \ln |X|^2\rangle \;&=\; \mu_1\left(\mathcal{N}_s(\sigma/H)^2\right)\,,\\
\partial_{Ht} {\rm Var}\left[ \ln |X|^2\right] \;&=\; \mu_2\left(\mathcal{N}_s(\sigma/H)^2\right)\,,
\end{aligned}}      &&  
\end{flalign}
where the functions $\mu_{1,2}$ are shown in Fig.~\ref{fig:nfrates_m}; therein the number of realizations as well as the scattering parameters are chosen as in its conformal counterpart, Fig.~\ref{fig:nfrates}. As it is clear, $\langle \ln|X|^2\rangle$ grows as $a^2$ outside the horizon for $\mathcal{N}_s(\sigma/H)^2\lesssim 0.3$, it increases at a reduced rate for $0.3 \lesssim \mathcal{N}_s(\sigma/H)^2\lesssim 50$, and grows exponentially with rate $\mu_1\sim \left[ \mathcal{N}_s(\sigma/H)^2 \right]^{0.25}-4>2$ for $\mathcal{N}_s(\sigma/H)^2\gtrsim 50$.

The time-rate dependence on the scattering parameter for the variance of $\ln|X|^2$ is shown in the lower panel of Fig.~\ref{fig:nfrates_m}. This dependence is very similar to that for the conformal mass case, with the variance growing as $\mu_2\sim  \mathcal{N}_s(\sigma/H)^2 $ in the weak scattering limit, while for strong scattering we have $\mu_2\sim \left[ \mathcal{N}_s(\sigma/H)^2 \right]^{0.32}$. In analogy with the conformal case, one can verify that $\langle \ln|X|^2 \rangle$ characterizes the typical member of the ensemble of realization, despite the rapidly growing variance (see Appendix~\ref{ap:typ} for details).\par\bigskip

\noindent{\bf Moments of $\ln (1+n)$ on sub-horizon scales}: 
The evolution of the mean and variance of $\ln(1+n)$ is shown in a log-scale in the second row of Fig.~\ref{fig:sigmeanvar_m}. Both moments grow exponentially ($\propto a^2$) inside the horizon, with values dependent on the scattering strength parameter $\mathcal{N}_s(\sigma/H)^2$, but with rates independent of it. The exponential growth continues until the mode leaves the horizon, for weak scattering, or shortly before horizon crossing, for strong scattering. Fig.~\ref{fig:kmeanvar_m} further demonstrates that the occupation number is dependent on the wavenumber, but the growth rate is independent of it. It is straightforward to check that the expressions (\ref{eq:subhnres}), valid in the conformal case, also correctly describe the sub-horizon evolution of the occupation number in the massless case. This is consistent with the fact that, in the $|k\tau|\gg 1$ limit, the mode functions in both cases have the same plane-wave form (\ref{eq:conf_fk}). \par\bigskip

\noindent{\bf Moments of $\ln (1+n)$ on super-horizon scales}: 
The third row of Figs.~\ref{fig:sigmeanvar_m} and \ref{fig:kmeanvar_m} displays the time evolution of the occupation number moments, but in this case in a linear scale. The linear growth of $\langle \ln(1+n)\rangle$ is evident, with a rate that is seemingly independent of $\mathcal{N}_s(\sigma/H)^2$ and $k$ for weak scattering, and independent only of $k$ for strong scattering, 
\begin{flalign}
& \text{($|k\tau|\ll 1$)} & \Cen{3}{
\begin{aligned}
\partial_{Ht} \langle \ln(1+n)\rangle \;&=\; \tilde{\mu}_1\left(\mathcal{N}_s(\sigma/H)^2\right)\,,\\
\partial_{Ht} {\rm Var}\left[ \ln(1+n)\right] \;&=\; \tilde{\mu}_2\left(\mathcal{N}_s(\sigma/H)^2\right)\,.
\end{aligned}}      &&  
\end{flalign}
The functional dependence parametrized by the $\tilde{\mu}_{1,2}$ functions in the massless case is shown in Fig.~\ref{fig:nfrates_m}. Note that for the massless case, $\tilde{\mu}_1\simeq \mu_1+4$ for any scattering strength, while $\tilde{\mu}_2\simeq \mu_2$ for $\mathcal{N}_s(\sigma/H)^2\gtrsim 10^{-1}$. The growth rate for the mean is approximately constant, $\partial_{Ht}\langle \ln(1+n)\rangle \simeq 6$, in the weak scattering regime, with $\mathcal{N}_s(\sigma/H)^2\lesssim 0.4$. When $0.4 \lesssim \mathcal{N}_s(\sigma/H)^2\lesssim 40$, the typical occupation number grows at a slower rate. Finally, for strong scattering, the mean grows with the scattering strength parameter, $\tilde{\mu}_1\sim \left[ \mathcal{N}_s(\sigma/H)^2 \right]^{0.25}$.

The time rate for ${\rm Var}\left[ \ln(1+n)\right]$ is shown in the lower panel of Fig.~\ref{fig:nfrates_m}. For very low values of the scattering strength parameter, $\mathcal{N}_s(\sigma/H)^2\lesssim 10^{-2}$, the rate  appears approximately constant, $\tilde{\mu}_2\sim 4\times 10^{-3}$. For larger values, $10^{-2} \lesssim \mathcal{N}_s(\sigma/H)^2\lesssim 1$, the growth rate becomes scattering strength-dependent and relatively steep, $\tilde{\mu}_2\sim \left[ \mathcal{N}_s(\sigma/H)^2 \right]^{1.4}$. In the strong scattering regime the power-law dependence is gentler with $\tilde{\mu}_2\sim \left[ \mathcal{N}_s(\sigma/H)^2 \right]^{0.32}$. Although this rate is steeper than that of the mean, we expect $\exp[\langle\ln(1+n)\rangle]$ to be a good measure of the typical number of particles produced, in analogy with the conformal scenario. We confirm this fact in detail in Appendix~\ref{ap:typ}.\par\bigskip

\noindent{\bf Moments of $\phi$ on sub-horizon scales}: 
The fourth row of Figs.~\ref{fig:sigmeanvar_m} and \ref{fig:kmeanvar_m} show the time evolution of the angular parameter $\phi$. For all values of $\mathcal{N}_s(\sigma/H)^2$ and $k$ we find that, inside the horizon,
\begin{flalign}
& \text{($|k\tau|\gg 1$)} & \Cen{3}{\langle \phi\rangle \simeq 0\,,\qquad {\rm Var}\left[ \phi \right] \simeq \frac{\pi^2}{12}\,,}      &&  
\end{flalign}
the expected results for a uniformly distributed random variable in $(-\pi/2,\pi/2)$. Similarly to the conformal case, both moments oscillate about these values as the mode leaves the horizon, with amplitudes dependent of the scattering strength parameter and independent of the wavenumber of the mode. \par\bigskip

\noindent{\bf Moments of $\phi$ on super-horizon scales}: 
Far outside the horizon, the mean and the variance of $\phi$ are driven to
\begin{flalign}\label{eq:phimvcn_m}
& \text{($|k\tau|\ll 1$)} & \Cen{3}{\langle \phi\rangle \simeq 0\,,\qquad {\rm Var}\left[ \phi \right] \rightarrow 0\,.}      &&  
\end{flalign}
Numerically we find that the final value of $\langle \phi\rangle$ deviates from zero for any $\mathcal{N}_s(\sigma/H)^2$ up to a numerical error $\lesssim \mathcal{O}(10^{-20})$. We also find that the decay of the variance may be parametrized as
\begin{flalign}\label{eq:phivrcn_m}
& \text{($|k\tau|\ll 1$)} & \Cen{3}{\partial_{Ht}\ln\left({\rm Var}\left[ \phi \right] \right) \;\simeq\; -3 \,,}      &&  
\end{flalign}
for any scattering strength.\par\bigskip

\noindent{\bf Moments of $\psi$ on sub-horizon scales}: The time-dependence of the moments of $\psi$ is shown in the bottom panels of Figs.~\ref{fig:sigmeanvar_m} and \ref{fig:kmeanvar_m}. In this case, the mean and the variance correspond to those of a uniformly distributed random variable in $(0,\pi)$ inside the horizon,
\begin{flalign}\label{eq:psimassmv}
& \text{($|k\tau|\gg 1$)} & \Cen{3}{\langle \psi\rangle \simeq \frac{\pi}{2}\,,\qquad {\rm Var}\left[ \psi \right] \simeq \frac{\pi^2}{12}\,.}      &&  
\end{flalign}\par\bigskip

\noindent{\bf Moments of $\psi$ on super-horizon scales}: Outside the horizon, both the mean and the variance of $\psi$ are dependent on the scattering strength, but independent of $k$. We find for their asymptotic values
\begin{flalign}\label{eq:psimv_m}
& \text{($|k\tau|\ll 1$)} & \Cen{3}{
\begin{aligned}
\langle \psi \rangle \;&\simeq\; \frac{\pi}{2}\,,\\
{\rm Var}\left[ \psi \right] \;&\simeq\;  \frac{\pi}{12}\times \begin{cases}
\mathcal{N}_s(\sigma/H)^2\,,\quad & \mathcal{N}_s(\sigma/H)^2\lesssim \pi\\[5pt]
\pi\,, & \mathcal{N}_s(\sigma/H)^2\gtrsim \pi
\end{cases}\,.
\end{aligned}}      &&   
\end{flalign}
Here we have ignored a mild dependence on the scattering parameter for $\langle\psi\rangle$, which corresponds to a $\lesssim 6\%$ variation. Note that the variance of $\psi$ depends on the scattering parameter only for weak scattering, and it is frozen at the same constant value as in (\ref{eq:psimassmv}) for strong scattering, consistent with a non-evolving probability distribution.
%

\subsubsection{Probability densities}\label{sec:pdfs_m}
\begin{figure}[t!]
\centering
    \includegraphics[angle=270,origin=c,width=\textwidth]{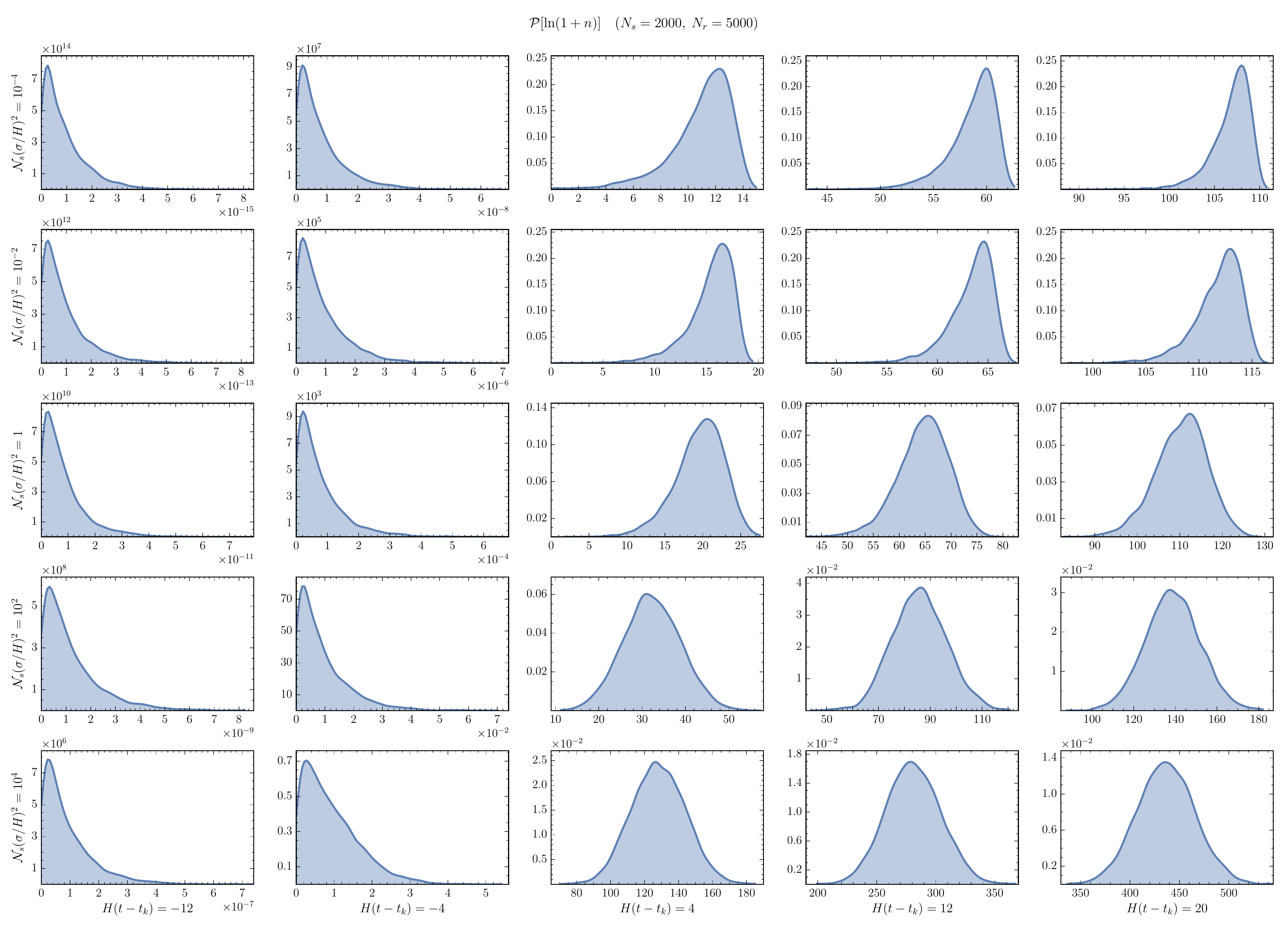}
    \caption{Pdf for $\ln(1+n)$ as a function of time and scattering strength.}
    \label{fig:ndist_m}\vspace{-50pt}
\end{figure}

\begin{figure}[t!]
\centering
    \includegraphics[angle=270,origin=c,width=\textwidth]{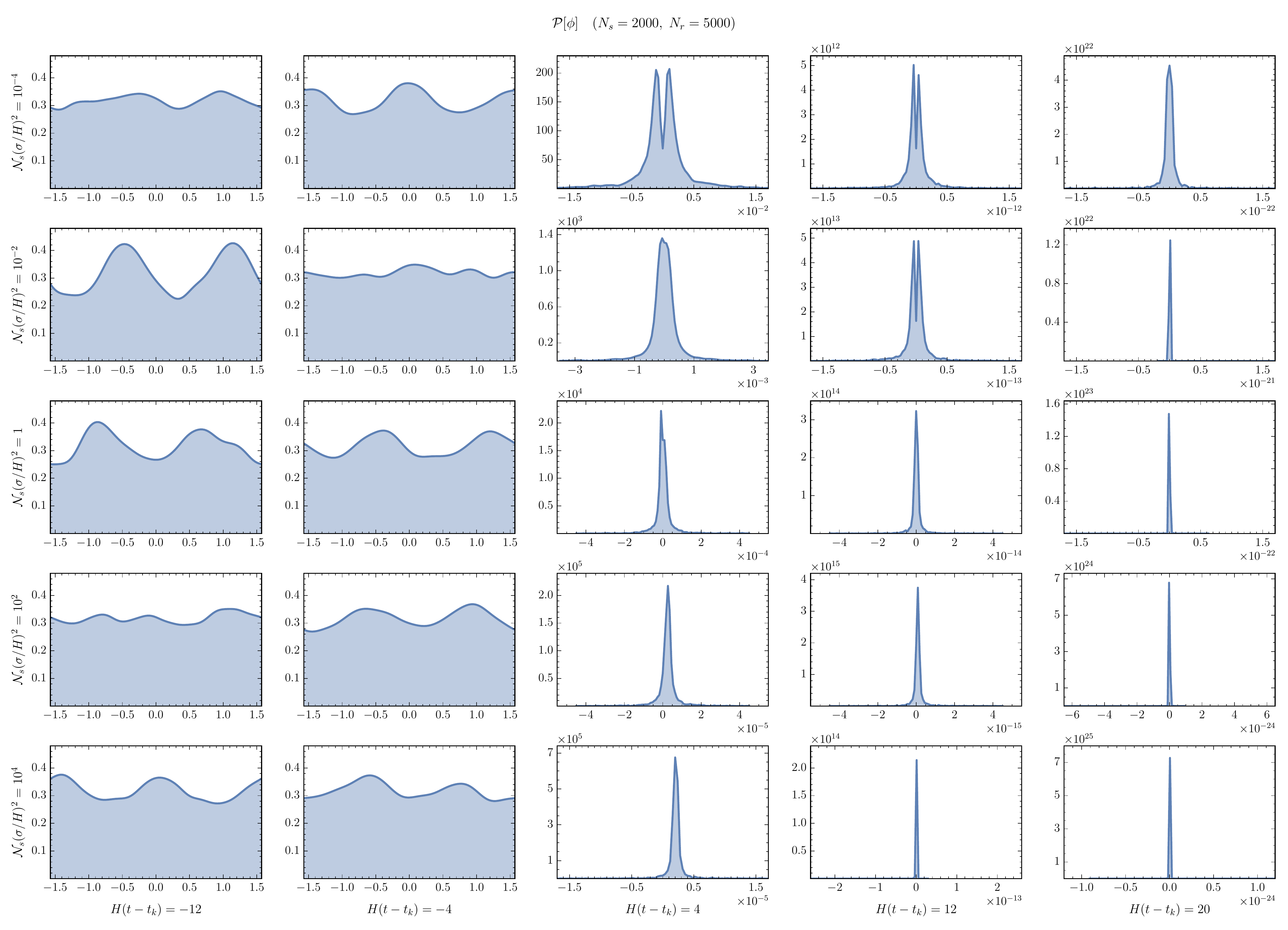}
    \caption{Pdf for $\phi$ as a function of time and scattering strength.}
    \label{fig:phdist_m}\vspace{-50pt}
\end{figure}

\begin{figure}[t!]
\centering
    \includegraphics[angle=270,origin=c,width=\textwidth]{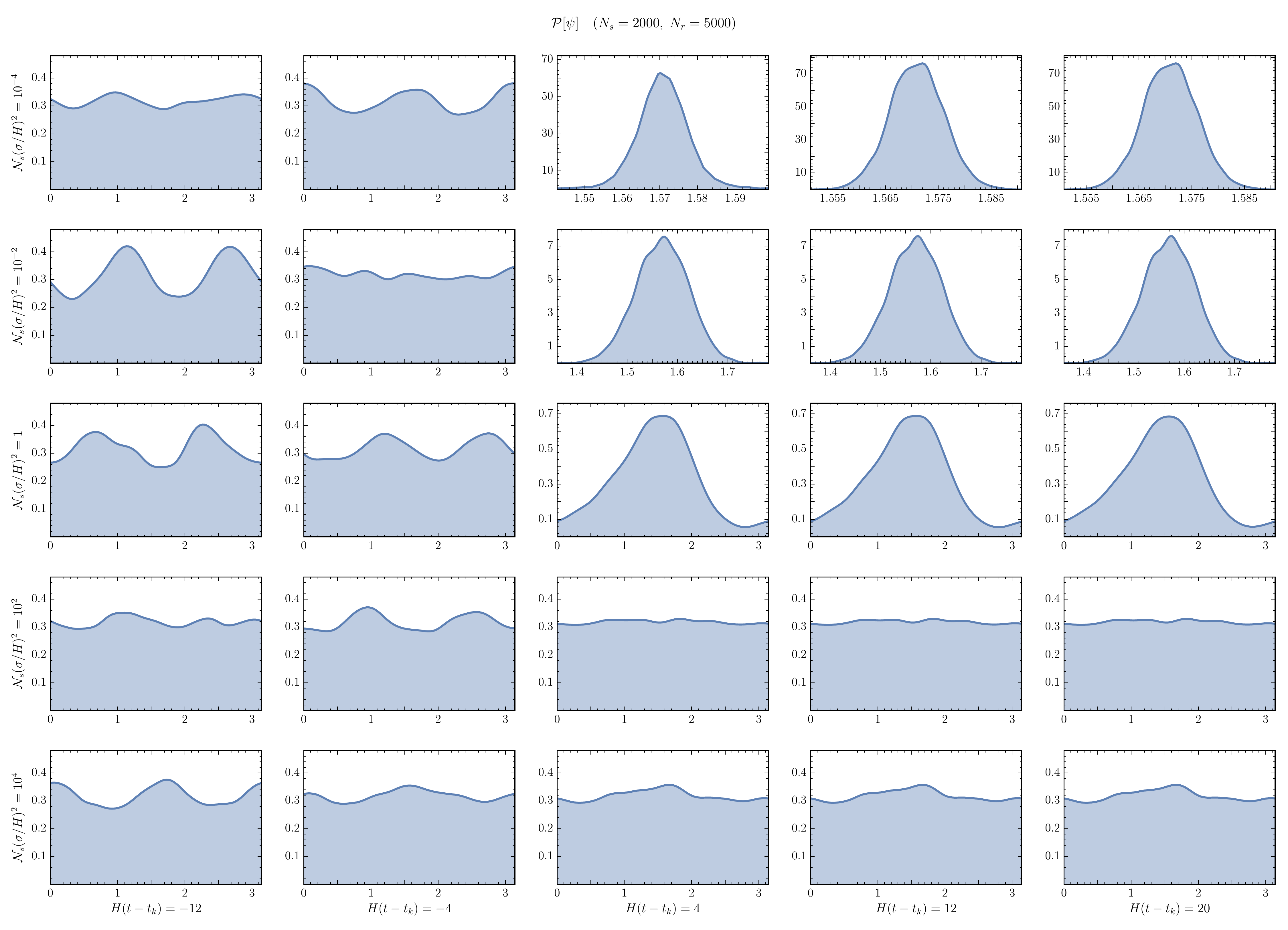}
    \caption{Pdf for $\psi$ as a function of time and scattering strength.}
    \label{fig:psdist_m}\vspace{-50pt}
\end{figure}

We now turn to the description of the evolution of the probability density functions for the random variables $\ln|X|^2$, $\ln(1+n)$, $\phi$ and $\psi$. We base our analysis on Figs.~\ref{fig:ndist_m}-\ref{fig:psdist_m}, which show snapshots of the instantaneous normalized pdfs for the field and transfer matrix parameters for selected values of $\mathcal{N}_s(\sigma/H)^2$. The scattering parameters are taken as those for the conformal case discussed in Section~\ref{sec:pdfs}, to allow for a simple comparison. As in that case, the pdfs are built using a Gaussian kernel density estimator of variable bin size, with a periodic extension of the data for the angular variables in the case where the pdf support is of width $\pi$.\\
\begin{figure}[t!]
\centering
    \includegraphics[angle=270,origin=c,width=\textwidth]{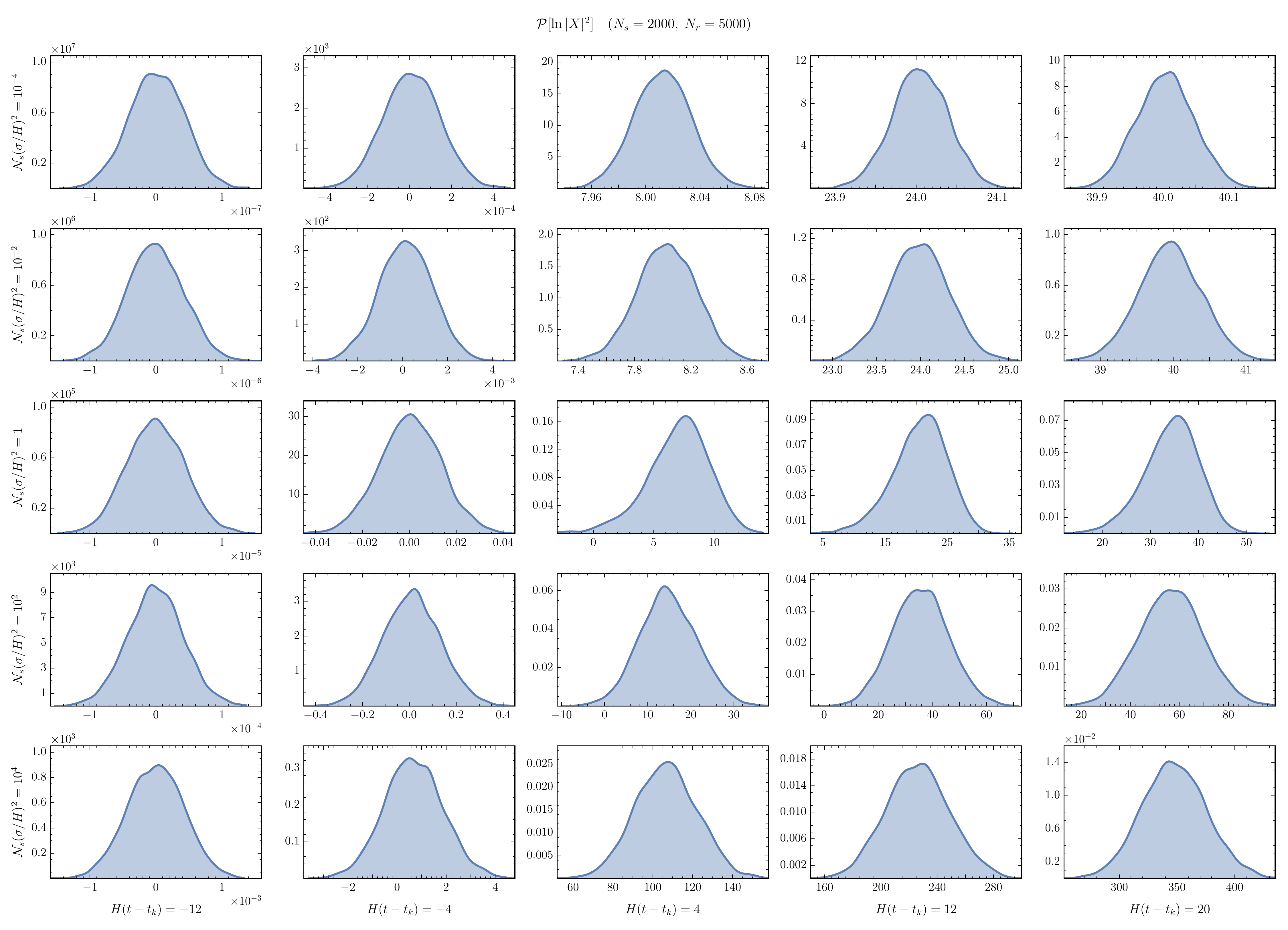}
    \caption{Pdf for $\ln|X|^2$ as a function of time and scattering strength.}
    \label{fig:fdist_m}\vspace{-50pt}
\end{figure}

\begin{figure}[t!]
\centering
    \includegraphics[width=0.94\textwidth]{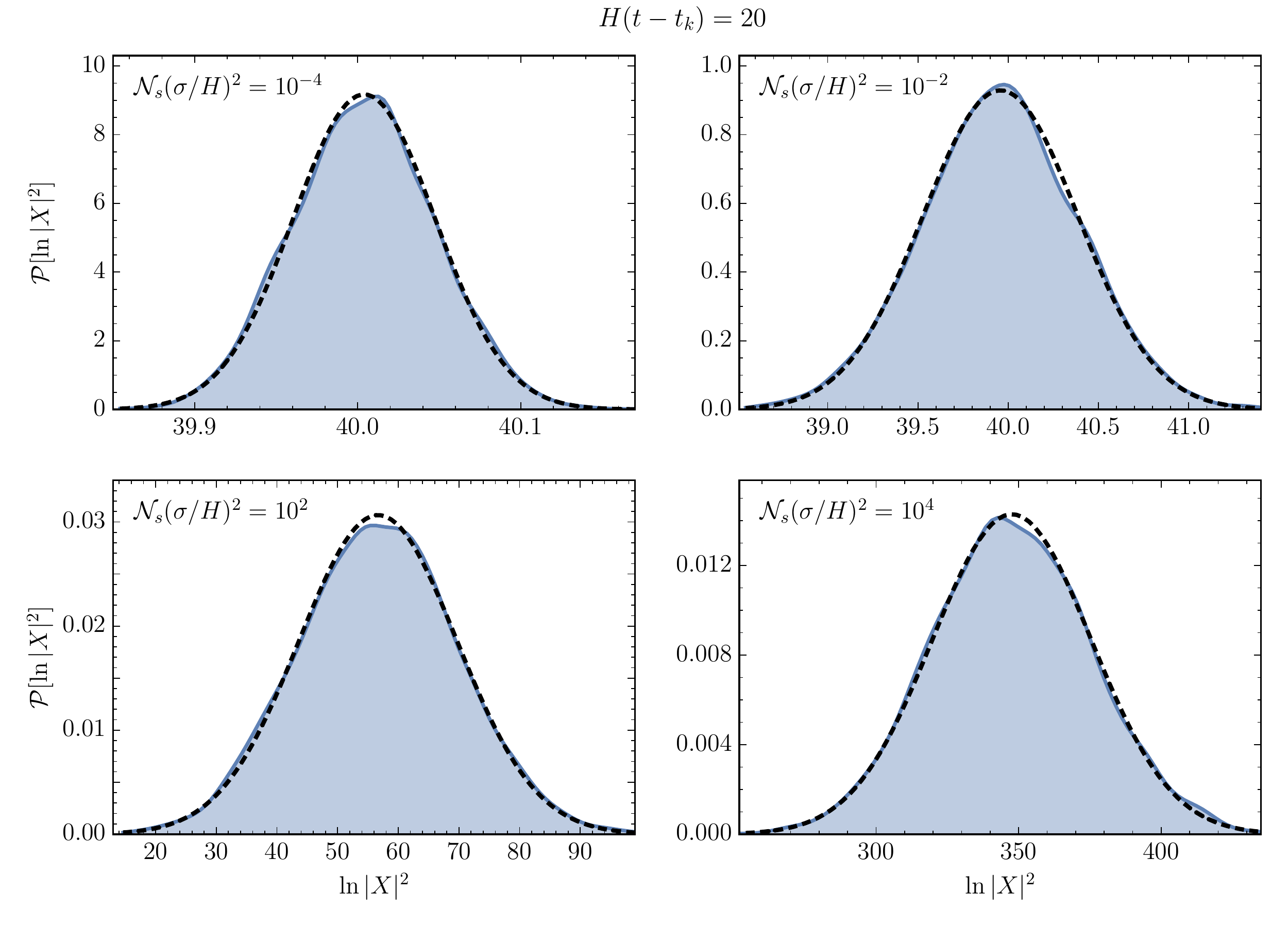}
    \caption{Pdf for $\ln|X|^2$ for selected values of $\mathcal{N}_s(\sigma/H)^2$ at $H(t-t_k)=20$ (massless case). Blue, continuous: numerical result. Black, dashed: Gaussian fit.}
    \label{fig:ffit_m}
\end{figure}

\noindent{\bf Pdf for $\ln|X|^2$}: 
The evolution of the numerical probability distribution for $\ln|X|^2$ is shown in Fig.~\ref{fig:fdist_m}. Similar to the conformal case, the description is straightforward, as the pdf exhibits the bell-shape for all times and values of scattering strength. For super-horizon modes, confirmation is provided in Fig.~\ref{fig:ffit_m}, where Gaussian fits are superimposed over the numerical pdf for four different values of the scattering parameter. In the case of sub-horizon modes, we prove the normality of $\ln|X|^2$ in Section~\ref{sec:analnom1} (see Fig.~\ref{fig:ffitg_m}). We therefore conclude in this case as well that \\
\noindent\rule{\textwidth}{1pt}\\
The squared-field amplitude, $|X|^2$, follows a log-normal distribution both inside and outside the horizon, for weak and strong scattering.\\
\noindent\rule{\textwidth}{1pt}
\par
\noindent{\bf Pdf for $\ln(1+n)$}: 
Fig.~\ref{fig:ndist_m} shows the evolving probability distribution for $\ln(1+n)$. The qualitative resemblance with the conformal case of Fig.~\ref{fig:ndist} is evident. The leftmost two columns show pdfs that flow in a manner almost independent of the value of $\mathcal{N}_s(\sigma/H)^2$, albeit with different numerical values. As anticipated in Section \ref{sec:meanvarconf_m}, the early time pdf is of a highly skewed, almost exponential form, with a coefficient of variation $\tau_{\ln(1+n)}\approx 1$ (see Appendix~\ref{ap:typ}). In the next section we will confirm this result analytically (see Fig.~\ref{fig:nfit_m}). The distributions grow in width and mean, maintaining their quasi-exponential shape until horizon crossing, albeit some deformation is clear for larger values of the scattering strength parameter.

\begin{figure}[t!]
\centering
    \includegraphics[width=0.94\textwidth]{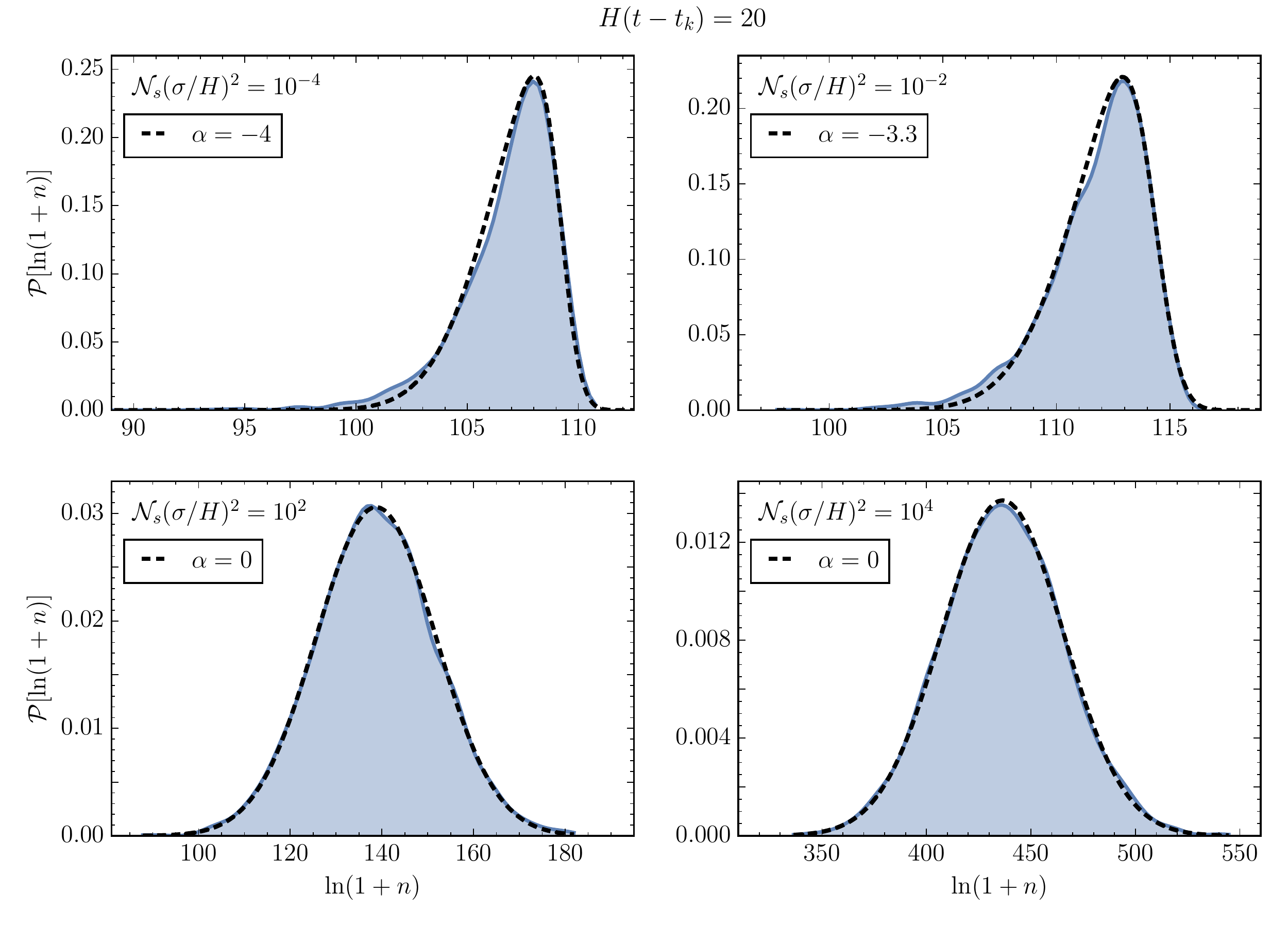}
    \caption{Pdf for $\ln(1+n)$ for selected values of $\mathcal{N}_s(\sigma/H)^2$ at $H(t-t_k)=20$ (massless case). Blue, continuous: numerical result. Black, dashed: skew-normal fit with shape parameter $\alpha$.}
    \label{fig:nfit_skw_m}
\end{figure}

In the middle column of Fig.~\ref{fig:ndist_m} we can see the form of the pdfs at the horizon exit transition. Unlike the conformal case, the transitional form for the pdfs is not evident here, as they have all clearly evolved toward their right- or center-lobed forms, depending on the value of $\mathcal{N}_s(\sigma/H)^2$. The last two columns of the figure in question show the deep super-horizon form of the distributions. Similarly to the conformally massive field, all pdfs are notably peaked about their means, with the weak scattering distributions retaining a tail of low $n$ realizations. Fig.~\ref{fig:nfit_skw_m} shows four of the five panels at $H(t-t_k)=20$ together with skew-normal fits to the data (c.f.~Eq.~(\ref{eq:skew})). For $\mathcal{N}_s(\sigma/H)^2\ll 1$ the distributions present significant skewness. In turn,  $\mathcal{N}_s(\sigma/H)^2\gg 1$ leads to a normally distributed $\ln(1+n)$, implying a log-normally distributed occupation number $n$.
\par\bigskip
\noindent{\bf Pdf for $\phi$}: 
Fig.~\ref{fig:phdist_m} displays the distribution of the transfer matrix phase $\phi$ at several times and scattering strengths. Comparison with Fig.~\ref{fig:phdist} reveals that the massless and conformal forms for the $\phi$ pdf have similar structures and evolution, although the central values and the narrowing rates differ. The two leftmost columns show approximately uniform distributions over $(-\pi/2,\pi/2)$ inside the horizon for all $\mathcal{N}_s(\sigma/H)^2$. Some structure is visible, although it is likely due to the finite number of realizations considered for the calculation. 

The central column of Fig.~\ref{fig:phdist_m} shows the form of $\mathcal{P}[\phi]$ at the horizon exit transition. It is clear here that the density evolves to a two-lobed distribution with a rate that increases with  $\mathcal{N}_s(\sigma/H)^2$. In the present case the two lobes are approximately symmetrically located with respect to $\phi=0$. This structure is clearly visible for weak scattering, but it is lost in the strong scattering case due to the rapid sharpening of the distribution. Unlike the conformal scenario, it is not clear that the height of both lobes is equal for $\mathcal{N}_s(\sigma/H)^2\gg 1$, as some asymmetry is clearly visible, consistent with the distribution of points in Fig.~\ref{fig:phigrid_m}.

The two rightmost columns of Fig.~\ref{fig:phdist_m} attempt to show the form for the $\phi$ pdf with $|k\tau|\ll 1$. In the top panels we can see the two lobes of the distribution approaching each other exponentially fast, with decreasing widths. As the rates for these processes are dependent on $\mathcal{N}_s(\sigma/H)^2$, this evolution cannot clearly be seen in the panels for $\mathcal{N}_s(\sigma/H)^2\gtrsim 1$; similarly to the conformal case our pdf estimator is not capable of showing clearly the structure of the distribution. We further discuss the super-horizon dynamics of the $\phi$ distribution in Section~\ref{sec:analnom2}, where we compute an analytical approximation by means of a reduced Fokker-Planck equation (see Fig.~\ref{fig:phfit_m}).\par\bigskip

\begin{figure}[t!]
\centering
    \includegraphics[width=0.95\textwidth]{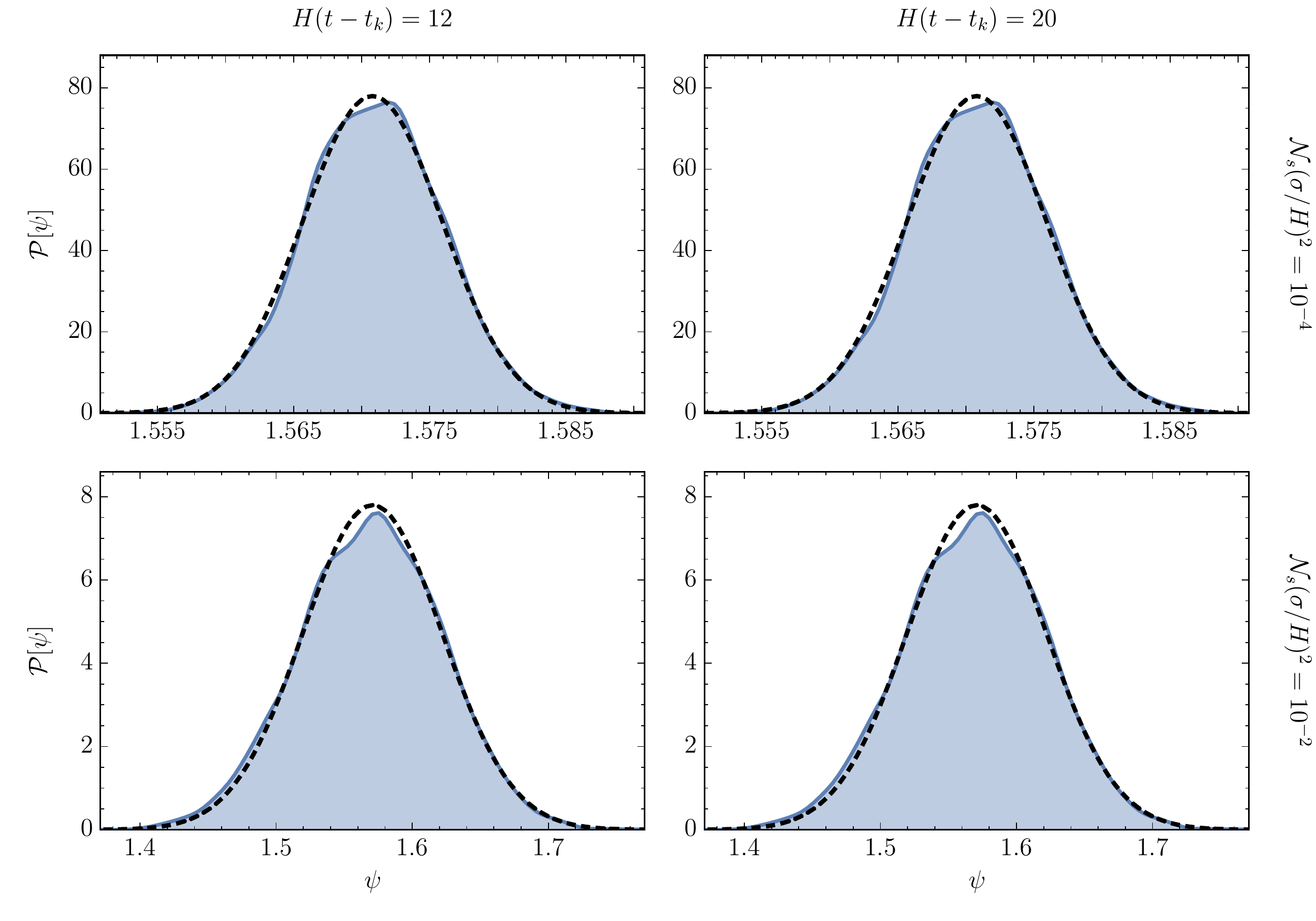}
    \caption{Pdf for $\psi$ for selected values of $\mathcal{N}_s(\sigma/H)^2\ll 1$ and time (massless case). Blue, continuous: numerical result. Black, dashed: normal distribution with mean and variance (\ref{eq:psimv_m}).}
    \label{fig:psfit_m}
\end{figure}

\noindent{\bf Pdf for $\psi$}: 
Fig.~\ref{fig:psdist_m} shows the instantaneous form of the pdf for $\psi$. Deep inside the horizon the distribution is approximately uniform at any time and for any scattering strength. Some fluctuating features are visible, likely an artifact of our finite sized ensemble. As in the conformal case, the uniformity of the distribution is preserved after horizon exit for $\mathcal{N}_s(\sigma/H)^2\gg 1$, and its freeze-out is clear from the rightmost three columns. For weak scattering the distribution evolves toward a bell shape centered at $\psi\simeq \pi/2$ outside the horizon. Fig.~\ref{fig:psfit_m} shows the four upper right panels of Fig.~\ref{fig:psdist_m} compared to a normal distribution with mean and variance (\ref{eq:psimv_m}); $\psi$ is therefore normally distributed in this limit. Finally, for $\mathcal{N}_s(\sigma/H)^2=1$ the limiting distribution appears intermediate between a uniform and a normal distribution.
\subsection{The field two-point function}\label{sec:twopconf}
\begin{figure}[t!]
\centering
    \includegraphics[width=\textwidth]{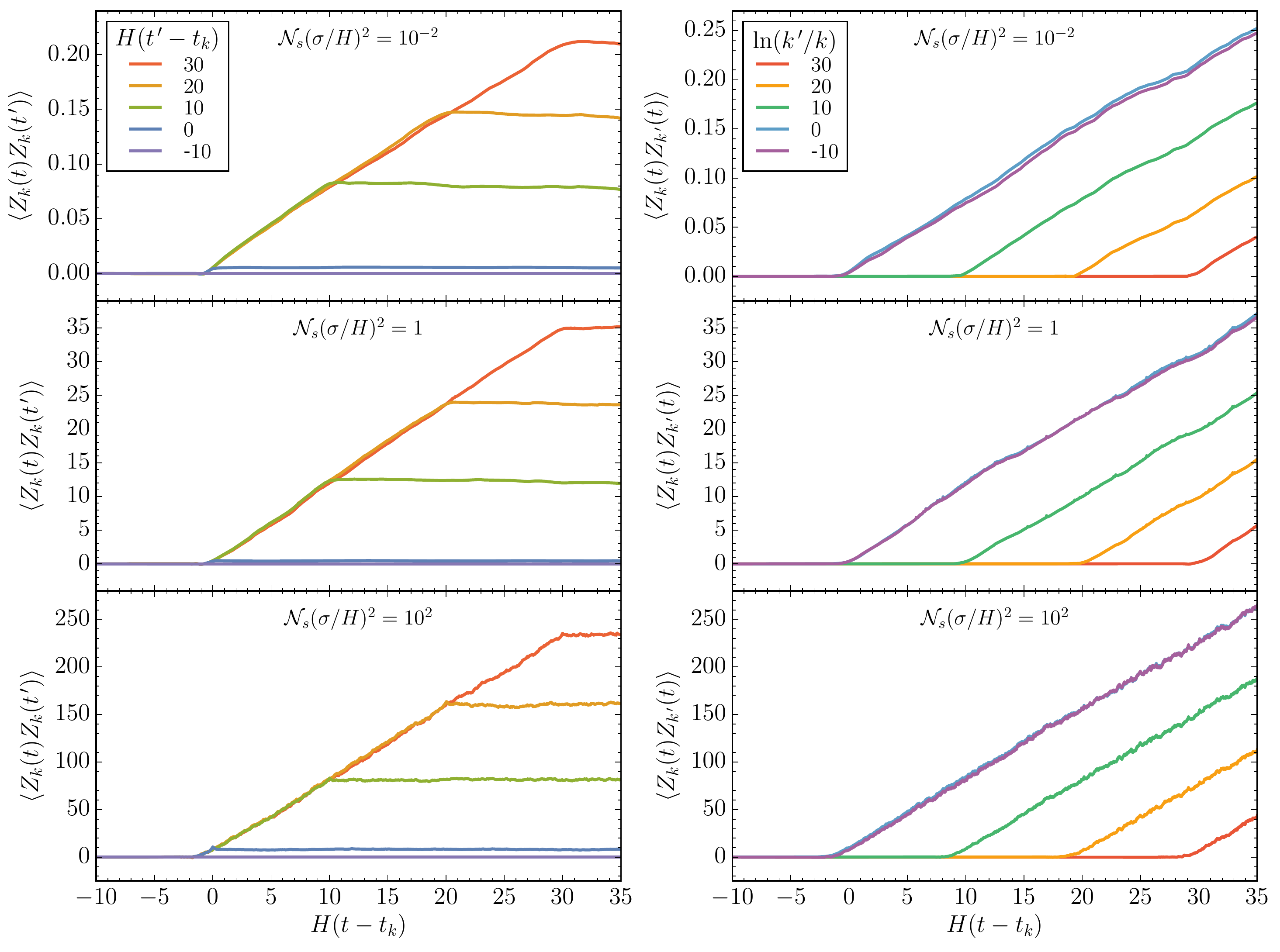}
    \caption{Sub- and super-horizon evolution of the field two-point function for equal momenta and unequal time (left), and unequal momenta and equal time (right), for different values of $\mathcal{N}_s(\sigma/H)^2$ in the massless case.}
    \label{fig:2point_mass}
\end{figure}

We now proceed to analyze numerically the two-point function of the logarithm of the field amplitude defined in (\ref{eq:twopdef}), in terms of the driftless variable $Z_{k}(t)$, defined in (\ref{eq:zdef}). Fig.~\ref{fig:2point_mass} shows the time-dependence of the two point function in two cases.\\

\noindent{\bf Unequal time}: 
The left column of Fig.~\ref{fig:2point_mass} corresponds to the unequal time ($t'\neq t$) but equal momenta ($k'=k$) case, for selected values of $t'$ and the scattering parameter. For all three values of $\mathcal{N}_s(\sigma/H)^2$ the behavior of the curves is very similar to that in the conformal case. Namely, for $t<t_k$ the two point function has negligible value for any $t'$. If $t' \lesssim t_k$, the two point function does not grow as the mode crosses outside the horizon (purple and blue curves). However, if $t'>t_k$, $\langle Z_k(t)Z_{k}(t')\rangle$ increases linearly with time after horizon crossing, with rate $\mu_2$, to posteriorly freeze at its value at $t=t'$ for $t>t'$ (green, orange and red curves). We can therefore summarize our results by Eq.~(\ref{eq:twoptime}), where $\mu_2$ denotes in this case the rate of growth of the variance of the logarithm of the massless field magnitude.\\

\noindent{\bf Unequal momenta}: 
The right column of Fig.~\ref{fig:2point_mass} shows the evolution of the field two-point function for equal time but unequal momenta. Here, if $k'\leq k$, the two-point function grows linearly with time after the mode leaves the horizon (purple, blue). If instead $k'>k$, the two-point function grows only after the mode with momentum $k'$ crosses the horizon (green, orange, red). This behavior is analogous to that in the conformal case, and therefore it is described by Eq.~(\ref{eq:twopk}).\\

\noindent{\bf Unequal time and momenta}: 
For the general case of unequal time and unequal momenta correlators, we have found that Eq.~(\ref{eq:twopointconf}) correctly approximates the super-horizon behavior of the field two-point function. This result suggests that $|X_{k}|^2$ describes a geometric Brownian motion with drift in cosmic time, in full analogy with the conformal case. Also analogous is the observation that a non vanishing correlation between unequal modes exists only if both modes are outside the horizon. Finally, note that Eq.~(\ref{eq:npointX}) for the $n$-point correlation function for the squared field magnitude is equally valid for the massless case.

\subsection{Analytical results}\label{sec:analnom}
In the previous sections we have discussed purely numerical results concerning the expectation values and distributions for the field amplitude and the transfer matrix parameters. Following the program established by our study of the conformally massive field, we now turn to the analytical results that are derivable for a massless spectator field using the Fokker-Planck formalism. As we previously discussed in Sections~\ref{sec:ssds} and \ref{sec:analconf}, we will only be able to probe analytically the very weak scattering regime $\mathcal{N}_s(\sigma/H)^2\ll 1$. However, as we have shown numerically, these results are universal in the sub-horizon limit, while in the super-horizon case we will be able to derive the form of the lowest moments of the occupation number, the squared field magnitude and the $\M$-parameter $\phi$, as well as the probability density of the later. We therefore split our discussion depending on the magnitude of $\mathcal{N}_s^{-1}|k\tau|$ (c.f.~Eq.~(\ref{eq:nsktau})).
\subsubsection{Sufficiently sub-horizon}\label{sec:analnom1}
\begin{figure}[t!]
\centering
    \includegraphics[width=0.94\textwidth]{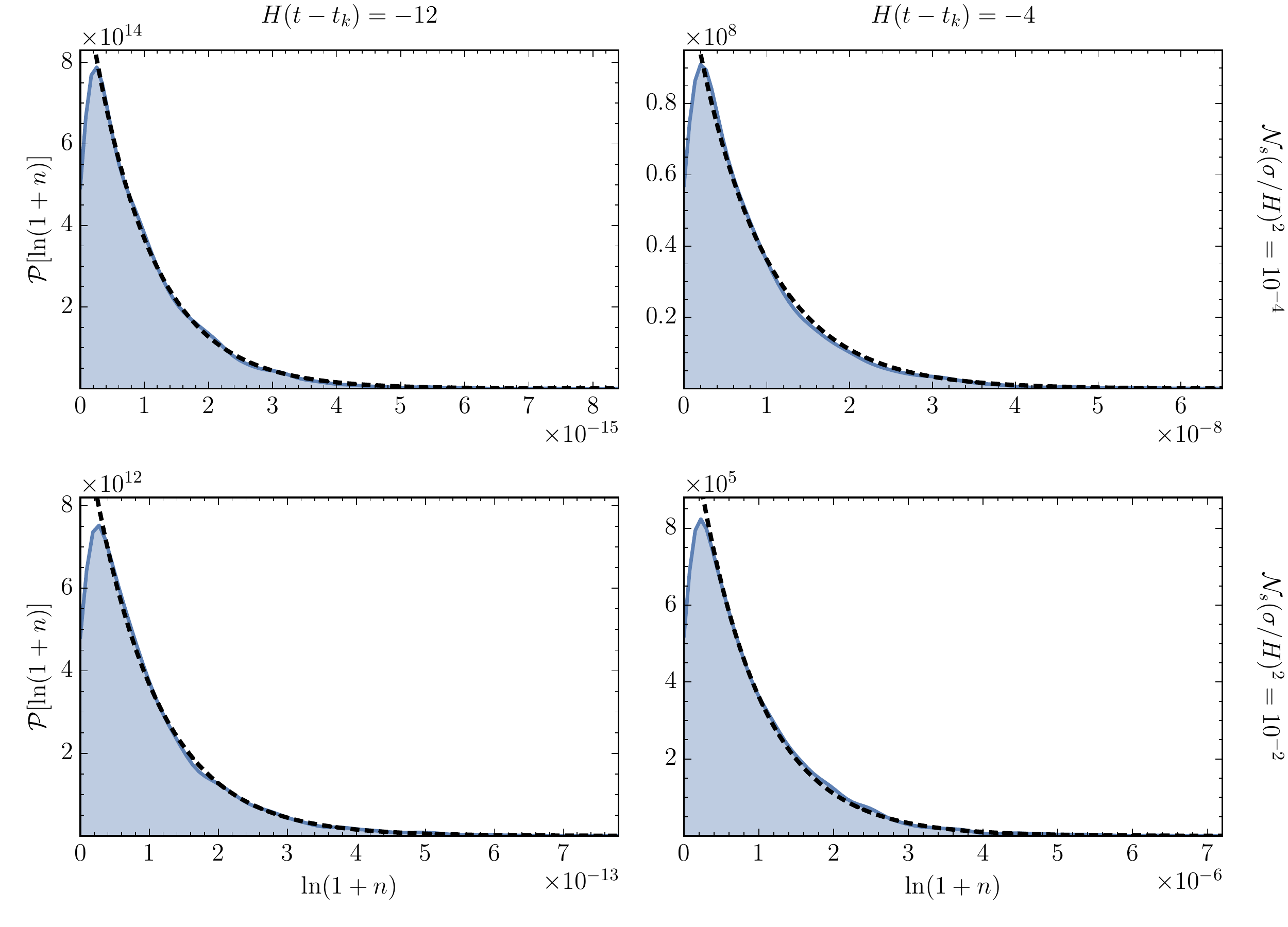}
    \caption{Pdf for $\ln(1+n)$ for selected values of $\mathcal{N}_s(\sigma/H)^2$ and time in sub-horizon scales (massless case). Blue, continuous: numerical result. Black, dashed: the approximation (\ref{eq:distappn_m}).}
    \label{fig:nfit_m}
\end{figure}

\begin{figure}[t!]
\centering
    \includegraphics[width=0.94\textwidth]{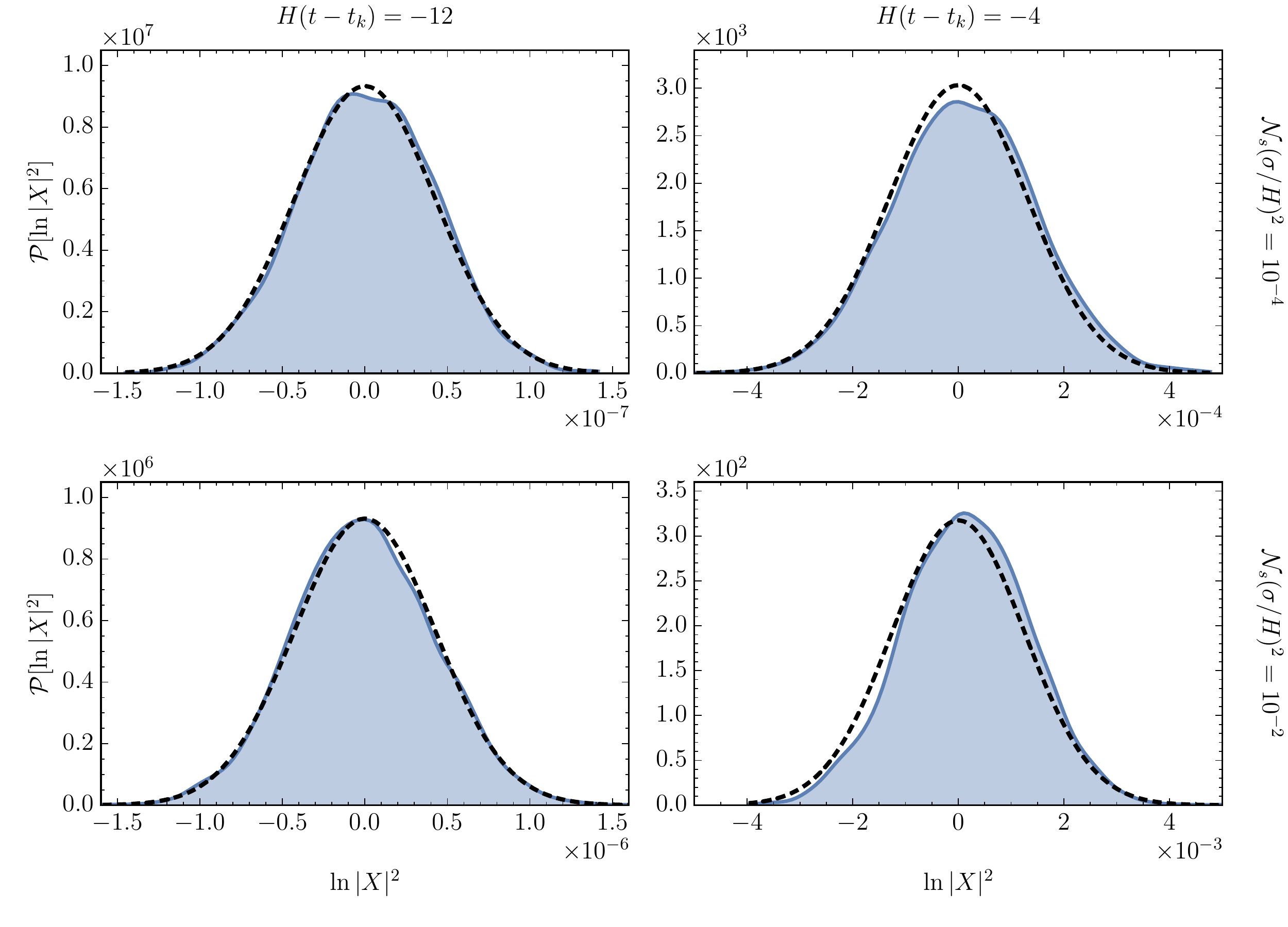}
    \caption{Pdf for $\ln|X|^2$ for selected values of $\mathcal{N}_s(\sigma/H)^2$ and time in sub-horizon scales (massless case). Blue, continuous: numerical result. Black, dashed: the approximation (\ref{eq:pchiin_m}). Note the re-scaling (\ref{eq:rescx}).}
    \label{fig:ffitg_m}
\end{figure}

Section~\ref{sec:FP} contains the general form for the coefficients of the FP equation given Dirac-delta scatterers with uncorrelated amplitudes of vanishing mean. In the present scenario with a massless spectator field, the coefficient functions (\ref{eq:g1g})-(\ref{eq:g4g}) take the form 
\begin{subequations} 
\begin{align} 
g^{(1)} & =  -\frac{i\tilde{\lambda} m_j}{ 2  H k \tau_j} \left[e^{-2 i (\phi+k \tau_j)}\left(1-\frac{i}{k \tau_j}\right)^2 - e^{2 i (\phi+k \tau_j)}\left(1+\frac{i}{k \tau_j}\right)^2 \right]\,,\\
\tilde{g}^{(1)} &= -\frac{i m_j}{H k \tau_j} \left[\tilde{\lambda}\left| 1 - \frac{i}{k \tau_j}\right|^2 + \lambda e^{-2 i (\phi + k \tau_j)} \left(1-\frac{i}{k \tau_j}\right)^2 \right]\,,\\ \notag
g^{(2)} & =  \frac{m_j^2}{4H^2 (k \tau_j)^2}\left| 1 - \frac{i}{k \tau_j}\right|^2\\
&\qquad \times \left[2\lambda \left| 1 - \frac{i}{k \tau_j}\right|^2 +  \tilde{\lambda} \left(e^{-2 i (\phi+k \tau_j)} \left(1-\frac{i}{k \tau_j}\right)^2  + e^{2 i (\phi + k \tau_j)} \left(1+\frac{i}{k \tau_j}\right)^2 \right) \right]\,,\\  \notag
\tilde{g}^{(2)} &= -\frac{m_j^2}{4H^2 (k \tau_j)^2} \left[ \tilde{\lambda} \left( \left| 1 - \frac{i}{k \tau_j}\right|^4 + e^{-4 i (\phi + k \tau_j)}\left(1-\frac{i}{k \tau_j}\right)^4 \right)\right.\\
&\qquad\qquad\qquad\qquad\qquad\qquad\qquad\qquad\quad \left.  +\, 2\lambda  e^{-2 i (\phi + k \tau_j)}\left| 1 - \frac{i}{k \tau_j}\right|^2\left(1-\frac{i}{k \tau_j}\right)^2 \right]\,.
\end{align}
\end{subequations}
It is clear that, in the deep sub-horizon regime $|k \tau|\ll1$, the previous expressions reduce simply to their conformal counterparts (\ref{eq:confgs1})-(\ref{eq:confgs4}). Therefore, all the analysis carried out in Section~\ref{sec:analconf1} is also valid in the massless case. This is a consequence of the Bunch-Davies (BD) initial condition imposed on any free-field mode function, which looks precisely like the free conformal mode function (\ref{eq:conf_fk})~\cite{Bunch:1978yq}. We can then immediately conclude that, in the massless scenario, $\phi$ is uniformly distributed inside the horizon, and
\beq\label{eq:distappn_m}
P(\ln(1+n),\xi) \;\simeq\; \frac{1}{\xi}e^{-\ln(1+n)/\xi}\,,
\eeq
\beq
\langle \ln(1+n)\rangle = \left({\rm Var}[\ln(1+n)] \right)^{1/2} = \xi\,,
\eeq
with $\xi = \frac{\mathcal{N}_s}{2}\left(\frac{\sigma}{2k_{\rm phys}}\right)^2$, as defined in (\ref{eq:xidef}). These expressions are in agreement with the numerical results for the moments of $\ln(1+n)$ discussed in Section~\ref{sec:meanvarconf_m}, and, as Fig.~\ref{fig:nfit_m} shows, they also agree with the numerically obtained pdfs at early times in the weak scattering regime. Similarly, for the scalar field amplitude we obtain a log-normal distribution inside the horizon,
\beq\label{eq:pchiin_m}
P(\ln|X|^2)\;\simeq\; \frac{1}{2 \sqrt{\pi  \xi }}\exp\left[-\frac{\left(\ln|X|^2 +\ln(2k) \right)^2}{4 \xi } \right]\,.
\eeq
This distribution agrees with the numerical results shown in Fig.~\ref{fig:ffitg_m}, which correspond to four selected weak scattering panels from Fig.~\ref{fig:fdist_m}. Note that the deviation between the numerical and the analytical results is maximal for the largest values of $\mathcal{N}_s(\sigma/H)^2$ and time, indicating that $X$ experiences some sub-horizon drift as the strength of the scattering is increased.
\subsubsection{Outside the horizon}\label{sec:analnom2} 
We now turn to the study of the FP equation outside the horizon. Following the recipe (\ref{eq:avrule2}), and assuming that $n\gg 1$, we obtain the following $\R$-matrix correlators,
\begin{subequations}  
\begin{align}\label{eq:confSHa_m}
\langle \delta \lambda^{(1)} \delta \lambda^{(1)}\rangle_{\delta t}  &\;\simeq\; \frac{\lambda^2\sigma^2}{k_{\rm phys}^2}|k\tau|^{-4}  \sin^2\left(2 (\phi + k \tau)\right)  \,,\\
\langle \delta \lambda^{(1)} \delta \phi^{(1)}\rangle_{\delta t} &\;\simeq\; -\frac{2 \lambda  \sigma^2}{k_{\rm phys}^2}|k\tau|^{-4}   \cos (\phi  + k \tau ) \sin^3 (\phi + k \tau ) \,, \\
\langle \delta \phi^{(1)} \delta \phi^{(1)}\rangle_{\delta t}  &\;\simeq\; \frac{\sigma^2}{k_{\rm phys}^2}|k\tau|^{-4}  \sin ^4(\phi + k \tau ) \,,\\
\langle \delta \lambda^{(2)}\rangle_{\delta t} &\;\simeq\; \frac{\lambda  \sigma^2}{k_{\rm phys}^2}|k\tau|^{-4}   \sin ^2(\phi + k \tau) \,,\\ \label{eq:confSHe_m}
\langle \delta \phi^{(2)} \rangle_{\delta t}  & \;\simeq\; \frac{\sigma^2}{k_{\rm phys}^2}|k\tau|^{-4}   \cos (\phi + k \tau) \sin^3 (\phi + k \tau)\,.
\end{align}
\end{subequations}

\begin{figure}[t!]
\centering
    \includegraphics[width=0.87\textwidth]{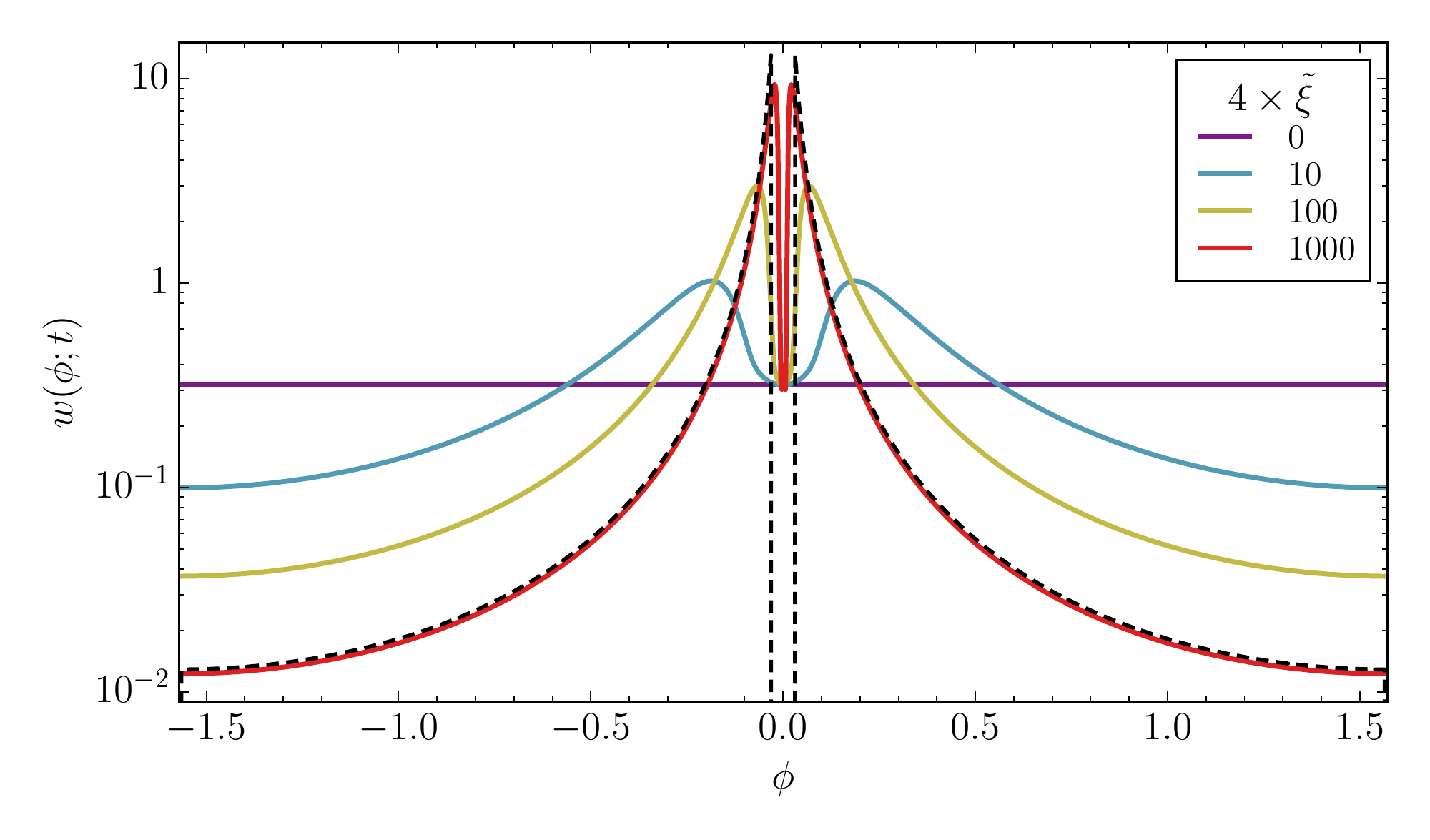}
    \caption{Numerical solution of the FP equation (\ref{eq:wmarc_m}) at different times, with initial condition $w(\phi;0)=1/\pi$. The late-time approximation (\ref{eq:latewc_m}) is shown as the dashed black curve.}
    \label{fig:phfit_m}
\end{figure}

\begin{figure}[t!]
\centering
    \includegraphics[width=0.94\textwidth]{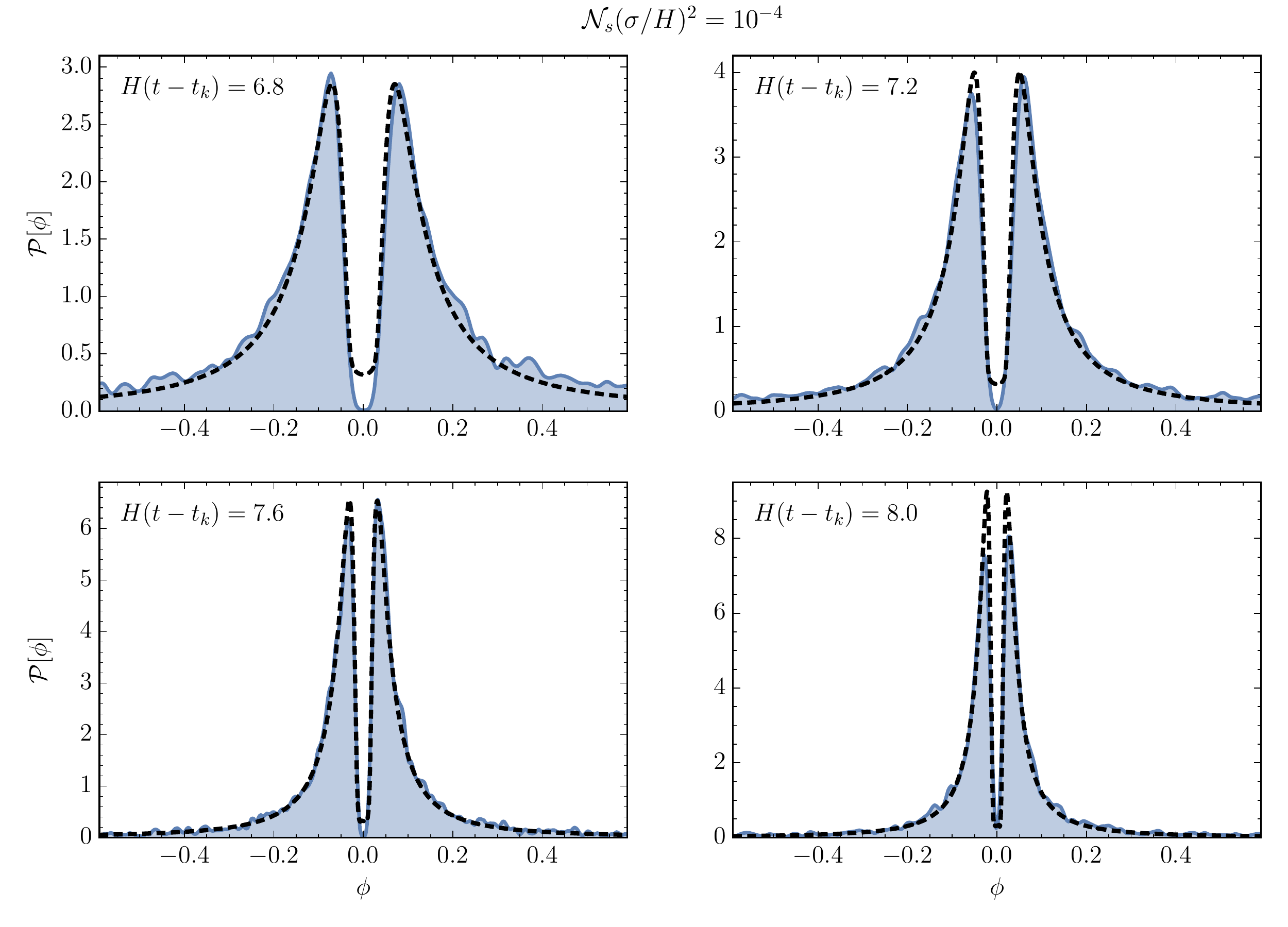}
    \caption{Pdf for $\phi$ for selected values of time in super-horizon scales; here $\mathcal{N}_s(\sigma/H)^2=10^{-4}$ (massless case). Blue, continuous: fully numerical result. Black, dashed: $w(\phi)$ from the numerical integration of (\ref{eq:wmarc_m}).}
    \label{fig:phfit2_m}
\end{figure}

\noindent{\bf Pdf for $\phi$}: Substitution of the correlators (\ref{eq:confSHa_m})-(\ref{eq:confSHe_m}) into (\ref{eq:FPgen}) leads to the FP equation
\begin{align} \notag
\frac{1}{\mathcal{N}_s}\left(\frac{k_{\rm phys}^6}{H^4\sigma^2}\right) \frac{\partial P}{\partial Ht} \;&= \; -\frac{\partial}{\partial \lambda}\Big[ \lambda \sin^2\varphi\,P \Big] - \frac{\partial}{\partial\phi}\Big[ \cos\varphi \sin^3 \varphi\,P\Big] + \frac{1}{2} \frac{\partial^2}{\partial\lambda^2} \Big[\lambda^2\sin^2 2\varphi\,P\Big]\\ \label{eq:FPsuphc_m}
&\qquad - 2\frac{\partial^2}{\partial\lambda\partial\phi} \Big[\lambda \cos\varphi \sin^3 \varphi\,P\Big] + \frac{1}{2}\frac{\partial^2}{\partial\phi^2}\Big[ \sin^4\varphi\,P\Big]\,.
\end{align}
where the shifted angular variable $\varphi$ was defined in (\ref{eq:phishift}). Integrating both sides of (\ref{eq:FPsuphc_m}) with respect to $\lambda$, we obtain the following equation for the marginal probability distribution $w$ defined in (\ref{eq:defmarg}),
\beq\label{eq:wmarc_m}
\frac{\partial w}{\partial (4\tilde{\xi})} \;=\; \frac{1}{2}\frac{\partial}{\partial\phi} \left[\sin^4\varphi \left( \frac{\partial w}{\partial\phi} + 2 w\,\cot\varphi \right)\right]\,,
\eeq
where
\beq
\tilde{\xi } \;\equiv\; \frac{\mathcal{N}_s}{6}\left(\frac{\sigma}{2k_{\rm phys}}\right)^2\left(\frac{H}{k_{\rm phys}}\right)^4 \;=\; \frac{1}{24}\mathcal{N}_s\left(\frac{\sigma}{H}\right)^2 |k\tau|^{-6} \;\propto\; a^6\,.
\eeq
In analogy with the conformal mass case, one would expect the temporal dependence to be negligible in the very-late time limit, with a definite limiting pdf. If this was the case, this distribution would correspond to the solution of the time-independent equation
\beq
w'(\phi)+2w(\phi) \cot\phi \;=\; 0\,,
\eeq
which has the solution $w(\phi)\propto \csc^2\phi$, that is divergent at $\phi=0$ and non-normalizable. Nevertheless, as the conformal case showed us, we can use this expression as an approximation to the time-dependent solution, provided that we choose a suitable time-dependent cutoff around $\phi=0$. The numerical solution of equation of (\ref{eq:wmarc_m}) is shown in Fig.~\ref{fig:phfit_m}, for different values of $\tilde{\xi}$, under the assumption of an initially uniform distribution. Also in this figure the approximation
\beq\label{eq:latewc_m}
w(\phi;t) \;\simeq\; \begin{cases}
\dfrac{\csc^2\phi}{2\cot\delta}\,, & \phi\in(-\frac{\pi}{2},-\delta)\cup(\delta,\frac{\pi}{2})\,,\\[5pt]
0\,,& \phi\in(-\delta,\delta)\,,
\end{cases}
\eeq
is shown as the dashed black curve. By comparison with the numerical result, we find the cutoff to be approximately equal to 
\beq
\delta \;\simeq\; \frac{\tilde{\xi}^{-1/2}}{2} \;\propto\; a^{-3}\,.
\eeq
This approximation implies that
\beq
\langle \phi\rangle = 0\,,\qquad {\rm Var}\,\phi \;\simeq\; \pi\ln(2)\,\delta \;=\; \pi\ln(2) \left(\frac{6}{\mathcal{N}_s(\sigma/H)^2}\right)^{1/2}|k\tau|^3\,,
\eeq
which coincide with the numerical results (\ref{eq:phimvcn_m}), (\ref{eq:phivrcn_m}). Fig.~\ref{fig:phfit2_m} shows the agreement between the numerical solution of (\ref{eq:wmarc_m}) and the fully numerical pdf computed at selected time slices, in the weak scattering case $\mathcal{N}_s(\sigma/H)^2=10^{-4}$.\\

\noindent{\bf Rates for $\langle \ln(1+n)\rangle$ and $\langle \ln|X|^2\rangle$}: 
We now calculate the super-horizon particle production rate in the weak scattering regime. Substituting the correlators (\ref{eq:confSHa_m})-(\ref{eq:confSHe_m}) into the general expression (\ref{eq:lngen}) gives
\begin{align} \notag
\frac{1}{\mathcal{N}_s}\frac{\partial}{\partial Ht}\langle \ln(1+n)\rangle \;&= \; \left( \frac{\sigma}{k_{\rm phys}} \right)^2\left(\frac{H}{k_{\rm phys}}\right)^4 \left\langle \frac{\lambda}{2(1+n)}  \,  \sin^2\varphi - \frac{\lambda^2}{8(1+n)^2} \,    \sin^2\left(2 \varphi \right) \right\rangle\\ \notag
&\simeq \; - \left( \frac{\sigma}{k_{\rm phys}} \right)^2 \left(\frac{H}{k_{\rm phys}}\right)^4 \left\langle \sin ^2\varphi  \cos\left(2 \varphi \right) \right\rangle \\ \label{eq:lnnmapp}
&\simeq \; \frac{6}{\mathcal{N}_s}\frac{\sin^2\delta}{\delta^2}\,.
\end{align}
Far outside the horizon $\delta\ll1$. Additionally, Eq.~(\ref{eq:chisuph2_m}) implies that
\beq
\partial_{Ht}\langle\ln|\chi|^2\rangle  \;\simeq\; \partial_{Ht}\langle \ln(n)\rangle + \partial_{Ht}\langle \ln|\varphi|^2\rangle + 2  \,.
\eeq
The approximation (\ref{eq:latewc_m}) yields $\langle \ln|\varphi|^2\rangle\simeq 2\ln\delta + 2$, which in turn implies that $\partial_{Ht}\langle \ln|\varphi|^2\rangle \simeq -6$. Together with (\ref{eq:lnnmapp}), this signifies that for a massless spectator field, in the super-horizon {\em and} weak-scattering limits,
\begin{align}
\partial_{Ht} \langle \ln(1+n)\rangle \;&\simeq\; 6\,,\\
\partial_{Ht} \langle \ln|X|^2\rangle \;&\simeq\; 2\,,
\end{align}
in agreement with the numerical result shown in Fig.~\ref{fig:nfrates_m}. 

\section{Conclusions}\label{sec:SU}

In this paper we have studied the non-adiabatic, stochastic production of particles due to the excitation of a spectator field in a de Sitter (inflating) background. To accomplish this, we have extended the framework of~\cite{Amin:2015ftc,Amin:2017wvc}, which was valid in Minkowski space, to an expanding universe. 

To simplify the analysis and focus on the impact of expansion, we considered the case with spectator fields being non-adiabatically excited by of a series of delta-function scatterers with random strengths.\footnote{We note that the formalism is general enough to accommodate any expansion history and different types of (localized in time) non-adiabatic interactions.}
We carried out detailed numerical calculations using the transfer matrix approach for a wide range of parameters (including strong and weak scattering), and we explored both the individual and the statistical properties of quantities related to particle production. We developed a Fokker-Planck equation to analytically understand the results of our numerical simulations in the weak scattering limit. In the limits tested we found excellent agreement between the analytical and numerical results, both in the sub- and super-horizon limits. While the sub-horizon behavior is consistent with previous work, the results for super-horizon behavior are new.

While we have already summarized our main results in Section~\ref{sec:Summary}, we re-emphasize three general results again: (1) In the limit of large number of scatterings per Hubble time $\mathcal{N}_s\gg 1$, a single parameter $\S$ determines most of the statistical properties of the spectator field outside the horizon. Here $\sigma$ characterizes the strength of the individual scatterers. (2) The field amplitude is log-normally distributed independent of the strength of scattering, the size of the physical momentum relative to the Hubble parameter, or the bare mass of the field. (3) The logarithm of the field amplitude satisfies the necessary properties for an approximately Wiener process outside the horizon.

We note that in Minkowski space, we had found a universal result that the occupation number is log-normally distributed (in the late time limit). Occupation numbers, however, lose their physical meaning outside the horizon. It is the field amplitude that connects more directly to observables, especially on super-horizon scales. 

In more detail, in the sufficiently sub-horizon regime  the evolution of the occupation number mimics the growth that would be observed in a Minkowski background, apart from the time-dependent rescaling $k/H \rightarrow k/aH$ (as expected). It is worth noting that this result is independent of the bare mass of the field when physical wavelength is sufficiently small inside the horizon. Since the growth rate for $ \ln(1+n)$ is now inversely proportional to the square of the physical momentum, we observe the expected exponential growth in the occupation number, albeit with $n\ll 1$ when $|k\tau|^2\gg \S$. Moreover, as $n$ is so small, the field decays on average as it would do in the absence of non-adiabatic events. 

Outside the horizon, the mass of the spectator scalar determines the form of the mode function, and therefore the growth rate for the occupation number and the field magnitude. Not surprisingly, when scattering is very weak, the field amplitude $|\chi|^2$ is impervious to the non-adiabatic excitation, decaying with the inverse of the scale factor in the conformal case or remaining frozen in the massless scenario. Nevertheless, it is not immediately obvious why this immunity to growth extends all the way to scattering strength parameters that are $\mathcal{O}(1)$, a domain far outside the reach of our Fokker-Planck formalism.

For strong scattering, field and occupation numbers grow exponentially with a rate dependent on $\S$. The dependence on $\S$ extends to $\M$-parameters $\phi$ and $\psi$. The phase $\phi$ approaches a double-delta function distribution with exponential speed, while the form of the distribution of $\psi$ depends strongly on $\mathcal{N}_s(\sigma/H)^2$, interpolating between a normal and a uniform distribution depending on the magnitude of this parameter. 

Our present work has been focused on describing stochastic particle production in an expanding universe; we have not addressed or calculated the observational implications here. We plan to make use of the results presented here to study the effects of a stochastically excited field acting as a source for curvature perturbations during inflation. Such sourcing (which can be taken as an indicator of a complex inflationary sector) could lead to features in the power spectrum \cite{Chluba:2015bqa}, as well as potentially universal scaling relations between higher point correlations functions \cite{Bartolo:2004if} of the curvature perturbations. The sourcing can also generate gravitational waves \cite{Cook:2011hg,Senatore:2011sp}. The stochastic framework can also be applied to the  early stages of non-perturbative reheating before non-linearity and thermalization take over. Finally, extending this framework to particle production in higher spin fields would be interesting to pursue. We postpone these studies for future publications.

\section*{Acknowledgements}
 We would like to thank Daniel Baumann, Horng-Sheng Chia (Amsterdam U.) for many stimulating and helpful discussions, Eva Silverstein (Stanford) for insightful discussions, and Jia-Liang Shen (Rice U.) for his involvement in calculating particle production in an expanding universe during the early stages of this project. Numerical results were obtained from a custom Fortran code utilizing the thread-safe arbitrary precision package MPFUN-For written by David H. Bailey.  MA and MG are supported by the US Dept.~of Energy grant DE-SC0018216. DG is supported by the US Dept.~of Energy grant DE-SC0019035.

\appendix
\section{Appendix}
\label{ap:nm}
\subsection{Typical vs.~average}\label{ap:typ}

As it is discussed in detail in Sections~\ref{sec:meanvarconf} and \ref{sec:meanvarconf_m}, the logarithms of the occupation number $n$ and the scalar field magnitude $|X|$ are characterized by rapidly growing variances both inside and outside the horizon. Such a large amount of dispersion could signify that the mean does not provide a good approximation to the behavior of a typical member of the ensemble. We address below these concerns.\\

\begin{figure}[t!]
\centering
    \includegraphics[width=0.8\textwidth]{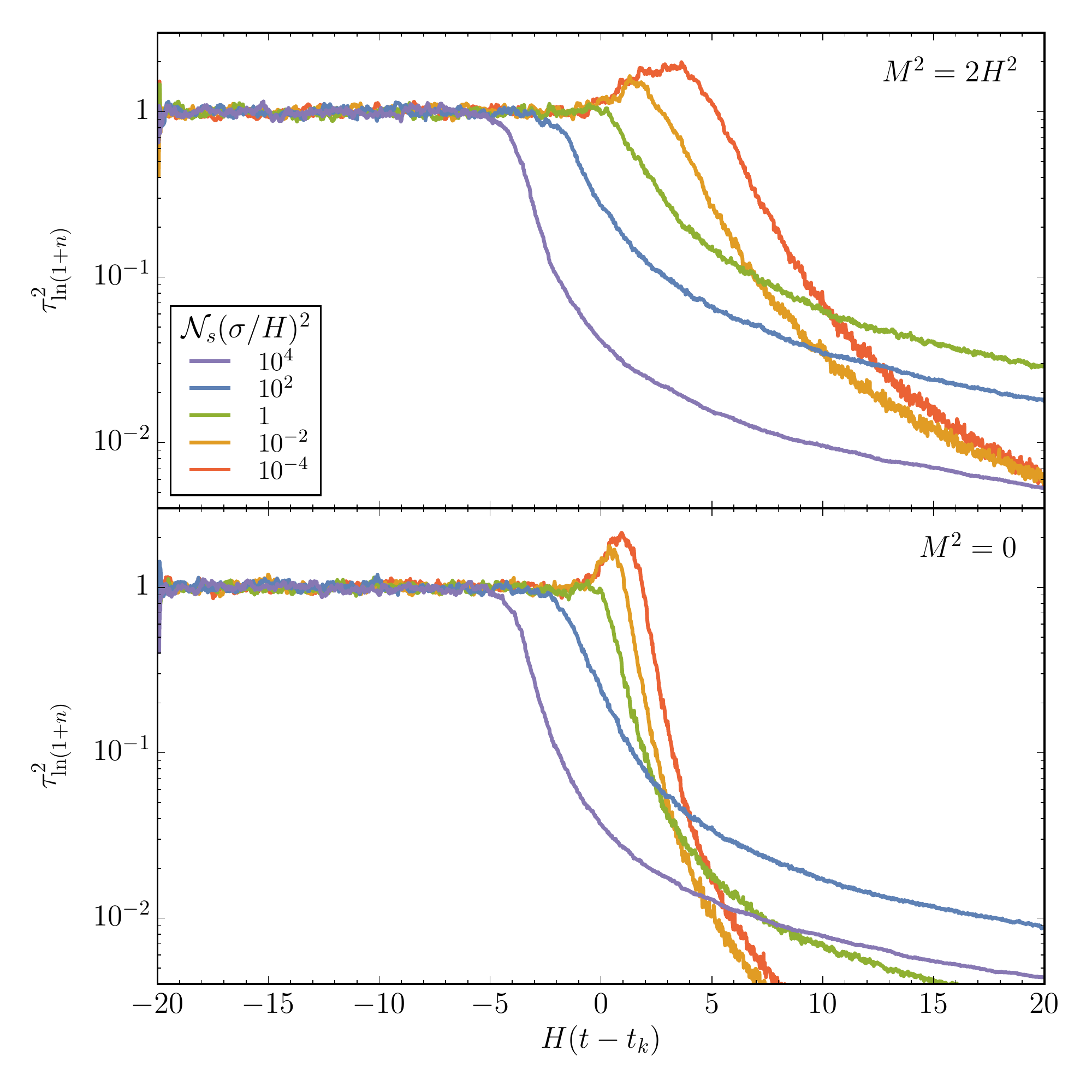}
    \caption{Square of the coefficient of variation as a function of time for the occupation number. Top: conformal case. Bottom: massless case. Scattering parameters are as in Figs.~\ref{fig:sigmeanvar} and \ref{fig:sigmeanvar_m}.}
    \label{fig:taun}
\end{figure}

In order to address the suitability of $\ln(1+n)$ as our variable of choice to describe the evolution of the occupation number over the ensemble of realizations, we consider the so-called coefficient of variation (or noise-to-signal ratio)
\beq
\tau_{\ln(1+n)} \;\equiv\; \frac{{\rm Var}\,[\ln(1+n)]^{1/2}}{\langle \ln(1+n)\rangle}\,,
\eeq
which determines the extent of variability in relation to the mean, assuming that the random variable in question is measured on a ratio scale (in this case $\ln(1+n)\geq 0$). The time dependence of $\tau_{\ln(1+n)}$ is shown in Fig.~\ref{fig:taun} for selected values of $\mathcal{N}_s(\sigma/H)^2$ for both the conformal (top) and massless (bottom) cases. In both scenarios, it is clear that, for any scattering strength, the coefficient of variation is approximately equal to one inside the horizon. This suggests an exponential distribution for $\ln(1+n)$, as confirmed in Sections~\ref{sec:pdfs} and \ref{sec:pdfs_m}. It is also worth noting a transient growth of $\tau_{\ln(1+n)}$ during horizon crossing for weak scattering, which is eventually overcome by the growth of the mean. Most importantly, in all cases we find that $\tau_{\ln(1+n)}$ decreases monotonically as a function of time far outside the horizon. The mean of $\ln(1 + n)$ is therefore a good measure of the number of particles produced in most regimes.\\

\begin{figure}[t!]
\centering
    \includegraphics[width=\textwidth]{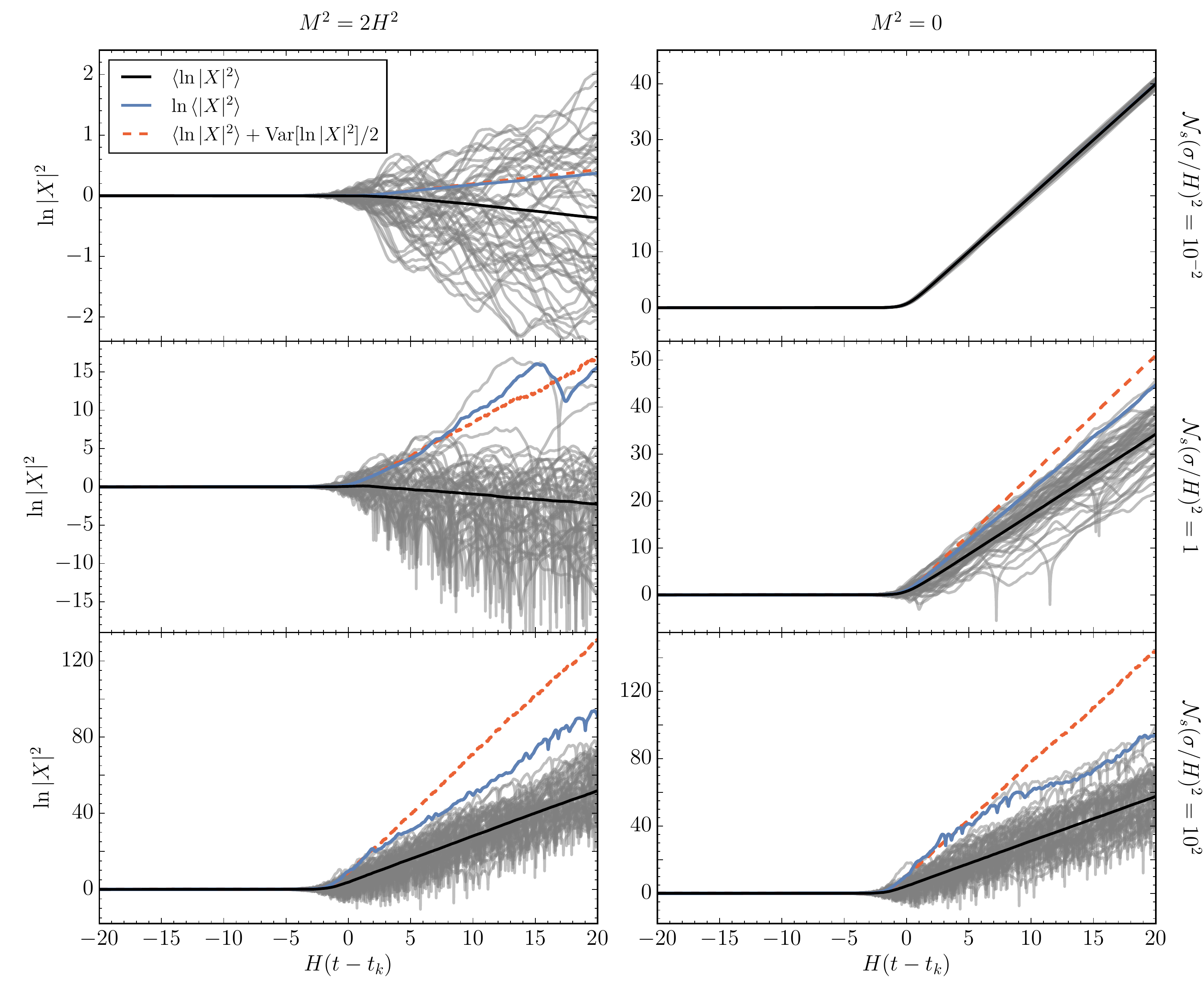}
    \caption{Time evolution of $\ln|X|^2$ for weak, moderate and strong scattering in the conformal and massless cases. Shown in gray are 50 representative trajectories, sampled from an ensemble of 5000 realizations of the scattering amplitudes and locations. Here $N_s=6000$. The black, continuous curve corresponds to the mean-of-the-log $\langle \ln|X|^2\rangle$, the typical (median) member of the ensemble. The blue, continuous curve is the log-of-the-mean $\ln\,\langle|X|^2\rangle$. Note that for moderate and strong scattering its value is dominated by the member of the ensemble with the largest values. The red, dashed curve is the logarithm of the right-hand side of (\ref{eq:logmean}), the infinite-ensemble limit of $\ln\,\langle|X|^2\rangle$. }
    \label{fig:trajsx}
\end{figure}

In the case of the scalar field magnitude, the coefficient of variation is not a suitable measure of the quality of $\langle \ln|X|^2\rangle$ as a descriptor of the behavior of a typical member of the ensemble, given that the value of this mean depends on the wavenumber $k$ and can be positive or negative. Nevertheless, a qualitative argument can be built by comparing the evolution in time of $\langle \ln|X|^2\rangle$ and several individual trajectories, for a given ensemble of amplitudes and locations of the non-adiabatic events. This comparison is made in Fig.~\ref{fig:trajsx}, where 50 individual trajectories, shown in gray, are displayed together with $\langle \ln|X|^2\rangle$, shown in black, for weak, moderate and strong scattering, and for both the conformal and massless cases. Also shown therein in blue is the logarithm of the expectation value of the squared field magnitude, $\ln\,\langle |X|^2\rangle$. As it is clear, the mean-of-the-log is in all cases a better descriptor of the behavior of a typical member of the ensemble.

The previous result is consistent with the fact that  $\ln|X|^2$ is normally distributed, as we have determined in sections~\ref{sec:pdfs} and \ref{sec:pdfs_m}. The normal distribution has zero skewness, and for it all average quantifiers (mean, median and mode) coincide. In contrast, $|X|^2$ has a lognormal distribution, which becomes heavily skewed as ${\rm Var}\,[\ln|X|^2]$ grows with time (it grows a ``fat tail'' of improbable trajectories). Its median coincides with the typical value, $|X|^2_{\rm typ}=e^{\langle \ln|X|^2\rangle}$, while its mean is dominated by unlikely events for which $|X|^2$ is large, as it is clear in Fig.~\ref{fig:trajsx}. In the limit of an infinite number of realizations, this mean can be related to the moments of $\ln|X|^2$ as follows,
\beq\label{eq:logmean}
\lim_{N_r\rightarrow \infty} \langle |X|^2\rangle \;=\; e^{\langle \ln|X|^2\rangle+\frac{1}{2}{\rm Var}\,[\ln|X|^2]}\,.
\eeq
The right hand side of the previous equation is shown as the red, dashed curves in Fig.~\ref{fig:trajsx}. Clearly, as the variance grows, $\langle |X|^2\rangle$ becomes less reliable as an estimate for the most probable ensemble member. Nevertheless, for a finite-sized ensemble, it is a reliable bound for the value of $|X|^2$, up to very improbable outliers.

\subsection{Convergence tests}\label{ap:conv}

We demonstrate here that, given a sufficiently high density of scatterers $\mathcal{N}_s$, the super-horizon evolution of the transfer matrix parameters and the scalar field magnitude is controlled solely by the scattering strength parameter $\mathcal{N}_s(\sigma/H)^2$. We also show that this evolution is independent of the underlying distribution of scatterer locations and amplitudes, assuming that they are randomly drawn from their corresponding ensembles. Our discussion here is restricted to the case of a conformally massive spectator field, but the conclusions are equally applicable to the massless field case.

\begin{figure}[t!]
\centering
    \includegraphics[width=0.87\textwidth]{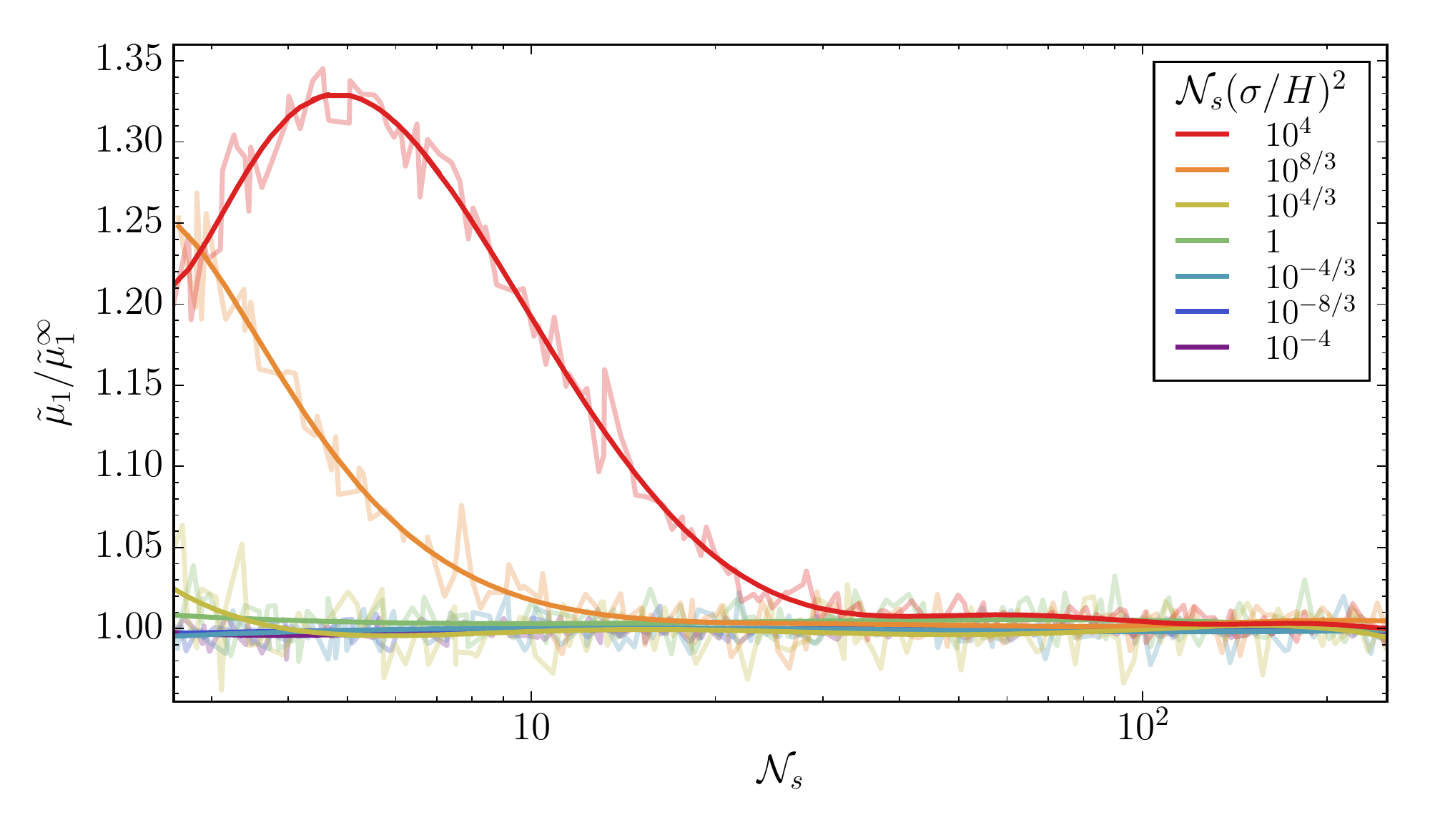}
    \caption{Dependence of the particle production rate on $\mathcal{N}_s$ for selected $\mathcal{N}_s(\sigma/H)^2$. Here $k/H=e^{20}$ over a total of 40 Hubble times. The plotted values correspond to the average of 400 realizations per value of $\mathcal{N}_s(\sigma/H)^2$ (transparent), further smoothed with a polynomial fit (solid). Amplitudes and locations are drawn from uniform distributions.}
    \label{fig:converg}
\end{figure}
Fig.~\ref{fig:converg} shows the dependence of the particle production rate $\tilde{\mu}_1 =  \partial_{Ht}\langle \ln(1+n)\rangle$ on the density of scatterers $\mathcal{N}_s$ for fixed values of $\mathcal{N}_s(\sigma/H)^2$. To construct this plot we have effectively varied $N_s$ from $10^2$ to $10^4$, which corresponds to varying $\mathcal{N}_s$ from 2.5 to 250, and we have plotted the ratio of the production rate to its asymptotic value corresponding to that with the largest $N_s$, $\tilde{\mu}_1^{\infty}\equiv \tilde{\mu}_1^{N_s=10^4}$. It is clear then that, for weak scattering, the rate is independent of both $\mathcal{N}_s$ and $\sigma^2$, since we found that in this regime $\tilde{\mu}_1 \simeq 2$. In the case of strong scattering, we observe a mild dependence on $\mathcal{N}_s$, which is enhanced as scattering becomes stronger. Nevertheless, for the values considered here, this deviation is at most of $\sim 33\%$, and it clearly decreases for $\mathcal{N}_s \gg 1$, implying a universal value of the rate for fixed $\mathcal{N}_s(\sigma/H)^2$. This justifies our parametrization.
\begin{figure}[t!]
\centering
    \includegraphics[width=0.87\textwidth]{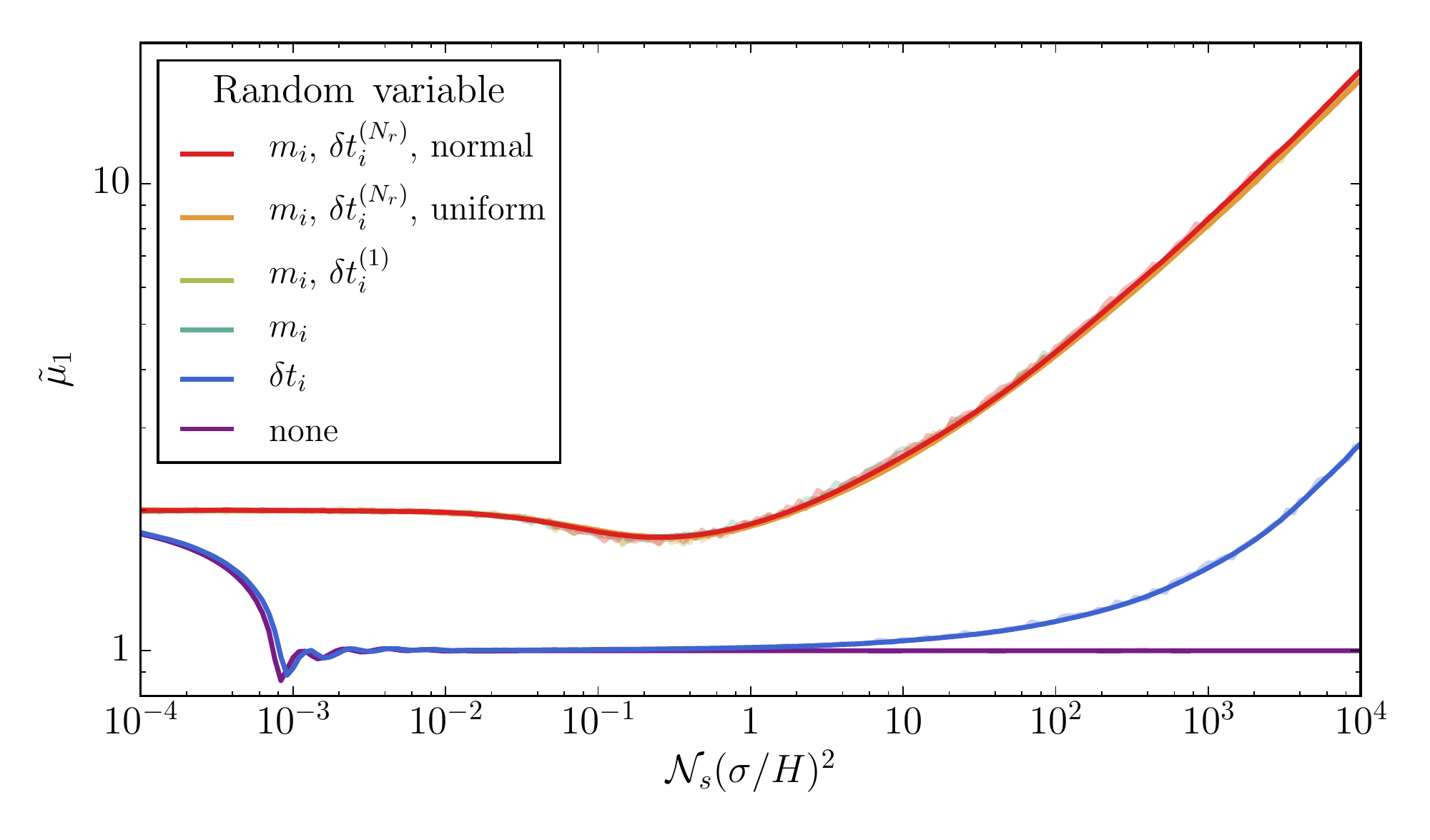}
    \caption{Dependence of the particle production rate on the statistics of $m_i$ and $\delta t_i$; distributions are assumed to be uniform in all cases, except for the red curve, for which $m_i$ and $\delta t_i$ have normal distributions. Here $\delta t^{(1)}$ denotes the case for which the same set of random locations is used for all the elements of the ensemble of realizations; $\delta t^{(N_r)}$ denotes that for which every realization has a different, random set of $\delta t_i$. Note that all cases for which $m_i$ vary randomly are indistinguishable. In order to scan over the scattering parameter we fix $\mathcal{N}_s\gg 1$ and vary $\sigma$. The plotted values correspond to the average of 200 realizations per value of $\mathcal{N}_s(\sigma/H)^2$ (transparent), further smoothed with a polynomial fit (solid).}
    \label{fig:scatdist}
\end{figure}\par\bigskip

The dependence of the particle production rate on the distribution of the scatterer locations and amplitudes is shown in Fig.~\ref{fig:scatdist}. To construct these results we have considered $N_r=200$ realizations per value of $\mathcal{N}_s(\sigma/H)^2$ for each of the following scenarios:
\begin{enumerate}[label=(\alph*)]
\item Neither $m_i$ or $\delta t_i$ are random (purple curve); their values are fixed to be equal to $+\sigma$ and $\langle \delta t_i\rangle /2$, respectively, of the corresponding random cases.
\item The amplitudes $m_i$ are fixed, while the locations $\delta t_i$ are random and uniformly distributed (blue curve). The set $\{\delta t_i\}$ is different between realizations for a given value of $\mathcal{N}_s(\sigma/H)^2$.
\item The amplitudes are random and uniformly distributed, while the locations are fixed and equispaced (teal curve). In this and all following cases, the ensemble of realizations is built out of $N_r$ distinct sets of $m_i$.
\item Both $m_i$ and $\delta t_i$ are uniformly distributed, but the ensemble of realizations is built out of a single set $\{ \delta t_i\}$, denoted as $\delta t_i^{(1)}$, for any given $\mathcal{N}_s(\sigma/H)^2$ (light green curve). This scenario is advantageous, as it guarantees that all realizations can be averaged at any given time step.
\item Both $m_i$ and $\delta t_i$ are uniformly distributed, and for each realization a different, random set $\{\delta t_i\}$ is considered; we denote this ensemble of $N_r$ different sets by $\delta t_i^{(N_r)}$ (orange curve). This scenario represents more faithfully the random ensemble of realizations, at the cost of an underlying coarse graining in time to allow for the calculation of expectation values.
\item The amplitudes $m_i$ follow a normal distribution centered at zero with standard deviation $\sigma$. The $\delta t_i$ are also normally distributed, with means and variances that coincide with those for case (e). Here the set of locations $\{\delta t_i\}$ is different between realizations (red curve).
\end{enumerate}
From Fig.~\ref{fig:scatdist}, it is clear that the stochasticity of the scattering amplitudes $m_i$ is what determines the time evolution of the transfer matrix parameters, here illustrated for the production rate $\tilde{\mu}_1$. In the absence of stochasticity in the amplitudes (cases (a) and (b)), a structure reminiscent of a band is visible. This band structure disappears if the $m_i$ are drawn from probability distributions, and the functional dependence of $\tilde{\mu}_1$ on the scattering strength parameter is independent of the details of these underlying distributions (cases (c), (d), (e) and (f)). We have verified that these results are independent of the value of the wavenumber $k$.

\subsection{Backreaction}\label{app:back}

In Sections~\ref{sec:meanvarconf} and \ref{sec:meanvarconf_m} we have found that the rate of growth of the mean and the variance of the scalar field magnitude can be $\mathcal{O}(1)$ or larger for $\mathcal{N}_s(\sigma/H)^2\gtrsim 1$. This rapid growth will inevitably lead to a $\chi$-dominated universe at sufficiently late times. That is, the mean energy density of the field would be comparable or larger than that of the background. In this Appendix we determine when the backreaction from the excitation of $\chi$ cannot be ignored. 

In order to avoid the discontinuous jumps in the time derivative of the scalar field due to the junction condition (\ref{eq:junct2}), we consider the energy density $\rho_{\chi}$ {\it in-between} scatterings. We can then write in general the mean energy density between scatterings as follows,
\begin{align}\notag
\langle \rho_{\chi}\rangle \;&=\; \frac{1}{2} \left\langle \dot{\chi}^2 + \frac{(\nabla\chi)^2}{a^2} + 2H^2\chi^2  \right\rangle\\ \label{eq:rhochi}
&=\; \frac{H^4\tau^2}{2(2\pi)^3}\int d^3 {\bf k}\, \left\langle \left|X_k+\tau \frac{d X_k}{d\tau}\right|^2 + \left(|k\tau|^2 + \frac{M^2}{H^2} \right) |X_k|^2 \right\rangle\,.
\end{align}
Note that this quantity depends on the mean squared-magnitude of the field, $\langle |X_k|^2\rangle$, and not on its ``typical'' (median) value, $\exp\langle \ln|X_k|^2\rangle$. In the present context, this is a sensible result: for a given ensemble of realizations, the average energy density will be dominated by those for which $|X_k|^2$ is large. Although the median would still provide a valuable measure of the energy density of the most probable member of the ensemble, for half of the members of the ensemble it will underestimate the onset of the backreaction regime. 

Note that sufficiently deep inside the horizon, (\ref{eq:rhochi}) will approximately lead to the (UV-divergent) energy density of the free scalar field. We can therefore restrict our calculation to modes that have left the horizon at the given (conformal) time $\tau$. In order to enforce this distinction, we will consider the ``adiabatically regulated'' energy density $\Delta \langle \rho_{\chi}\rangle$, which we define as
\beq\label{eq:deltarho}
\Delta \langle \rho_{\chi}\rangle \;\equiv\; \langle \rho_{\chi}\rangle - \rho_{\chi}^{\rm free}\,.
\eeq
We now proceed to evaluate this regulated energy density for the conformally massive and massless cases. The backreaction regime will then be defined as that for which this mean energy density becomes comparable to the background, $\Delta \langle \rho_{\chi}\rangle \simeq 3 M_P^2 H^2$.

\subsubsection{Conformally massive field}

In light of our definition (\ref{eq:deltarho}), we can immediately disregard the contribution to the mean energy density from sufficiently deep sub-horizon modes. More specifically, Eq.~(\ref{eq:subhlnch}) shows that the mean of $\ln|X_k|^2$ is always well approximated by its free, vacuum value, while (\ref{eq:subhvarlnch}) implies that the variance is $\ll 1$ provided that
\beq
|k\tau|^{2} \;\gg\; \mathcal{N}_s\left(\frac{\sigma}{H}\right)^2  \,.
\eeq
In order to simplify our following estimate of $\langle \rho_{\chi}\rangle$, we will assume that the deviation of the mean from its free value and the magnitude of the variance can be disregarded until horizon crossing, $|k\tau|\simeq 1$. Note that this in general will somewhat overestimate the computed energy density in the case of weak scattering, while it may underestimate it for strong scattering.

Let us now note that, in between scatterings, Eq.~(\ref{eq:genchi}) implies that for super-horizon modes with $n\gg1$,
\beq\label{eq:Xdertco}
X_k + \tau \frac{d X_k}{d\tau} \;\simeq\; X_k\,,
\eeq
up to a $\mathcal{O}(|k\tau|^2)\ll 1$ correction. This allows us to write the regulated energy density as
\beq\label{eq:drhoconf}
\Delta \langle\rho_{\chi}\rangle \;\simeq\; \frac{3 H^4\tau^2}{4 \pi^2}\int_{0}^{-\tau^{-1}} dk\,k^2\, \left[\langle |X_k|^2 \rangle - |X_k^{\rm free}|^2 \right]\,.
\eeq

In the strong scattering regime, the scalar field grows exponentially fast outside the horizon with a rate that can even exceed that determined by the scale factor. It is therefore natural to expect that the backreaction regime will be reached within a few $e$-folds of inflation if $\mathcal{N}_s(\sigma/H)^2$ is sufficiently large. Hence, we now assume that scatterings have not taken place for an arbitrarily long period of time but instead started when $\tau=\tau_0$. When this is the case, the form of $\langle |X_k|^2 \rangle$ will depend on the moment in which the mode $k$ leaves the horizon. If the mode leaves the horizon during the non-adiabatic epoch, while scatterings are active ($|k\tau_0|>1$), we can write (see Eq.~(\ref{eq:logmean}))
\begin{flalign}
& \text{($|k\tau_0|> 1>|k\tau|$)} & \Cen{3}{
\begin{aligned}
\langle |X_k(\tau)|^2 \rangle \;&\simeq\; |X_k(\tau_k)|^2 e^{(\mu_1 + \frac{1}{2} \mu_2)H(t-t_k)}\\
&\simeq\; \frac{1}{2k}\left(\frac{-1}{k\tau}\right)^{\mu_1 + \frac{1}{2} \mu_2}\,,
\end{aligned}}      &&  
\end{flalign}
where we have used the lognormality of $|X_k|^2$, $\langle |X_k|^2 \rangle=e^{\langle \ln|X_k|^2\rangle + \frac{1}{2}\langle Z_k^2\rangle}$. In passing, note that this implies that the ``typical'' energy density can be recovered from our expressions by formally taking the limit $\mu_2\rightarrow 0$. If instead the mode with wavenumber $k$ crosses outside the horizon before scatterings are active, we have 
\begin{flalign}
& \text{($|k\tau_0|<1$)} & \Cen{3}{
\begin{aligned}
\langle |X_k(\tau)|^2 \rangle \;&\simeq\; |X_k(\tau_0)|^2 e^{(\mu_1 + \frac{1}{2} \mu_2)H(t-t_0)}\\
&\simeq\; \frac{1}{2k}\left(\frac{\tau_0}{\tau}\right)^{\mu_1 + \frac{1}{2} \mu_2}\,.
\end{aligned}}      &&  
\end{flalign}
Combining these results we can rewrite (\ref{eq:drhoconf}) as
\begin{align} \notag
\Delta \langle\rho_{\chi}\rangle \;&\simeq\; \frac{3 H^4\tau^2}{8 \pi^2} \Bigg[ \left(\frac{\tau_0}{\tau}\right)^{\mu_1 + \frac{1}{2} \mu_2} \int_{0}^{-\tau_0^{-1}} dk\,k +  \int_{-\tau_0^{-1}}^{-\tau^{-1}} dk\,k\, \left(\frac{-1}{k\tau}\right)^{\mu_1 + \frac{1}{2} \mu_2} -  \int_{0}^{-\tau^{-1}} dk\,k \Bigg]\\ \label{eq:rhoinconf}
&\simeq\; \frac{3 H^4}{16 \pi^2}  \left(\frac{\mu_1+\frac{1}{2}\mu_2}{\mu_1+\frac{1}{2}\mu_2-2}\right)\Bigg[ \left(\frac{\tau_0}{\tau}\right)^{\mu_1 + \frac{1}{2} \mu_2 - 2 } -1 \Bigg]\,.
\end{align}
Therefore, the mean energy density in $\chi$ will become comparable to the background if, during scatterings, $\Delta\langle \rho_{\chi}\rangle \simeq 3 M_P^2 H^2$, or equivalently, if
\beq
N_{e}(\tau) \;\equiv\; \ln\left(\frac{\tau_0}{\tau}\right) \;\simeq\; \frac{1}{\mu_1 + \frac{1}{2} \mu_2 - 2} \ln\left[1 + 16 \pi^2\left(\frac{\mu_1+\frac{1}{2}\mu_2-2}{\mu_1+\frac{1}{2}\mu_2}\right) \frac{ M_P^2 }{H^2}\right]\,,
\eeq
where $N_e(\tau)$ denotes the number of e-folds between $\tau_0$ and $\tau$. \par\medskip

\begin{figure}[!t] 
   \centering
   \includegraphics[width=0.75\textwidth]{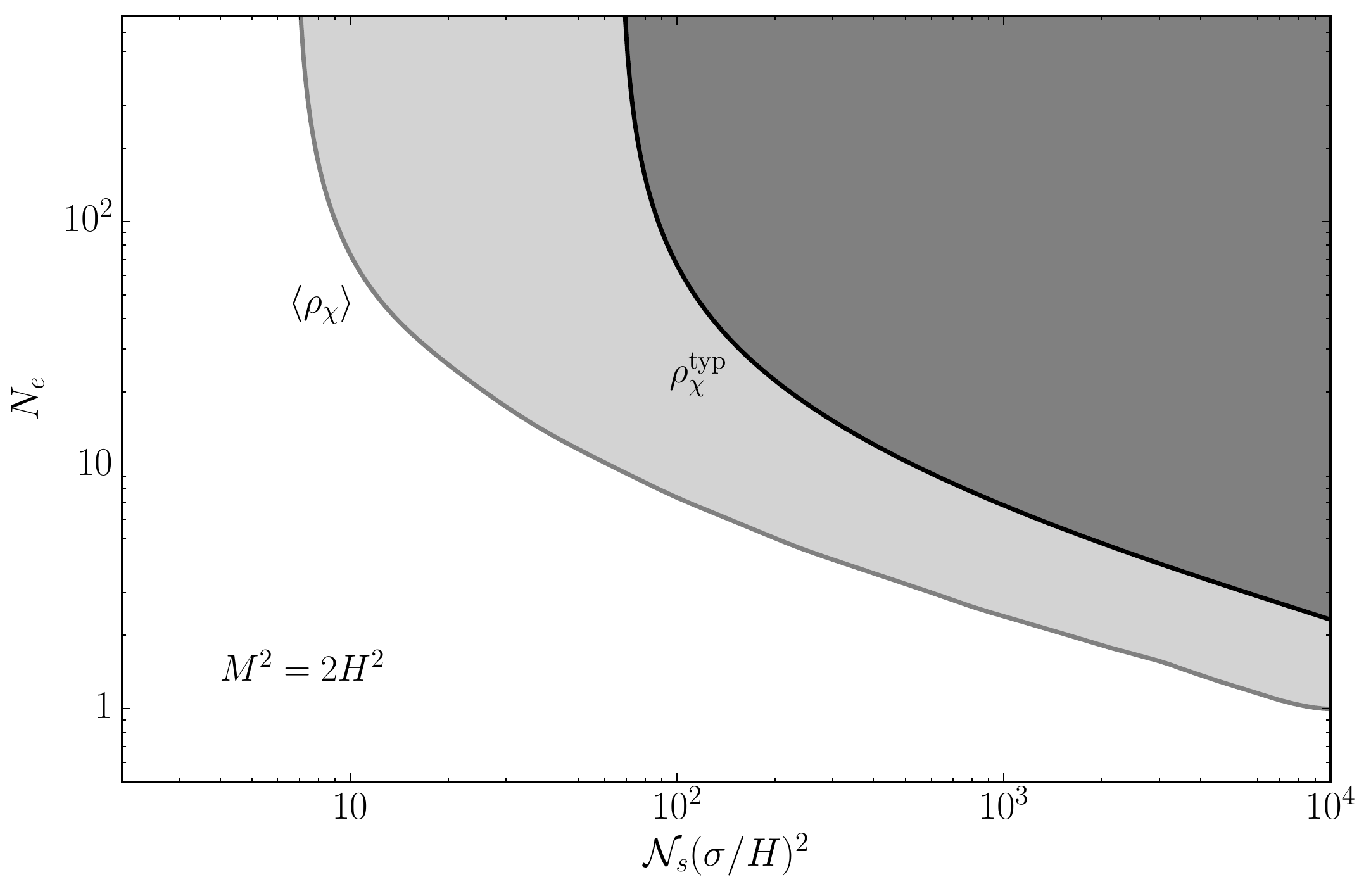} 
   \caption{Backreaction constraints for the conformally massive spectator field.  The horizontal axis corresponds to the scattering strength parameter, while the vertical axis denotes the number of $e$-folds from the beginning of scatterings. The light gray exclusion region is determined by the constraint on the average energy density of $\chi$, $\langle \rho_{\chi}\rangle$. For a finite-sized ensemble, this average is dominated by the member with the largest energy density, see Fig.~\ref{fig:trajsx}. The dark gray region is excluded by the $\mu_2\rightarrow 0$ limit of the mean energy density, denoted by $\rho_{\chi}^{\rm typ}$, which corresponds to the energy density of the typical (median) member of the ensemble. Here $H=10^{13}\,{\rm GeV}$.}
   \label{fig:br_curves_conf}
\end{figure}

Fig.~\ref{fig:br_curves_conf} shows the region of parameter space excluded by the backreaction constraint $\langle\rho_\chi\rangle \lesssim 3H^2 M_P^2$, for $H=10^{13}\,{\rm GeV}$ (light gray). The boundary curve has been constructed from the numerical results shown in Fig.~\ref{fig:nfrates}. Note that for $\mathcal{N}_s(\sigma/H)^2\lesssim 6.8$ the duration of the non-adiabatic epoch is not bounded by backreaction. For strong scattering, the bound depends on the scattering strength parameter, decreasing from $N_e\lesssim 10^2$ for $\mathcal{N}_s(\sigma/H)^2 \sim 10$ to $N_e\lesssim 1$ for $\mathcal{N}_s(\sigma/H)^2 \sim 10^4$. 

Also shown shaded in the figure in dark gray is the backreaction constraint on the typical member of the ensemble. As we discussed above, it may be obtained as the $\mu_2\rightarrow 0$ limit of the mean energy density, and we have denoted it by $\rho_{\chi}^{\rm typ}$. Note that for it, as expected, the constraint is milder. It is inexistent for $\mathcal{N}_s(\sigma/H)^2\lesssim 67$ and becomes dependent on the scattering strength parameter in an almost parallel way to the constraint on $\langle\rho_{\chi}\rangle$, for strong scattering. Also note in this case that $N_e\lesssim 2$ for $\mathcal{N}_s(\sigma/H)^2 \sim 10^4$.

\subsubsection{Massless field}

Let us now consider the massless spectator field. Similarly to the conformal case, we will assume that the free field solution is a valid approximation to the dynamics of the excited field up to horizon crossing. From Eq.~(\ref{eq:genchi_m}) we obtain that outside the horizon, with $n\gg 1$,\footnote{Note that (\ref{eq:Xdertco}) and (\ref{eq:Xdertnom}) imply that 
\beq
\frac{d \chi_k}{d\tau} \;\simeq\; \frac{\chi_k}{\tau} \times \begin{cases}
1\,, & M^2=2H^2\,,\\
|k\tau|^2\,, \quad & M^2=0\,.
\end{cases}
\eeq }
\beq\label{eq:Xdertnom}
X_k + \tau \frac{d X_k}{d\tau} \;\simeq\; |k\tau|^2X_k\,.
\eeq
The mean energy density takes then the approximate form
\beq\label{eq:drhonom}
\Delta \langle\rho_{\chi}\rangle \;\simeq\; \frac{H^4\tau^2}{4 \pi^2}\int_{0}^{-\tau^{-1}} dk\,k^2\, \left[|k\tau|^2\langle |X_k|^2 \rangle - \frac{1}{2k} \right]\,,
\eeq
where we have approximated the free mode function outside the horizon.

Following the same arguments as for the conformal case, for a mode $k$ that leaves the horizon when scatterings are active, we can immediately write
\begin{flalign}\label{eq:xktt}
& \text{($|k\tau_0|> 1>|k\tau|$)} & \Cen{3}{
\begin{aligned}
\langle |X_k(\tau)|^2 \rangle \;&\simeq\; |X_k(\tau_k)|^2 e^{(\mu_1 + \frac{1}{2} \mu_2)H(t-t_k)}\\
&\simeq\; \frac{1}{2k}\left(\frac{-1}{k\tau}\right)^{\mu_1 + \frac{1}{2} \mu_2}\,,
\end{aligned}}      &&  
\end{flalign}
where $\mu_{1,2}$ are now given by their massless values, shown in Fig.~\ref{fig:nfrates_m}. If instead the mode crosses the horizon before scatterings are active, we now have
\begin{flalign}\label{eq:xkto}
& \text{($|k\tau_0|<1$)} & \Cen{3}{
\begin{aligned}
\langle |X_k(\tau)|^2 \rangle \;&\simeq\; |X_k(\tau_0)|^2 e^{(\mu_1 + \frac{1}{2} \mu_2)H(t-t_0)}\\
&\simeq\; \frac{1}{2k}(k\tau_0)^{-2}\left(\frac{\tau_0}{\tau}\right)^{\mu_1 + \frac{1}{2} \mu_2}\,.
\end{aligned}}      &&  
\end{flalign}
Substitution of (\ref{eq:xktt}) and (\ref{eq:xkto}) into (\ref{eq:drhonom}) leads to 
\begin{align} \notag
\Delta \langle\rho_{\chi}\rangle \;&\simeq\; \frac{H^4\tau^2}{8 \pi^2} \Bigg\{ \left(\frac{\tau_0}{\tau}\right)^{\mu_1 + \frac{1}{2} \mu_2 - 2} \int_{0}^{-\tau_0^{-1}} dk\,k + \int_{-\tau_0^{-1}}^{-\tau^{-1}} dk\,k\, \left(\frac{-1}{k\tau}\right)^{\mu_1 + \frac{1}{2} \mu_2 - 2} - \int_{0}^{-\tau^{-1}} dk\,k\Bigg\}\\
&\simeq\; \frac{H^4}{16 \pi^2}  \left(\frac{\mu_1+\frac{1}{2}\mu_2 -2 }{\mu_1+\frac{1}{2}\mu_2-4}\right)\Bigg[ \left(\frac{\tau_0}{\tau}\right)^{\mu_1 + \frac{1}{2} \mu_2 - 4 } -1 \Bigg]\,.
\end{align}
Note the similarity between this result and (\ref{eq:rhoinconf}), with the identification $\mu_1\rightarrow \mu_1-2$. We can therefore immediately write the number of $e$-folds that the excitation of the massless field $\chi$ can last before its mean energy density becomes comparable to that of the background:
\beq
N_{e}(\tau)\;\simeq\; \frac{1}{\mu_1 + \frac{1}{2} \mu_2 - 4} \ln\left[1 + 16 \pi^2\left(\frac{\mu_1+\frac{1}{2}\mu_2-4}{\mu_1+\frac{1}{2}\mu_2 - 2}\right) \frac{ M_P^2 }{H^2}\right]\,.
\eeq

\begin{figure}[!t] 
   \centering
   \includegraphics[width=0.75\textwidth]{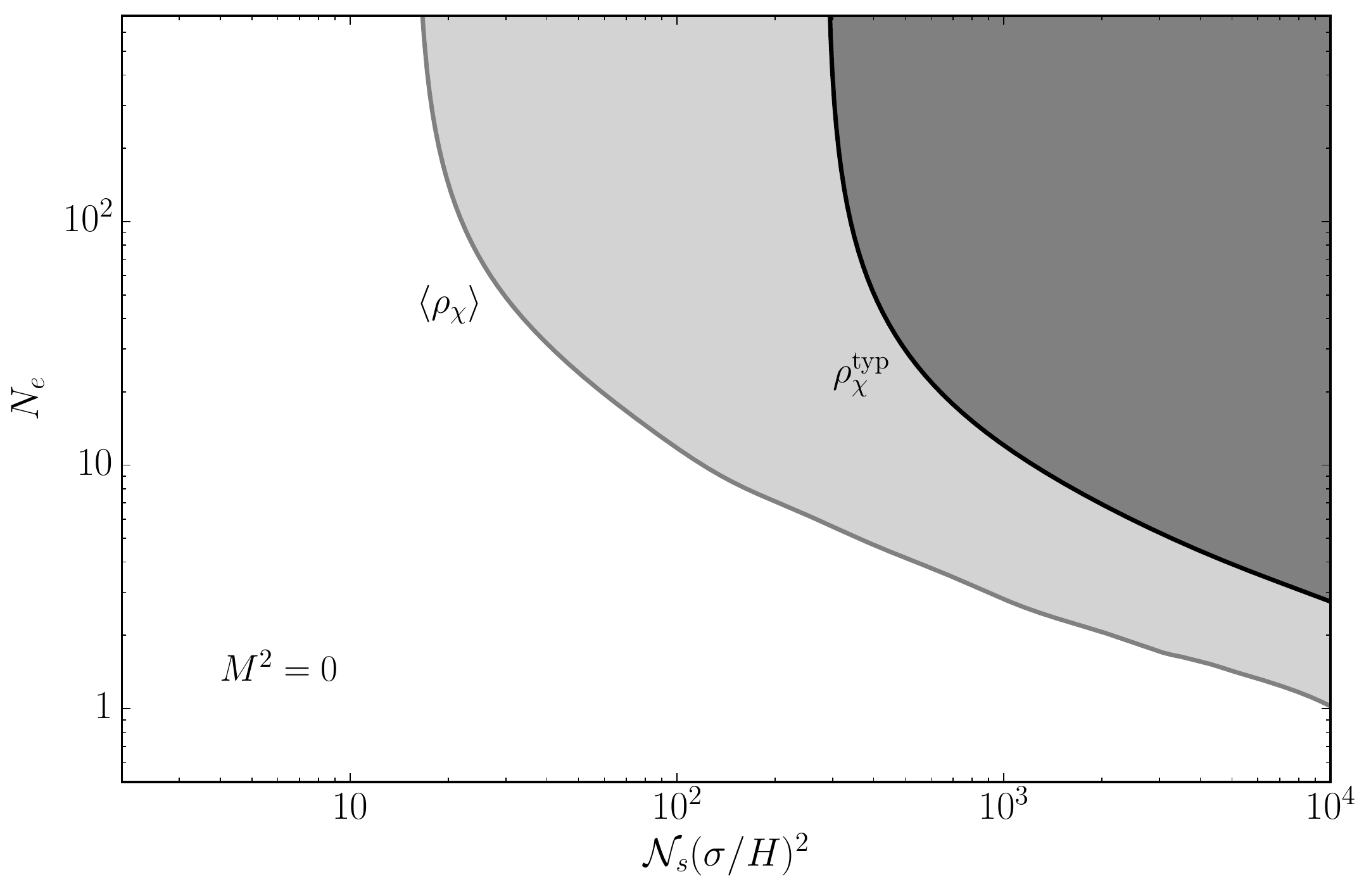} 
   \caption{Backreaction constraints for the massless spectator field.  The horizontal axis corresponds to the scattering strength parameter, while the vertical axis denotes the number of $e$-folds from the beginning of scatterings. For a finite-sized ensemble, this average is dominated by the member with the largest energy density, see Fig.~\ref{fig:trajsx}. The dark gray region is excluded by the $\mu_2\rightarrow 0$ limit of the mean energy density, denoted by $\rho_{\chi}^{\rm typ}$, which corresponds to the energy density of the typical (median) member of the ensemble. Here $H=10^{13}\,{\rm GeV}$.}
   \label{fig:br_curves_mass}
\end{figure}

Fig.~\ref{fig:br_curves_mass} shows the exclusion regions determined by $\langle \rho_{\chi}\rangle$ (light gray) and $\rho_{\chi}^{\rm typ}$ (dark gray) as functions of $N_e$ and the scattering strength parameter. Notice that, unlike the conformal case, the number of $e$-folds is unconstrained by $\langle \rho_{\chi}\rangle$ for $\mathcal{N}_s(\sigma/H)^2 \lesssim 16$, and by $\rho_{\chi}^{\rm typ}$ for $\mathcal{N}_s(\sigma/H)^2 \lesssim 287$. Nevertheless, the boundary contours are steep functions of the scattering strength parameter and, similarly to the conformal case, only $N_e\sim \mathcal{O}(1)$ is allowed for the largest value of the scattering strength considered, $\mathcal{N}_s(\sigma/H)^2 \sim 10^4$.

\addcontentsline{toc}{section}{References}
\bibliographystyle{utphys}
\bibliography{Refs} 

\end{document}